\documentclass[onecolumn,10pt,final]{asme2ej}
\pdfoutput=1
\usepackage{graphicx} %% for loading jpg figures
\usepackage{hyperref}   % to set up hyperlinks
\hypersetup{
	colorlinks=true,
	linkcolor=blue,
	citecolor=blue,
	urlcolor=blue,
}

\let\minute\undefined
\let\hour\undefined
\usepackage[square,numbers]{natbib}

\title{An ASME Journal Article Created Using 
\LaTeX2\raisebox{-.3ex}{$\epsilon$}
in ASME Format for Testing Your Figures}

\usepackage[utf8]{inputenc}
\usepackage{textcomp}

\usepackage{amsmath}
\usepackage{longtable,tabularx}
\input{settings}
\title{Pressure gain combustion for gas turbines: Analysis of a fully coupled engine model}
\author{Rupert Klein
    \affiliation{
	Mathematik \& Informatik\\
	Freie Universit\"at Berlin\\ 
	Arnimallee 6\\
	14195 Berlin, Germany\\
    Email: rupert.klein@math.fu-berlin.de
    }	
}

\author{Maikel Nadolski
    \affiliation{
	Mathematik \& Informatik\\
	Freie Universit\"at Berlin\\ 
	Arnimallee 6\\
	14195 Berlin, Germany\\
    Email: guttula@zedat.fu-berlin.de
    }	
}

\author{Christian Zenker
    \affiliation{
	Fakultät 1, Institut für Mathematik\\
	Brandenburgische Technische Universit\"at\\ Cottbus-Senftenberg\\ 
	Platz der Deutschen Einheit 1\\
	03046 Cottbus, Germany\\
    Email: Christian.Zenker@b-tu.de
    }	
}

\author{Michael Oevermann
    \affiliation{
	Fakultät 1, Institut für Mathematik\\
	Brandenburgische Technische Universit\"at\\ Cottbus-Senftenberg\\ 
	Platz der Deutschen Einheit 1\\
	03046 Cottbus, Germany\\
    Email: Michael.Oevermann@b-tu.de
    }	
}

\author{Christian Oliver Paschereit
    \affiliation{
	Chair of Fluid Dynamics\\
	Hermann-Föttinger-Institut (HFI)\\
	Technische Universit\"at Berlin\\ 
	M\"uller-Breslau Str.\ 8\\
	10623 Berlin, Germany\\
    Email: oliver.paschereit@tu-berlin.de
    }	
}

\begin{document}

\maketitle

% !TEX root = main_ASME.tex

\begin{abstract}
{\itshape%
The ``Shockless Explosion Combustion" (SEC) concept for gas turbine combustors, introduced in 2014, approximates constant volume combustion (CVC) by harnessing acoustic confinement of autoigniting gas packets. The resulting pressure waves simultaneously transmit combustion energy to a turbine plenum and facilitate the combustor's recharging against an average pressure gain. Challenges in actualizing an SEC-driven gas turbine include i) the creation of charge stratifications for nearly homogeneous autoignition, ii) protecting the turbo components from combustion-induced pressure fluctuations, iii) providing evidence that efficiency gains comparable to those of CVC over deflagrative combustion can be realized, and iv) designing an effective one-way intake valve. This work addresses challenges i)-iii) utilizing computational engine models incorporating a quasi-one-dimensional combustor, zero- and two-dimensional compressor and turbine plena, and quasi-stationary turbo components. Two SEC operational modes are identified which fire at roughly one and two times the combustors' acoustic frequencies. 
Results for SEC-driven gas turbines with compressor pressure ratios of 6:1 and 20:1 reveal 1.5-fold mean pressure gains across the combustors. Assuming ideally efficient compressors and turbines, efficiency gains over engines with deflagration-based combustors of 30\% and 18\% are realized, respectively. With absolute values of 52\% and 66\%, the obtained efficiencies are close to the theoretical Humphrey cycle efficiencies of 54\% and 65\% for the mentioned pre-compression ratios. Detailed thermodynamic cycle analyses for individual gas parcels suggest that there is room for further efficiency gains through optimized plenum and combustor designs.%
}
\end{abstract}

%\begin{keywords}
%Gas dynamics, combustion
%Pressure gain combustion, quasi-1D reactive gasdynamics, acoustic resonance
%\end{keywords}

% !TEX root = main_ASME.tex

%----------------------------------------------------------------------------------------------------------------------------
\bigskip

\noindent
%----------------------------------------------------------------------------------------------------------------------------

\section{Introduction}
\label{intro}

The prospect of utilizing combustion cycles in gas turbines that involve significant pressure increases in the course of reaction progress has triggered intense research in the past decades, with renewed interest in recent years. From a theoretical thermodynamics perspective, efficiency gains in excess of 30\%\ are conceiveable \citep{HeiserPratt2002}. Tapping into a sizeable fraction of this potential would be highly beneficial in many respects, from economic to environmental. Combustor designs involving pulsating \citep{HeiserPratt2002,RoyEtAl2004,Wolanski2013} and rotating \citep{LuEtAl2011,LuBraun2014,AnandGutmark2019} detonation waves are under investigation worldwide, whereas subsonic near constant volume combustion (CVC) concepts that do not internally rely on deflagrative combustion, and hence near constant pressure, are as yet rare. The ``shockless explosion combustion'' (SEC) process discussed in this paper is of that type. 

\begin{figure*}
\begin{center}
\includegraphics[width=1.0\textwidth]{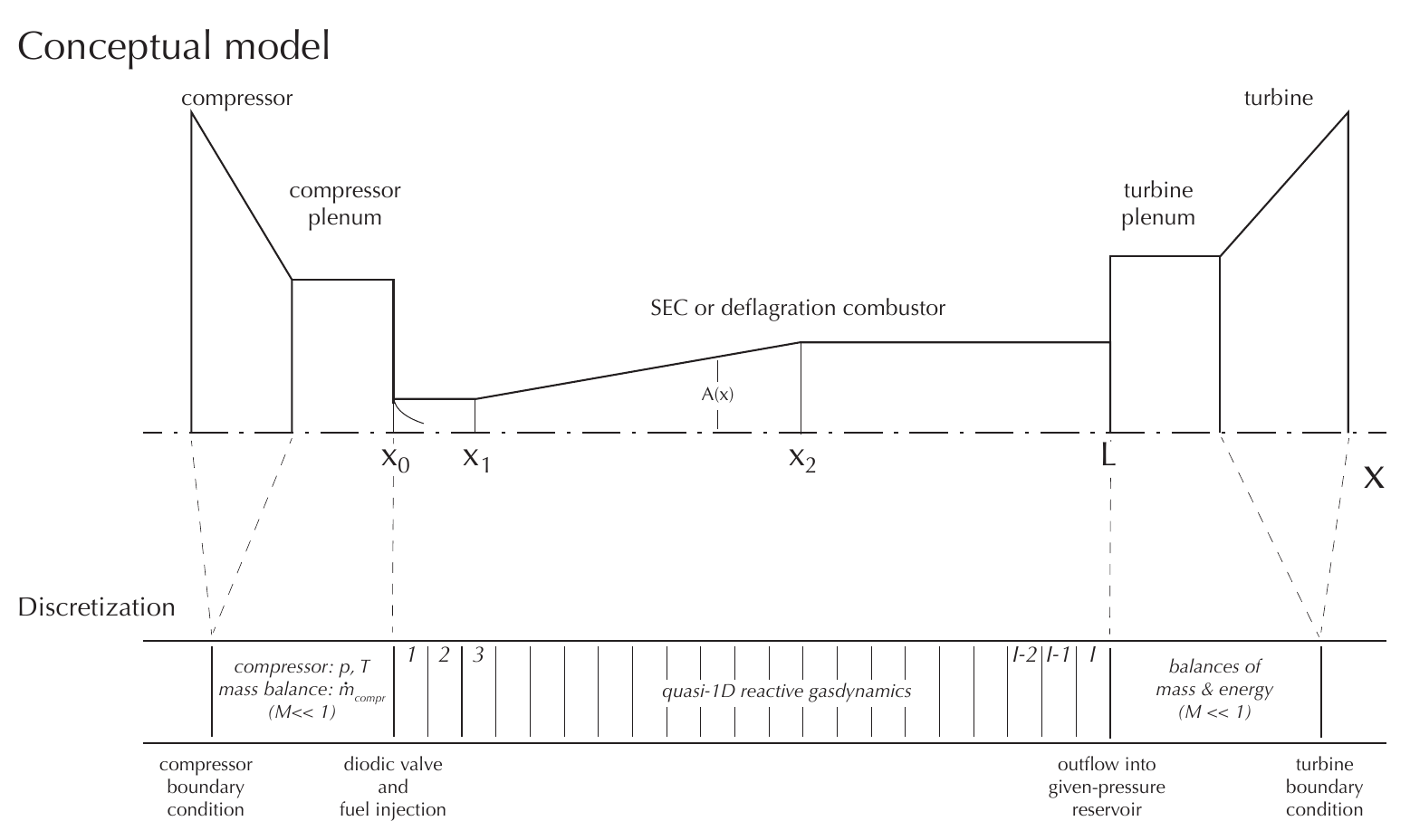}
\end{center}
\caption{Upper part: Design of the one-dimensional conceptual engine model, consisting of (from left to right): a quasi stationary compressor model; a compressor plenum represented as a single large control volume with a time dependent homogeneous mean thermodynamic state; the SEC combustor involving a unidirectional restrictor (diodic) valve, a short straight ignition section, a diffusor, and a final straight section; the turbine plenum, again with a time dependent mean state; and a quasi stationary turbine model. The computational model also includes the inertia of the rotor that mechanically couples turbine and compressor, and a power output controller. For comparison with classical gas turbines based on near constant pressure combustion, and to model idling and low-load operation conditions, the combustion section optionally represents quasi-steady deflagrative or pulsating SEC-combustion involving cyclic autoignitions. Lower part: Graphical representation of the spacial discretization used in the quasi one-dimensional computational implementation of the model.}
\label{fig:SystemDesign}
\end{figure*}

The SEC concept has been introduced by two of the authors and co-workers at the 24th ICDERS conference in 2013 \citep{BobuschEtAl2014}. It involves nearly homogeneous autoignition of limited-size premixed charge packets in a combustor that maintains large-amplitude resonant pressure waves as an important part of the design. Near constant volume combustion is achieved in this process by acoustic confinement of the autoigniting charge packets. A local pressure increase prior to the reaction progress is a primary desired effect as it increases the thermodynamic efficiency of the chemical energy conversion \citep{Nalim2002}. Such a post-compressor/pre-ignition pressure increase is achieved in the SEC. However, the large-amplitude pressure waves needed may be destructive when they impinge on the turbine entry unmediated. The conceptual engine design, presented here for the first time, addresses this challenge by exploiting the properties of resonant large-amplitude pressure waves in a combustor that consists of a short and narrow initial straight-tube section connected to the compressor plenum, a diffusor, and a wider second straight section which forms the turbine plenum entry, cf.\ Fig.\,\ref{fig:SystemDesign}. This design and mode of operation have three beneficial consequences: (i) Upon passage through the diffusor, the combustion-generated compression waves create strong backward propagating suction waves that enable reliable recharging even against a significant mean pressure increase along the combustor. (ii) Simultaneously, the compression wave amplitudes decrease upon downstream propagation in the diffusor, thereby significantly weakening the flow pulsations that eventually impinge upon the turbine entry. (iii) Backward propagating compression waves, created as a result of the combustor-turbine plenum interactions, amplify upon backward passage through the diffusor and significantly pre-compress the next cycle's fresh charge before it auto-ignites. This pre-compression beyond the compressor plenum level significantly contributes to the efficiency gain of the present SEC design over a classical gas turbine \citep{Stathopoulos:2018}. An illustration of the SEC combustor design is provided in Fig.\,\ref{fig:SystemDesign} as part of the full gas turbine model.

The structure and main results of this paper are as follows: Section~\ref{sec:EngineModel} provides a detailed description of the mathematical model representing the conceptual full-engine model of Fig.\,\ref{fig:SystemDesign} together with a summary of its numerical discretization. Section~\ref{sec:PerformanceResults} documents reference simulations obtained with the - thus far - most efficient model configurations, one with a compressor pressure ratio of $p_2/p_1 = 6:1$, one with a ratio of $p_2/p_1 = 20:1$. Within this modelling framework, the SEC-based engine outperforms the deflagration-based engine at the same compressor pressure ratio and with the same compressor and turbine characteristics with efficiency gains around $+20\%$ and with significantly increased power output at comparable turbine plenum mean temperatures. Section~\ref{sec:ThermodynamicCycle} provides thermodynamic cycle analyses from different perspectives. Close inspection of the flow states in the immediate vicinity of the combustor entry provides first thermodynamic estimates corroborating our explanation of the engine model's efficiency. That interpretation is based on a Humphrey cycle with two-stage precompression, the compression stage being realized by the turbo-compressor, the second by resonant pressure waves in the combustor. A more detailed Lagrangian analysis of the evolution of the thermodynamic states of individual gas parcels on their way through the system follows, and it compares the resulting effective thermodynamic cycle with the idealized  Brayton/Joule and Humphrey cycles. From our analysis, several key observations emerge: (1) The objective of achieving autoignition within an acoustically confined environment is largely met. (2) The combustor experiences notable increases in mean pressure as it diverges, and this is further accentuated by powerful pressure waves. These waves lead to a pronounced acoustic pre-compression of the reactants prior to their autoignition. (3) As a result, there is a time-averaged pressure rise across the SEC combustor by factors of 1.5 or more.

Our main conclusions, in the sense of a proof of concept by computational modelling, are:  
\begin{enumerate}
\item A ``shockless explosion combustor'' (SEC), as proposed originally by \citet{BobuschEtAl2014}, can be embedded in a complete gasturbine model built from standard turbo-components. 
\item Such an engine should deliver roughly the efficiency gains over classical gasturbines of 15\% to 30\% as are expected on theoretical grounds from the shift to constant volume combustion.
\item Appropriate geometrical designs of the combustor and of two plena that separate it from the turbo-components essentially shield the latter from excessive unsteady pressure fluctuations. 
\item How much of the efficiency gains predicted by our idealized computational model will be realizeable in the end will depend upon (i) the availability of a reliable and efficient uni-directional inlet valve, on (ii) a successful synchronization of acoustic and chemical time scales in the system, and on (iii) the losses from turbulent dissipation within the pulsating high-speed flow in the combustor.
\end{enumerate}

Section~\ref{sec:Conclusions} draws further conclusions and discusses next steps and future perspectives. 
 
% ===================================================================================
% ===================================================================================
% ===================================================================================

\section{Full-engine model and its computational representation}
\label{sec:EngineModel}

Refering to the quasi-one-dimensional gas turbine model in Fig.\,\ref{fig:SystemDesign}, its compressor and turbine components are represented using standard stationary turbo-machinery characteristics. These characteristics are  used to define boundary conditions for both the inflow into the compressor plenum and the outflow from the turbine plenum. For simplicity, we assume compressor and turbine to be on one shaft.

% ===================================================================================
% ===================================================================================

\subsection{Turbine}
\label{sec:TurbineModel}

For the turbine we assume operation at constant reduced mass flow rate, i.e.
\begin{equation}\label{eq:TurbineEquation}
  \frac{{\dot m_{\turb}} \, \sqrt{T_{\tilde{3}}}}{p_{\tilde{3}}} = C_{\turb} = \text{const.}\,,
\end{equation}
where ${\dot m_{\turb}}$ is the mass flow rate per unit area, and $p_{\tilde{3}}$ and $T_{\tilde{3}}$ are mean values of pressure and temperature in the turbine plenum, respectively, see section\,\ref{sec:ThermodynamicCycleNearTubeEntry} below for a detailed discussion. $C_{\turb}$ is a constant that depends on the detailed turbine design which will be chosen as needed in the present work. Essentially, the product $C_{\turb} A_{\turb}$, where $A_{\turb}$ is the turbine inlet crossection, corresponds to selecting a turbine that provides an appropriate power output for the purpose of the entire engine. A constant reduced mass flow rate is a legitimate assumption for sonic or near sonic flow conditions inside the turbine and allows to determine the mass flow rate from the turbine inlet conditions only.

Following standard thermodynamic modelling assumptions for quasi-stationary turbo component operation, the temperature ratio across the turbine is given by
\begin{equation}
  \label{eq:TurbineRatioT}
  \frac{T_4}{T_{\tildethree}}
   = 1 - \etaT \left(1 - \left(\frac{p_4}{p_{\tildethree}}\right)^{\frac{\kappa-1}{\kappa}} \right),
\end{equation}
where $\eta_{\turb}$ is the turbine's thermodynamic efficiency, index 4 denotes conditions at turbine outlet, and $\kappa = c_p/c_v$ is the working fluid's isentropic exponent with $c_p$ and $c_v$ the constant specific heat capacities at constant pressure and volume, respectively. We assume the turbine exit pressure to be equal to that of the environment, \ie,
\begin{equation}
   p_4 = p_1.
\end{equation}
The power generated by the turbine is determined from
\begin{equation}
  \label{eq:Pturb}
P_{\turb} 
   = \dot m_{\turb} A_{\turb} \, c_p (T_{\tildethree} - T_4)\,.
\end{equation}
%

% ===================================================================================
% ===================================================================================

\subsection{Compressor}
\label{sec:CompressorModel}

Since we wish to model not only stationary machine characteristics but also demonstrate that engine run-up from low speeds to full SEC operation can be realized straightforwardly within the framework of our model, we need to describe the coupling of engine shaft rotation rate and compressor compression ratio. This relation is modelled here qualitatively following \citet{CaresanaEtAl2014} (see their Fig.~1 and the middle panel of Fig.\,\ref{fig:Verdichterkennfeld} below). 

For the compressor we assume, accepting some approximations, that its rotor's rotational speed $n$ and pressure ratio $\Pi(n) = (p/p_1)(n)$ are related according to 
\begin{equation}\label{eq:Kompressordruck}
\begin{array}{c}
% \dss \frac{p_{2}(n)}{p_1} = 
\dss \Pi(n) 
= \Pibar + C_{\text{C}} \DelPi \frac{2}{\pi}\arctan\left(\pi \left[\frac{n}{n_0} - 1\right]\right) \,,
\\[10pt]
\dss \Pibar = \frac{\Pi_{\text{max}} + \Pi_{\text{min}}}{2}\,,
\quad
\DelPi = \frac{\Pi_{\text{max}} - \Pi_{\text{min}}}{2}\,,
  \\[10pt]
\dss \Pi_{\text{min}} = 1\,, 
\quad
\Pi_{\text{max}} = 50\,, 
\quad
C_{\text{C}} = \frac{22}{17} \,,
\end{array}
\end{equation}
cf.\ the dimensionless representation in the top panel of Fig.\,\ref{fig:Verdichterkennfeld}. The parameters have been fitted to qualitatively capture the fact that towards low and high rotation speeds the rate of change of the compression with variations of the shaft rotation rate is lower than it is in the middle range.

The middle panel of Fig.\,\ref{fig:Verdichterkennfeld} is modelled after Fig.\,1 of \citet{CaresanaEtAl2014}. The important observation from this graph is that the dependence of the compression ratio on the mass flux is weak so that in an unsteady situation the mass flux will quickly adjust to maintain the instantaneous pressure ratio prescribed by the compressor. The bounding lines to the left and right of the (nearly) horizontal characteristic curves indicate the surge (left) and choke (right) limits. When the mass flux reaches these lines, the flow through the compressor stalls and the system fails. A control mechanism that will prohibit such critical states is yet to be developed and implemented.

The diagram in the bottom panel of Fig.\,\ref{fig:Verdichterkennfeld} describes this, admittedly somewhat extreme, setting of an exactly massflux-independent pressure ratio. The horizontal lines in the graph correspond, from bottom to top, to increasing rotor revolution rates spanning the compressor's design range of operation.
\begin{figure}
\centering
\includegraphics[width=0.75\textwidth]{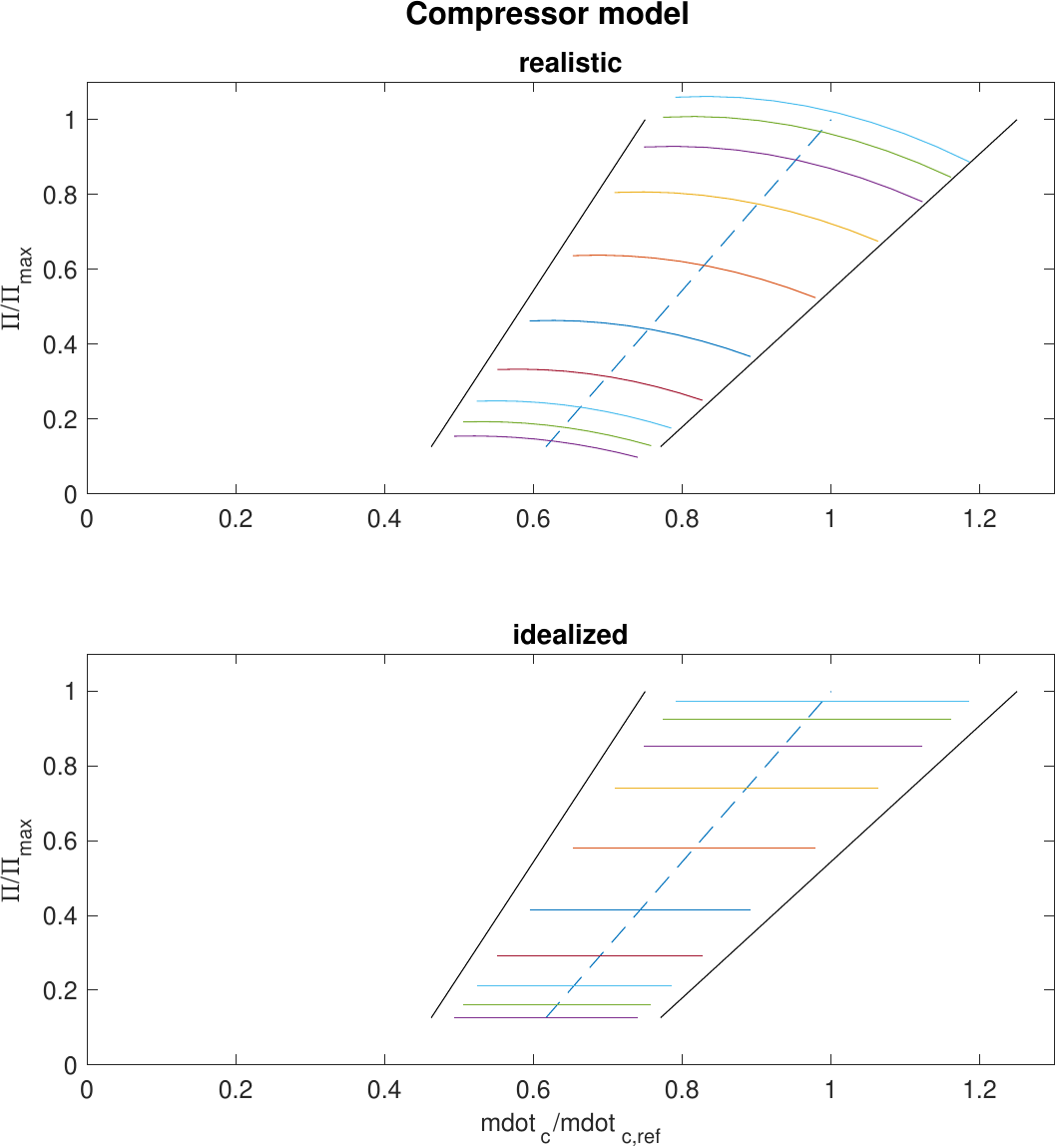}

\bigskip
\includegraphics[width=0.75\textwidth]{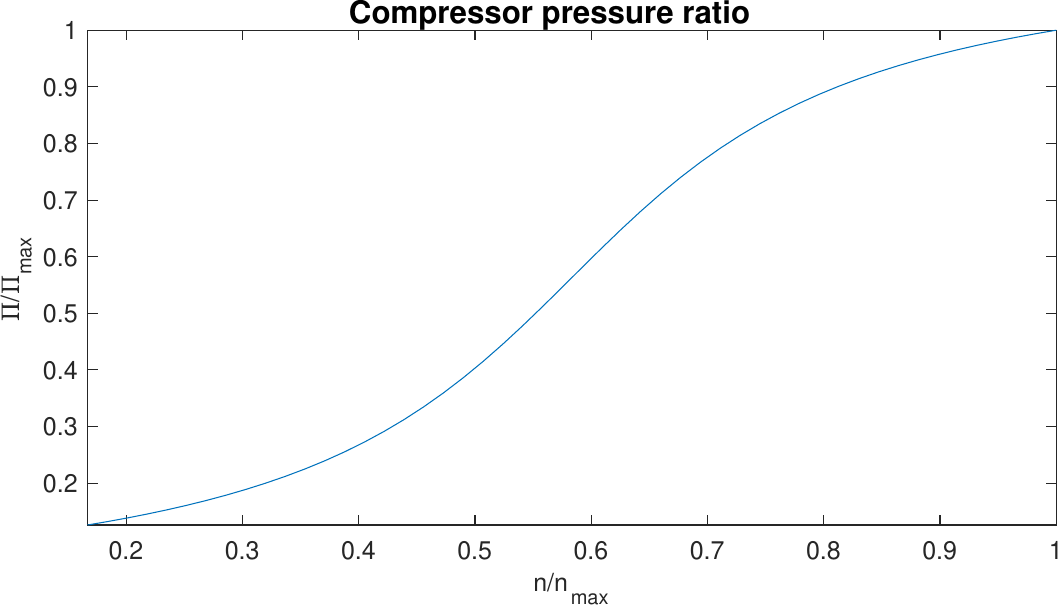}
\caption{Top: Compression ratio as a function of compressor rotation rate; qualitatively correct (middle) and idealized (bottom) compressor characteristics. For the two reference cases with compressor compression ratios of $\Pi = 6:1$ and $20:1$ we chose $\Pi_{\text{max}} = 1.25\, \Pi$, and minimum and maximum engine rotation rates of $n_{\text{min}} = 10\, 000$ and $n_{\text{max}} = 60\,000 \ \text{rpm}$.
\label{fig:Verdichterkennfeld}}
\end{figure}

Under the assumption that the compressor mass flux quickly adjusts as described above, the pressure in the compressor plenum is instantaneously set to the compression ratio prescribed by the shaft rotation rate. The temperature is then set to 
\begin{equation}\label{eq:Verdichtertemperatur}
\frac{T_2}{T_1} 
= 1 + \frac{1}{\eta_{\verd}} \left( \left(\frac{p_2}{p_1}\right)^{\frac{\kappa-1}{\kappa}} - 1\right)\,.
\end{equation}
This corresponds to a non-ideal compression with thermodynamic efficiency $\eta_{\verd}$. With pressure and temperature set, we obtain the density from the ideal gas equation of state, \emph{i.e.}, $\rho_2/\rho_1 = (p_2/p_1) / (T_2/T_1)$, and the total mass in the compressor plenum then follows from its volume. In the course of a simulation, this sets the mass flux $\dot m_{\verd} A_{\verd}$ that is assumed to enter the plenum from the compressor side in each time step once the new time level compressor pressure and the amount of mass that escaped into the combustor during the time step are known. In analogy with (\ref{eq:Pturb}), the power input to the compressor is
\begin{equation}
  \label{eq:PowerComp}
P_{\verd} 
   =  -\dot m_{\verd} A_{\verd}\, c_p (T_2 - T_1),
\end{equation}
where $A_{\verd}$ is the compressor exit crossection.

% ===============================================================================================
% ===============================================================================================

\subsection{Compressor-turbine coupling}

We consider a single-shaft engine with compressor, turbine, and driven load on one common shaft. Then, 
coupling between compressor and turbine is provided via the dynamics of the engine rotational speed determined by the energy balance
\begin{equation}
  \label{eq:engineDynamics}
I \frac{\textnormal{d}n}{\textnormal{d}t} = P_{\turb} - P_{\verd} - P_{\nutz}\,, 
\qquad 
n(0) = n_0\,,
\end{equation}
where $I$ is the mass moment of inertia of all rotating components, $n$ the rotational speed and $n_0$ its initial value at time $t=0$. $P_{\nutz}$ is the power extracted from the turbine shaft.

The unknowns introduced by the turbo-component models are 
\begin{equation}
p_2, T_2,\ \ p_{\tildethree}, T_{\tildethree},\ \ T_4, \ \ (\dot m, P)_{\verd}, (\dot m, P)_{\turb}, \ \ n(t)\,,
\end{equation}
while the parameters used to control engine operation are the time traces of 
\begin{equation}
P_{\nutz}(t), \ Y_{F,2}(t), \ T_2(t)\,,
\end{equation}
where $Y_{F,2}$ and $T_2$ are the fuel mass fraction and temperature at the combustor entry. 

% ===================================================================================
% ===================================================================================

\subsection{Combustor entry and fuel injection}
\label{sec:CombustorEntryModel}

The combustor entry is controlled by an ideal unidirectional restrictor (or diodic) valve. As sketched in Fig.\,\ref{fig:SystemDesign} (bottom panel), the combustor is discretized by a onedimensional finite volume grid, and the simulated conditions within the first grid cell are used to control the valve depending on the operating mode: In deflagration mode combustion, the valve is open permanently, whereas in SEC mode the valve closes when the pressure in this cell exceeds the compressor plenum pressure. When the pressure in the cell decreases below the plenum pressure, the compressed air enters the combustor at a flow rate calculated assuming adiabatic expansion from the current plenum state. 

Fuel injection is simulated to occur at the interface between the plenum and the first grid cell. The rate of fuel injection over time varies based on the specific combustion mode. In deflagration mode, fuel is injected continuously after valve opening to guarantee that fuel-air mixture with a given equivalence ratio is entering the first grid cell. In contrast, in SEC mode, fuel injection is delayed after valve opening for a short interval of time to set up a cold air buffer between the previously burnt gas in the combustor and the new charge to be established next. After this delay, the fuel load or entropy of the incoming stream are stratified within part of the initial straight section of the combustor tube. The stratification is controlled in such a way that all mass elements of the charge will autoignite approximately at the same time. The control algorithm takes into account that at the time of autoignition, different mass elements of the charge have seen different residence times within the tube, so that their autoignition delay times \emph{measured from the time of entry into the tube} must differ correspondingly. In contrast to earlier approaches proposed by \citet{BobuschEtAl2014,BerndtKlein2017,SchaepelEtAl2018} we do not implement this here by layering the fuel mass fraction, but rather by manipulating the inflow temperature as a function of time as described in detail in App.~\ref{app:InjectionControl}.

% ===================================================================================
% ===================================================================================

\subsection{Combustor}
\label{sec:CombustorModel}
The combustor tube's (time-independent) cross-sectional area is given by
\begin{equation}\label{eq:CombustorAreaVariation}
A(x) = A_0 \begin{cases} 
           1 
             & x \in [0,x_1]  
               \\ 
           1 + (\alpha-1)\frac{\xi-\xi_1}{\xi_2-\xi_1}
             & x \in [x_1,x_2]  
               \\ 
           \alpha 
             & x \in [x_2,L]  
         \end{cases}\,, 
\end{equation}
where $L$ is the tube's length, $A_0$ is the crossection at the inflow, and $A_L = \alpha A_0$ that of the outflow section. The area ratio, $\alpha$, and the positions $x_1$ and $x_2$ of the beginning and end of the diffusor section are given below in Tables~\ref{tab:CombustorModel1} and \ref{tab:CombustorModel2}, respectively. To account for this area variation, we solve the reactive quasi one-di\-men\-sion\-al Navier-Stokes equations:
\begin{subequations}\label{eq:Q1DNS}
  \begin{align}
  (\rho A)_t + (\rho u A)_x
    & = 0\,,
      \\ 
  \label{eq:Q1NSMomemtum}
  (\rho u A)_t + ([\rho u^2 +p] A)_x
    & = p \, A_x + \frac{4}{3} (\rho\nu_{\text{eff}} A u_x)_x \,,
      \\ 
  (\rho E A)_t + ([\rho E + p] u A)_x
    & = (\rho \lambda_{\text{eff}} A T_x)_x \,,
      \\
  (\rho A Y_F)_t + ([\rho Y_F u A)_x &
   = (\rho D_{\text{eff}} A Y_F)_x + \rho A \dot{Y}_F. 
  \end{align}
\end{subequations}
Here, $\rho$ denotes density, $u$ velocity, $p$ pressure, $E$ total specific energy and $Y_F$ the fuel mass fraction with source term $\dot{Y}_F$. We use constant effective kinematic transport coefficients $\nu_{\text{eff}}$, $\lambda_{\text{eff}}$ and $D_{\text{eff}}$ to account for turbulence.
A calorically perfect gas is assumed with 
\begin{subequations}\label{eq:Thermo}
  \begin{align}
    p &= \rho R T, \label{eq:Thermo_p}\\
    E &= U + \dfrac{1}{2}\rho u^2, \label{eq:Thermo_E}\\
    U &= c_v T + Q\, Y_F = \frac{1}{\kappa-1} \dfrac{p}{\rho} + Q\, Y_F, \label{eq:Thermo_e}
  \end{align}
\end{subequations}
where $Q$ is the heat of formation of the mixture at standard conditions.
A simple one-step reaction $ \textnormal{F} \rightarrow \textnormal{P}$
for the conversion of fuel into products is assumed with reaction rate
\begin{equation}
  \label{eq:SwitchedArrheniusKinetics}
  \dot{Y}_F = - H(T-\Tsw) B\, Y_F\, \exp{\left( - \frac{E}{RT}\right) }\,.
\end{equation}
Here, $H$ is the Heaviside step function and the related term suppresses the reaction for temperatures below a threshold value $\Tsw$ to avoid fake detonations induced by numerical diffusion \citep{BerkenboschEtAl1998}. Parameters for the chemistry modeling  are adjusted to represent typical gas turbine fuels such as kerosene or DME (dimethylether), see Tabs.\ \ref{tab:CombustorModel1} and \ref{tab:CombustorModel2}.

\begin{table}
  \begin{center}
  \caption{\label{tab:CombustorModel1} Mathematical models used for participating physical processes and parameter settings for the small-to-medium size turbine with compressor pressure ratio 6:1.}
  \begin{tabular}{l|l}
  physical process \rule[-8pt]{0pt}{25pt}
    & mathematical model 
      \\
  \hline
  inviscid gasdynamics \rule[-8pt]{0pt}{25pt}
    & perfect gas \\
    & $\kappa = 1.4$, $c_p = \unit{1005.9}{J/(kg\ K)}$, 
    $c_v = \unit{718.5}{J/(kg\ K)}$, $R = \unit{287.4}{J/(kg\ K)}$
      \\[10pt]
  reference values 
    & $\rfr{T} = \unit{300}{K}$, $\rfr{p} = \unit{10^5}{Pa}$, 
      $\rfr{\rho}=\rfr{p}/(R\, \rfr{T}) = \unit{1.16}{kg/m^3}$
  \\[5pt]
  & $\rfr{l} = \unit{1}{m}$, $\rfr{u} = \sqrt{\rfr{p}/\rfr{\rho}}=\unit{294}{m/s}$, $\rfr{t} = \rfr{l} / \rfr{u} = \unit{3.4 \times 10^{-3}}{s}$
  \\[5pt]
      &
      combustor tube: $D = \unit{5}{cm}$,  $A_0 = \pi \, D^2 / 4$
      \\[5pt]
  turbulent transport \rule[-4pt]{0pt}{15pt}
    & 
    \begin{minipage}[t]{0.4\textwidth}
      \begin{tabbing}
        deflagration mode: \= $\nu_{\text{eff}} = \unit{0.44}{m^2/s}$, constant 
        \\[5pt]
        SEC mode: \> $\nu_{\text{eff}} \in \unit{[0.44, \, 0.13]}{m^2/s}$, decreasing
        \\[5pt]
        \> (see Fig.\,\ref{fig:EnginePerformanceAndPlenumStates} and Section\,\ref{ssec:PerformanceResultsSmallEngine}, eq.\,\eqref{eq:DecreasingViscosityCase1})
        \\[7pt]
        $\Pr = \Sc = 1$: $D_\text{eff} = \nu_\text{eff}$, $\lambda_\text{eff} = \nu_\text{eff} \rfr{\rho} c_p$
      \end{tabbing}
    \end{minipage}
    \\[5pt]
  chemical model \rule[-4pt]{0pt}{20pt}
    & switched Arrhenius kinetics, see eqs.\ \eqref{eq:Thermo_e} and \eqref{eq:SwitchedArrheniusKinetics} and
      \\
    & \citep{BerkenboschEtAl1998} with
      \\[5pt]
    & $\dss \frac{\Tsw}{\rfr{T}} = 1.6, \ \ \frac{E}{R} = \unit{4\,000}{K},  \ \ 
      \frac{Q}{Y_{\text{st}}} = \unit{1.875\cdot 10^3}{kJ/kg}$,
      \\[15pt]
     &
       \begin{minipage}[t]{0.4\textwidth}
         \begin{tabbing}
         ``thickened flame'' Deflagration mode:  \=  $B = \unit{6.166\cdot 10^5}{s^{-1}}$
         \\[5pt]
         SEC-mode:   \> $B = \unit{1.250\cdot 10^5}{s^{-1}}$
         \\[10pt]
         {equivalence ratio: \ $\Phi_{\text{DEFL}}^{\text{6:1}} = 0.69$}
         \\[5pt]
         {equivalence ratio: \ $\Phi_{\text{SEC}}^{\text{6:1}} = 0.7$, \ \ air buffer time: $\unit{0.136}{\milli\second}$}
         \\[-5pt]
         \end{tabbing}
       \end{minipage}
      \\[15pt]
  compressor  \rule[-4pt]{0pt}{20pt}
    & compressor parameters
      \\ 
    &  $\dss \text{area:} \ \frac{A_{\text{exit,C}}}{A_0} = 8.0; \ \  \text{efficiency:} \ \eta_{\verd} = 1.0$
      \\[15pt]
  compressor plenum
    & modelled as a torus with
      \\ 
    &  $\dss \text{cross-sectional area}: \ \frac{A_{\text{pl,C}}}{A_0} = 4.0; 
            \ \ \text{length:} \ \frac{L_{\text{pl,C}}}{\rfr{l}} = 5.0$
      \\[15pt]
  
  \text{combustor}
  & 
  cross-sectional area: $\xi = x/\rfr{l}, \ a = A/A_0$
  \\
    & 
      $a(x) 
        = \begin{cases} 
             1 
               & \xi \in [0,\xi_1]  
                 \\ 
             1 + (\alpha-1)\frac{\xi-\xi_1}{\xi_2-\xi_1}
               & \xi \in [\xi_1,\xi_2]  
                 \\ 
             \alpha 
               & \xi \in [\xi_2,1]  
           \end{cases}\,, 
           \quad\text{with}\quad
           \begin{array}{r@{\ }c@{\ }l} 
             \alpha 
               & = 
                 & 4.0
                   \\
             \xi_1
               & =
                 & 0.2
                   \\ 
             \xi_2 
               & = 
                 & 0.65
           \end{array}\,.   
       $
       \\[25pt]
  turbine plenum
    & modelled as a torus with
      \\ 
    &  $\dss \text{cross-sectional area:} \ \frac{A_{\text{pl,T}}}{A_0} = 8.0; 
            \ \ \text{length:} \ \frac{L_{\text{pl,T}}}{\rfr{l}} = 5.0$
      \\[15pt]
  turbine 
    & modelled as quasi-stationary and locally near-sonic (see eq.\,\eqref{eq:TurbineEquation})
      \\[5pt]
    & area: $\dss \frac{A_{\text{entry}}}{A_0} = 0.3052$; \ \ efficiency: $\eta_{\turb} = 1.0$; \ \ $C_{\turb} \sqrt{R} = 0.5505$  
  \end{tabular}
  \end{center}
  \end{table}

The underlying mathematical models and parameter settings are summarized in Tabs.~\ref{tab:CombustorModel1} and \ref{tab:CombustorModel2} for two reference engine setups with compressor pressure ratios of 6:1 and 20:1, respectively, to be studied in this paper. The numerical methods employed to solve the set of one-dimensional equations and to  represent the contributing physical processes are summarized in Appendix \ref{app:Quasi1DEulerDiscretization}.

  \begin{table}[h]
  \begin{center}
  \caption{\label{tab:CombustorModel2} Mathematical models used for participating physical processes and parameter settings for the high-performance reference turbine with primary compression ratio 20:1. Parameters under ``inviscid gasdynamics'' and ``reference values'' are the same as in Table*~\ref{tab:CombustorModel1} and not listed again.}

  \begin{tabular}{l|l}
  physical process \rule[-8pt]{0pt}{25pt}
    & mathematical model 
      \\
  \hline
  turbulent transport \rule[-4pt]{0pt}{15pt}
    & 
    \begin{minipage}[t]{0.38\textwidth}
      \begin{tabbing}
        deflagration mode: \= $\nu_{\text{eff}} = \unit{17.6}{m^2/s}$, constant \\[3pt]
        SEC mode: \> $\nu_{\text{eff}} \in \unit{[0.44, \, 0.44\times10^{-5}]}{m^2/s}$, decreasing:
        \\[3pt]
        \> (see caption of Fig. \ref{fig:EnginePerformanceAndPlenumStates_20-1} and Section\,\ref{ssec:PerformanceResultsHPEngine})
        \\[3pt]
        $\Pr = \Sc = 1$: $D_\text{eff} = \nu_\text{eff}$, $\lambda_\text{eff} = \nu_\text{eff} \rfr{\rho} c_p$
      \end{tabbing}
    \end{minipage}
    \\[5pt]
  chemical model \rule[-4pt]{0pt}{15pt}
    & switched Arrhenius kinetics, see eqs.\ \eqref{eq:Thermo_e} and \eqref{eq:SwitchedArrheniusKinetics} and
      \\
    & \citep{BerkenboschEtAl1998} with
      \\[5pt]
    & $\dss \frac{\Tsw}{\rfr{T}} = 2.25, \ \ \frac{E}{R} = \unit{7\,084}{K}, \ \ 
     \frac{Q}{Y_{\text{st}}} = \unit{1.875\cdot 10^3}{kJ/kg}$, 
      \\[15pt]
     & 
       \begin{minipage}[t]{0.4\textwidth}
         \begin{tabbing}
         ``thickened flame'' Deflagration mode:  \=  $B = \unit{1.492\cdot 10^6}{s^{-1}}$
         \\[5pt]
         SEC-mode:   \> $B = \unit{2.766\cdot 10^5}{s^{-1}}$
         \\[10pt]
         {equivalence ratio: \ $\Phi_{\text{DEFL}}^{\text{20:1}} = 0.58$}
         \\[5pt]
         {equivalence ratio: \ $\Phi_{\text{SEC}}^{\text{20:1}} = 0.8$, \ \ air buffer time: $\unit{0.612}{\milli\second}$}
         \\[-5pt]
         \end{tabbing}
       \end{minipage}
      \\[15pt]
  compressor \rule[-4pt]{0pt}{20pt}
    & compressor parameters
      \\ 
    &  $\dss \text{area:} \ \frac{A_{\text{exit,C}}}{A_0} = 8.0; \ \  \text{efficiency:} \ \eta_{\verd} = 1.0$
      \\[15pt]
  compressor plenum
    & modelled as a torus with
      \\ 
    &  $\dss \text{cross-sectional area}: \ \frac{A_{\text{pl,C}}}{A_0} = 4.0; 
            \ \ \text{length:} \ \frac{L_{\text{pl,C}}}{\rfr{l}} = 5.0$
      \\[15pt]
  
  \text{combustor}
  & 
  cross-sectional area: $\xi = x/\rfr{l}, \ a = A/A_0$
  \\
    & 
      $a(x) 
        = \begin{cases} 
             1 
               & \xi \in [0,\xi_1]  
                 \\ 
             1 + (\alpha-1)\frac{\xi-\xi_1}{\xi_2-\xi_1}
               & \xi \in [\xi_1,\xi_2]  
                 \\ 
             \alpha 
               & \xi \in [\xi_2,1]  
           \end{cases}\,, 
           \quad\text{with}\quad
           \begin{array}{r@{\ }c@{\ }l} 
             \alpha 
               & = 
                 & 7.8
                   \\
             \xi_1
               & =
                 & 0.28
                   \\ 
             \xi_2 
               & = 
                 & 0.91
           \end{array}\,.   
       $
       \\[25pt]
  turbine plenum
    & modelled as a torus with
      \\ 
    &  $\dss \text{cross-sectional area}: \ \frac{A_{\text{pl,T}}}{A_0} = 10.0; 
            \ \ \text{length:} \ \frac{L_{\text{pl,T}}}{\rfr{l}} = 5.0$
      \\[15pt]
  turbine 
    & modelled as quasi-stationary and locally near-sonic  (see eq.\,\eqref{eq:TurbineEquation})
      \\ 
    & area: $\dss \frac{A_{\text{entry}}}{A_0} = 1.089$; \ \ efficiency: $\eta_{\turb} = 1.0$; \ \ $C_{\turb} \sqrt{R} = 0.3782$
  \end{tabular}
  
  \end{center}
  \end{table}
  %
  
% ===================================================================================
% ===================================================================================

\subsection{Turbine plenum and turbine}
\label{sec:TurbinePlenumAndTurbineModel}
A plenum volume, situated between the combustor and the turbine, serves to dampen the intense pressure pulses from the combustor before they reach the turbine inlet. This plenum is modelled as another cell of the one-dimensional computational grid, albeit with a volume that is much larger than those of the combustor grid cells. The fluxes across its interfaces are determined separately. Owing to the plenum's large volumetric capacity, the gas in the plenum is assumed to be approximately at rest, so that $u_{\turb} \equiv 0$. At the interface between combustor and turbine plenum we aim to model a sudden large increase of the flow cross section. To this end, the following processes are approximated in the numerical discretization: Let subscripts $I$ and $\turb$ denote the last grid cell of the combustor model and the turbine plenum, respectively. Then, 
\begin{itemize}

\item[--] For $p_I > p_{\turb}; \ u_I > 0$, \ie, when a pressure pulse exits the combustor pushing gas into the plenum, the gas will quickly and adiabatically expand to the current turbine plenum pressure $p_{\turb}$ thereby accelerating the gas flow accordingly. 

\item[--] When $p_I > p_{\turb}; \ u_I \leq 0$, \ie, when the pressure near the end of the combustor $p$ exceeds $p_{\turb}$ as before, but the flow velocity is directed into the combustor, the result of the ensuing wave interactions will depend on the particular values of all the state variables next to the interface. In this case we have the typical situation of a Riemann problem formed by the combustor end state and the current plenum state. The approximate solution to that Riemann problem as given by Einfeldt's HLLE-M numerical flux function, which includes his original low-dissipation correction for the entropy characteristic, \cite{Einfeldt1988b}, provides the fluxes across the interface. 

\item[--] When $p_I < p_{\turb}$ and $u_I > 0$, the gas exiting the combustor is decelerated adiabatically until it attains either the plenum pressure or undergoes a velocity reversal in the process. In the latter case, \ie, when $p_I < p_{\turb}$ and $u_I \leq 0$, the gas entering the combustor from the plenum is accelerated by an adiabatic expansion wave connecting the two pressure levels, and this again determines the fluxes. 

\end{itemize}
See App.~\ref{app:Quasi1DEulerDiscretization} for more details regarding the numerical scheme.

% ===================================================================================
% ===================================================================================
% ===================================================================================

\section{Performance of the SEC engine in two reference configurations}
\label{sec:PerformanceResults}

Here we discuss simulations corresponding to two reference engine model configurations. The first represents a small-to-medium sized turbine, as detailed in Table,\ref{tab:CombustorModel1}, while the second represents a high-performance turbine, as outlined in Table,\ref{tab:CombustorModel2}.

% ===================================================================================
% ===================================================================================

\subsection{Reference configuration 1: moderate engine size with ${p_2/p_1 = 6:1}$}
\label{ssec:PerformanceResultsSmallEngine}

\begin{figure*}
\begin{center}

\includegraphics[height=0.93\textwidth]{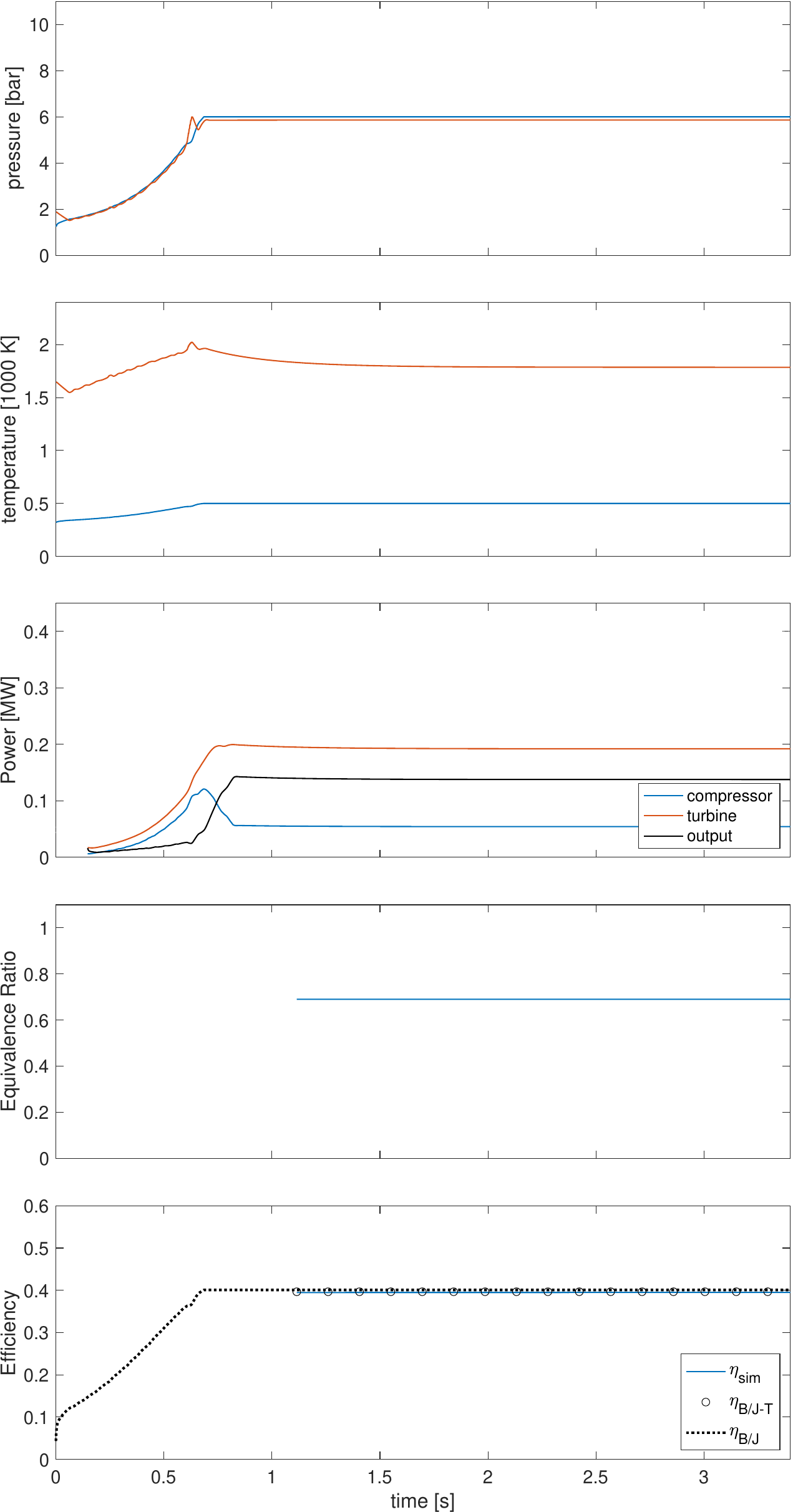}
\includegraphics[height=0.93\textwidth]{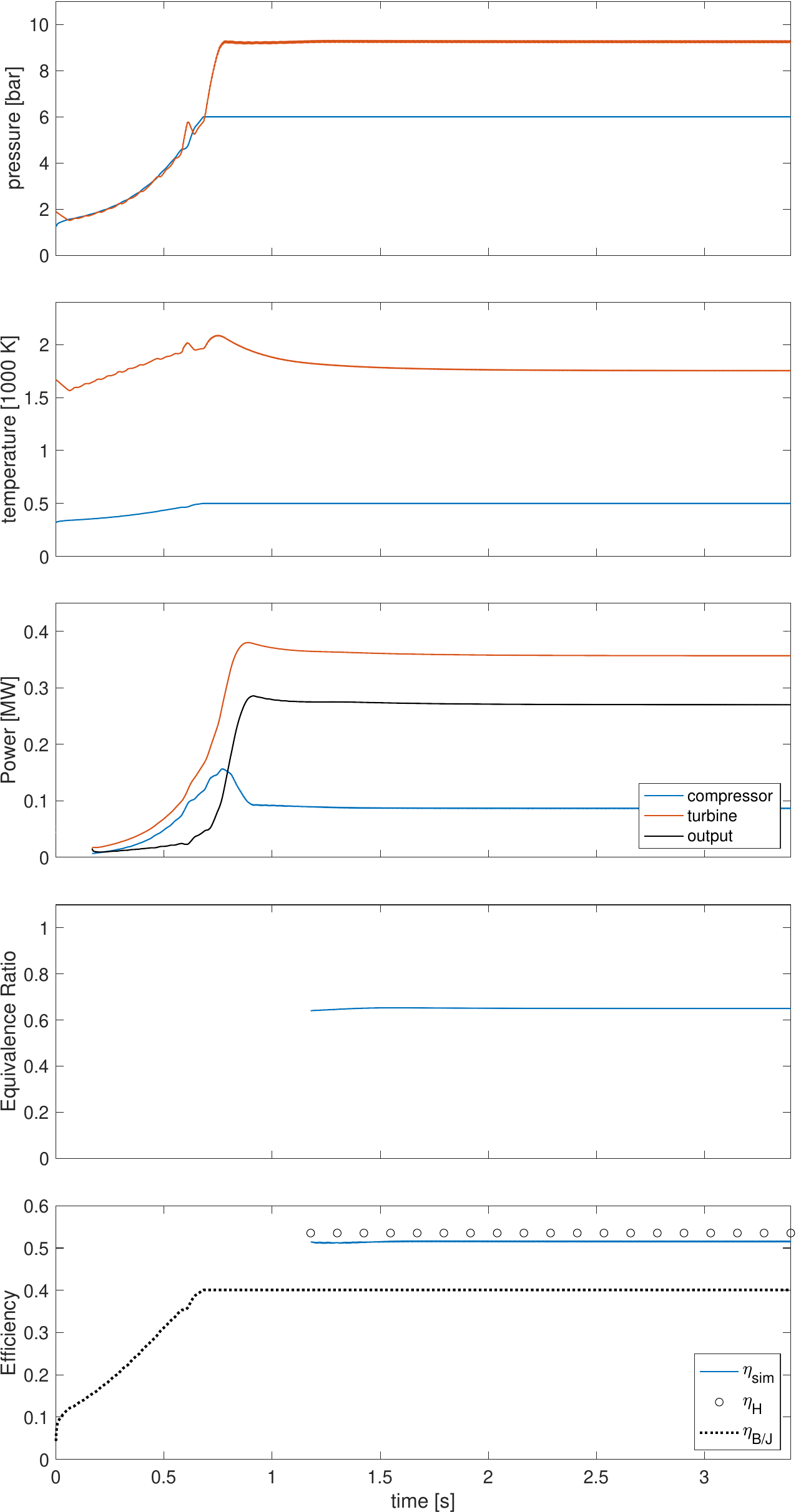}

\caption{Selected time evolutions for the classic engine model with deflagration combustion (left) and for the SEC reference configuration from Table\,\ref{tab:CombustorModel1} (compression ratio 6:1). The resolution in both cases is 1024~cells and for the SEC we consider decreasing diffusive transport so that the kinematic viscosity is 
$\nu_\text{eff} = \unit{0.44}{m^2/s} \cdot 0.82^k$ for $t \in [t_k,t_{k+1})$, where $t_k = (0.884 + 0.34 k)\, \second$ for $k = 0, ... 5$ and $t_6 = \unit{3.4}{\second}$.}
\label{fig:EnginePerformanceAndPlenumStates}
\end{center}
\end{figure*}

Consider first the comparison of the small-to-medium size/lower budget engine configuration, yet driven by a classical deflagration-based combustor (Fig.\,\ref{fig:EnginePerformanceAndPlenumStates}, left column of panels), with its SEC-based setup (right column of panels). Both variants of the engine are operated under -- on average -- lean conditions, adjusted such that the mean turbine plenum temperatures reach about $\unit{1\,780}{\kelvin}$. For the deflagration-based configuration, this is achieved by chosing an equivalence ratio of $\phi_{\text{DEFL}}^{\text{6:1}} = 0.69$, whereas for the SEC a lean mixture with $\phi_{\text{SEC}}^{\text{6:1}} = 0.7$ is injected. In combination with the inert air buffer that shields fresh from burnt gas in SEC-mode, this results in an averaged equivalence ratio of $\overline{\phi}_{\text{SEC}}^{\text{6:1}} = 0.65$ for the SEC. (See fourth row of panels in Fig.\,\ref{fig:EnginePerformanceAndPlenumStates}.)

The top row of panels in Fig.\,\ref{fig:EnginePerformanceAndPlenumStates} shows the evolution of the compressor (blue) and turbine (orange) plenum pressures. For $t \leq \unit{0.4}{\second}$, the graphs overlap because the transition to SEC operation is only possible when the compressed gas reaches a temperature that facilitates rapid autoignition. Consequently, we utilize the deflagration combustor mode during the initial spin-up for both simulations. With the engine in deflagration mode (left column), a moderate pressure drop along the combustor is observed. This is expected owing to the low but non-zero mean flow Mach number. In contrast, a pressure gain in excess of $p_{\tilde 3} / p_2 = 9.5:6$ establishes shortly after the SEC mode is enabled. 

The second row of Fig.\,\ref{fig:EnginePerformanceAndPlenumStates} shows the time evolution of the compressor (blue) and turbine (orange) plenum temperatures. By design of the compressor model used, the compressor plenum temperatures remain identical for both the classical and SEC-based engines, while the turbine plenum temperatures at steady operation conditions are within $\unit{1\,775 \pm 5}{\kelvin}$ for both engines by adjustment of the net equivalence ratios and of the fresh air buffer size for the SEC as mentioned above. 
%Only during the spin-up phase of the SEC, during which the combustor is operated in deflagration mode with a near stoichiometric mixture for simplicity of the set-up, does the turbine plenum temperature exceed this value for a short period of time. Development of a more elaborate control strategy for the spin-up phase that avoids this temperature overshoot is deferred to future work.  

In the third row of Fig.\ref{fig:EnginePerformanceAndPlenumStates}, the power demand by the compressor is illustrated in blue, the turbine's power output in orange, and their difference — representing the engine's net power output — is shown in black. The compressor pressure ratio is maintained at $6:1$ for both engine setups, and the compressor, turbine, and plenum parameters are identical. As stated above, the net mean turbine plenum temperatures are controlled within $\unit{1\,775 \pm 5}{\kelvin}$ to ensure that the thermal loads on the turbines are comparable. Under these conditions, the SEC-based engine outperforms the classical gas turbine in terms of net power output by $100\%$.

The fourth row of panels reveals that under the deflagrative and SEC combustion modes the mean equivalence ratios are  $\overline{\Phi}_{\text{DEFL}}^{\text{6:1}} = \Phi_{\text{DEFL}}^{\text{6:1}} = 0.69$ for the deflagration based and $\overline{\Phi}_{\text{SEC}}^{\text{6:1}} = 0.65$ for the SEC-based engines, respectively. For the latter, this mean equivalence ratio results from averaging over the small fresh air buffers with $\Phi = 0$ and the fresh mixture loads with $\Phi_{\text{SEC}}^{\text{6:1}} = 0.7$. 

The fifth row shows the measured and estimated engine efficiencies (see section~\ref{sec:ThermodynamicCycle} for details). The compressor and turbine efficiencies are set to unity here to allow comparing only the combustor efficiencies in these reference runs. Under this premise, the estimated efficiencies ($\eta_{\text{sim}}$, blue solid) of the engines are calculated as the ratios between moving time averages of the net power output and the mean chemical energy injection over time windows of $\unit{0.34}{\second}$, which corresponds to about $150$ acoustic cycles (see Fig.~4). The estimated efficiency of the classical turbine model (left column) comes out just $0.6\%$ below that of the Brayton-Joule cycle based on the same compressor pressure ratio ($\eta_{B/J}$, dotted). We interpret this as corroborating that our non-stationary quasi-onedimensional model of the flow through the plena and the combustor reproduces well-known thermodynamic relationships with good accuracy. Also shown in this graph is the estimated efficiency of the Brayton-Joule cycle when we assume, instead of the compressor pressure, the slightly lower turbine plenum pressure to be relevant to the cycle efficiency ($\eta_{\text{B/J-T}}$, open circles), see section~\ref{sec:ThermodynamicCycle} for a detailed discussion and \eqref{eq:IdealCycleEfficiencies} for the analytical efficiency formulae. An analogous efficiency estimate for the SEC engine ($\eta_{\text{sim}}$, blue solid) in the fifth row, right column of panels reveals an efficiency gain of about $30\%$ over the deflagration-based engine and of $29\%$ over the Brayton/Joule cycle ($\eta_{\text{B/J}}$, dotted). Also shown is the efficiency of the Humphrey cycle ($\eta_{\text{H}}$, open circles) for the same pre-compression ratio and based on the mean equivalence ratio $\overline{\Phi}_{\text{SEC}}$. A substantial fraction of its gain over the Brayton/Joule-cycle is actually realized by the SEC-based engine. 

\begin{figure*}
\begin{center}

\includegraphics[width=\textwidth]{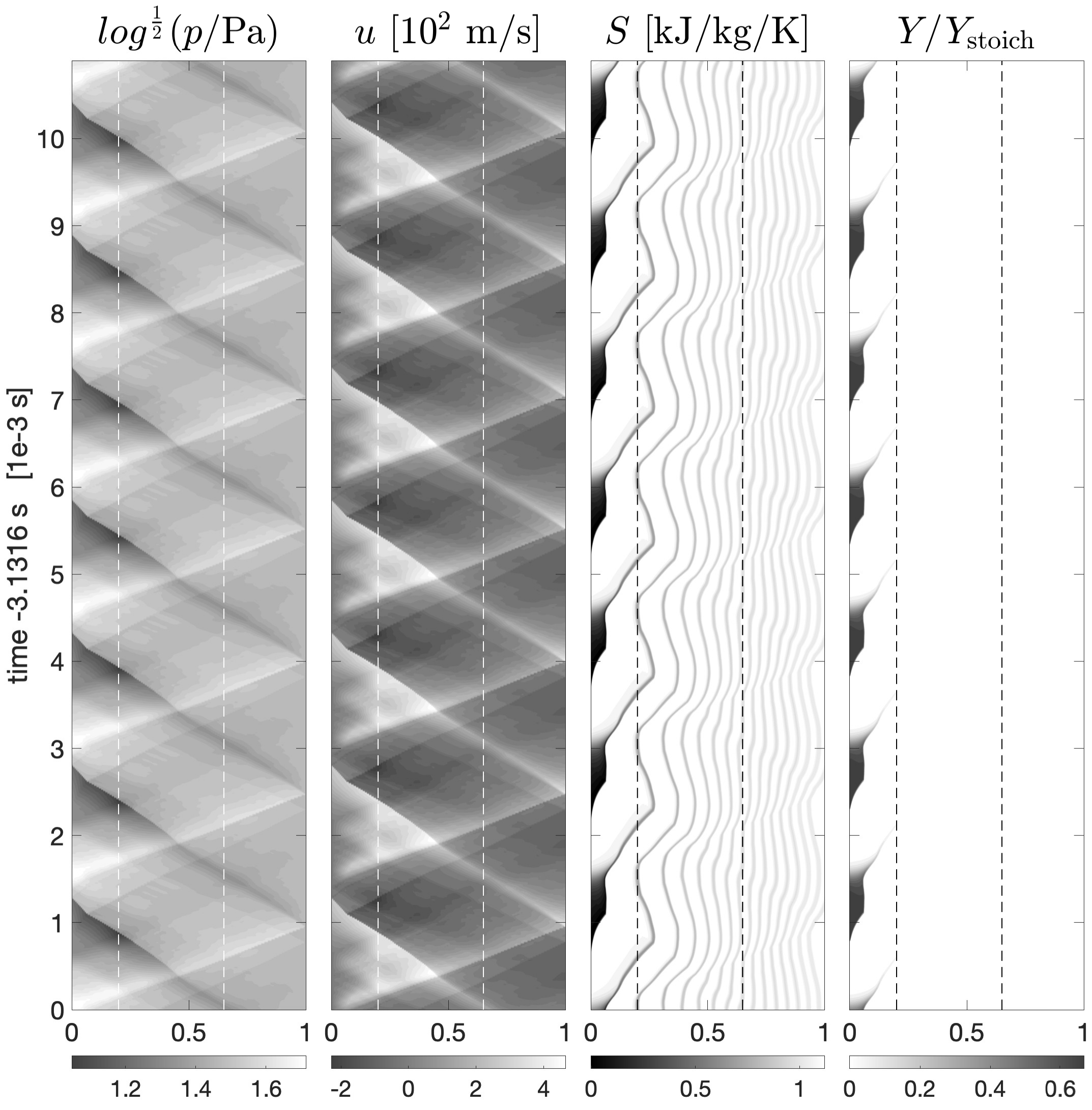}

\caption{Space-time diagrams for the SEC reference simulation from Table\,\ref{tab:CombustorModel1} (compression ratio 6:1) at 1024 gridpoint resolution. The vertical dashed lines indicate the beginning and end locations of the diffusor section.}
\label{fig:SEC_Evolutions_6-1}
\end{center}
\end{figure*}
Figure\,\ref{fig:SEC_Evolutions_6-1} reveals the spatio-temporal gasdynamic evolution during the short time interval from $t = \unit{3.13}{\second}$ to $t = \unit{3.145}{\second}$. The left-most panel in Fig.\,\ref{fig:SEC_Evolutions_6-1} represents the space-time evolution of the pressure field. To alleviate the very large dynamic range of the pressure field, which would render the plot unintelligible except for showing zones where the peak pressures are attained, we display the scaled field $\sqrt{\ln(p/\rfr{p})}$ instead of pressure $p$ itself. The strong downstream propagating \hbox{(over-)}pressure waves (sharp light traces) emerging from the local explosions near the entry of the tube are clearly visible. They will be called ``explosion waves'' in the sequel. Similarly, one observes strong gas expansions (dark shades) near the left boundary that are generated when the closed-valve boundary condition prohibits the gas from moving downstream in response to the departure of the explosion waves. These pressure drops are further enhanced by reflected suction waves that are triggered when the explosion waves enter the diffusor section of the combustor, which is located between the two vertical white dashed lines in the graph. Said expansions are superimposed in addition with weaker expansion waves that travel upstream from the downstream end of the combustor and are generated there when the explosion waves (from two explosions back in time) exit into the turbine plenum.

The second panel from the left shows an analogous plot for the axial flow velocity. The overall pattern resembles that of the pressure evolution as expected. In addition, the relative darkness of the shades in the downstream half of the tube indicates relatively low flow velocities. One important consequence is that the gas carries relatively little kinetic energy when it exits into the turbine plenum. This effect is beneficial for the overall efficiency of the engine for the following reason: By the time the gas has passed the diffusor section of the tube, much of the kinetic energy associated with the strong explosion waves is already recovered and turned into internal energy by the near-adiabatic mean pressure gain across the combustor. 

The third and fourth panels of Fig.\,\ref{fig:SEC_Evolutions_6-1} display the space-time traces of the excess entropy relative to the reference state and the fuel mass fraction relative to the stoichiometric one, \ie,
\begin{equation}\label{eq:EntropyDefinition}
S = c_v \ln\left(\frac{p/\rfr{p}}{(\rho/\rfr{\rho})^\kappa}\right)\,
\qquad\text{and}\qquad
Y = \frac{\rho Y}{\rho Y|_{\text{stoich.}}}\,,
\end{equation}
respectively. The time intervals of inflow valve opening are  indicated by the intermittent elongated dark spots near the upstream edge of the domain in the entropy map. A dark shade means low entropy, \ie, it signals cold gas to be introduced into the combustor. Close inspection shows that the similar black spots in the equivalence ratio map start later in time and do not extend quite as far into the domain. This simply documents the fact that a small inert air buffer separates the fresh charge from the high-entropy burnt gas (see also the related entry under ``chemical model/equivalence ratio'' in Table~\ref{tab:CombustorModel1}). The black spots are terminated by an approximately horizontal transition zone, \ie, by a nearly homogeneous autoignition event with rapid fuel consumption and entropy increase. The entropy map, however, also shows a dark grey streak emerging from the top right edge of the cold gas black spot which represents the air buffer being slowly advected downstream along an oscillatory path. The wavy stripes seen in the domain result from previous resonant cycles, and their relatively slow motion down the tube documents that the mean flow Mach number in the combustor is rather low. This corroborates our earlier observation of the efficient conversion of kinetic into thermodynamic potential energy. 

The fact that diffusion smears out the cold air packets as time proceeds is a consequence of turbulent transport  which is modelled here roughly by constant coefficient diffusivities of mass, thermal energy, and the chemical species. Assuming turbulent Prandtl and Schmidt numbers of unity, we estimate a realistic turbulent Reynolds number as
\begin{subequations}\label{eq:TurbulenceEstimates}
\begin{equation}
Re_t = \frac{\rfr{u} L}{\nu_t} \approx \frac{\rfr{u}\, \rfr{l}}{c_\mu u' D} = 1111\,,
\end{equation}
where we assumed a scaling factor $c_\mu$, taken from the standard $k-\varepsilon$ model, a turbulence intensity $u'/\rfr{u}$, and a combustor aspect ratio, $D/\rfr{l}$ of, rounded, 
\begin{equation}
c_\mu = 0.09\,,
\quad
\frac{u'}{\rfr{u}} = 0.2\,,
\quad\text{and}\quad
\frac{D}{\rfr{l}} = 0.05\,,
\end{equation}
\end{subequations}
respectively. This amounts to $\nu_t \approx \unit{0.26}{\meter^2\per\second}$ and lies within the range of values adopted in the first reference case (see \eqref{eq:DecreasingViscosityCase1} below). Note that this is to be understood as a reasonable estimate rather than a precise value for a specific case.

To demonstrate that, within bounds, the SEC combustion mode is independent of the turbulent diffusivity, we reduce the effective viscosity step by step during the SEC run according to 
\begin{equation}\label{eq:DecreasingViscosityCase1}
\nu_\text{eff} = \unit{0.44}{m^2/s} \cdot 0.82^k \ \ \text{for} \ \ t/\rfr{t}
\in [t_k,t_{k+1}),
\end{equation}
where 
$t_k = (0.68 + 0.34 k) \second 
\ \ \text{for} \ \ k = 0, ... 5 \ \ \text{and} \ \ t_6 = \unit{3.4}{\second}$,
while the Prandtl and Schmidt numbers remain constant (see also Table\,\ref{fig:EnginePerformanceAndPlenumStates}).  Clearly, the SEC run does not reveal any visible imprint of this change of viscosities in the graphs of Fig.\,\ref{fig:EnginePerformanceAndPlenumStates}. Thus, SEC operation in this case is stable at least in the range $\unit{0.16}{\meter^2\per\second} \leq \nu_{\text{eff}} \leq \unit{0.44}{\meter^2\per\second}$ of the turbulent transport parameter. Note, however, that with much smaller viscosities and for the set-up of the present reference case ($p_2/p_1 = 6:1$) the SEC combustor fails to lock into a stable resonant cycle and ``misfires'' repeatedly (not shown). This is different for the second reference case ($p_2/p_1 = 20:1$) which maintains stable operation even with negligible turbulent transport (cf.~section\,\ref{ssec:PerformanceResultsHPEngine} and discussions in section\,\ref{sec:Conclusions}). More detailed investigations of these stability properties are referred to future work.

% ===================================================================================
% ===================================================================================

\subsection{Reference configuration 2: high-performance engine with ${p_2/p_1 = 20:1}$}
\label{ssec:PerformanceResultsHPEngine}

\begin{figure*}
\begin{center}

\includegraphics[width=0.49\textwidth]{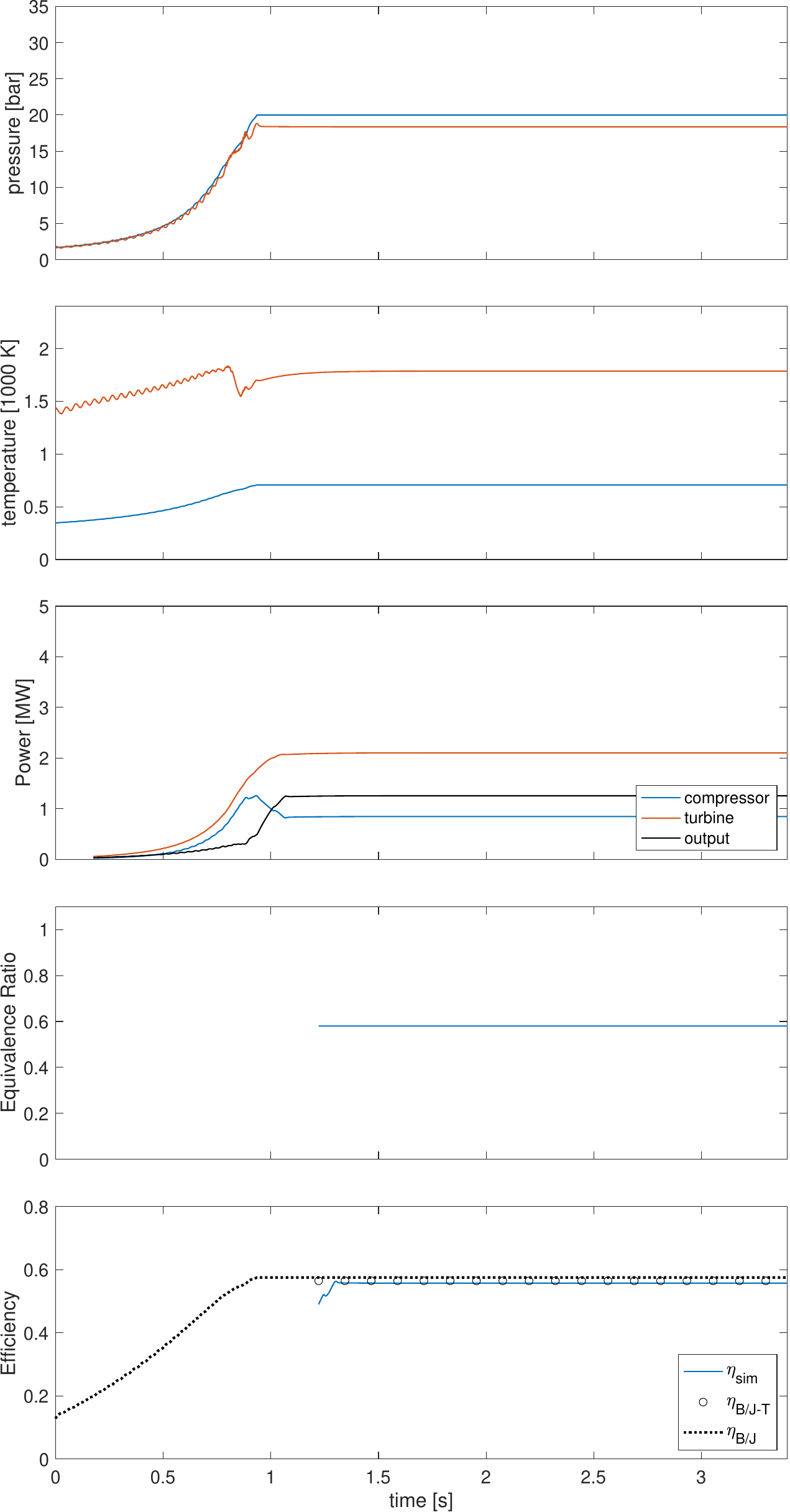}
\includegraphics[width=0.49\textwidth]{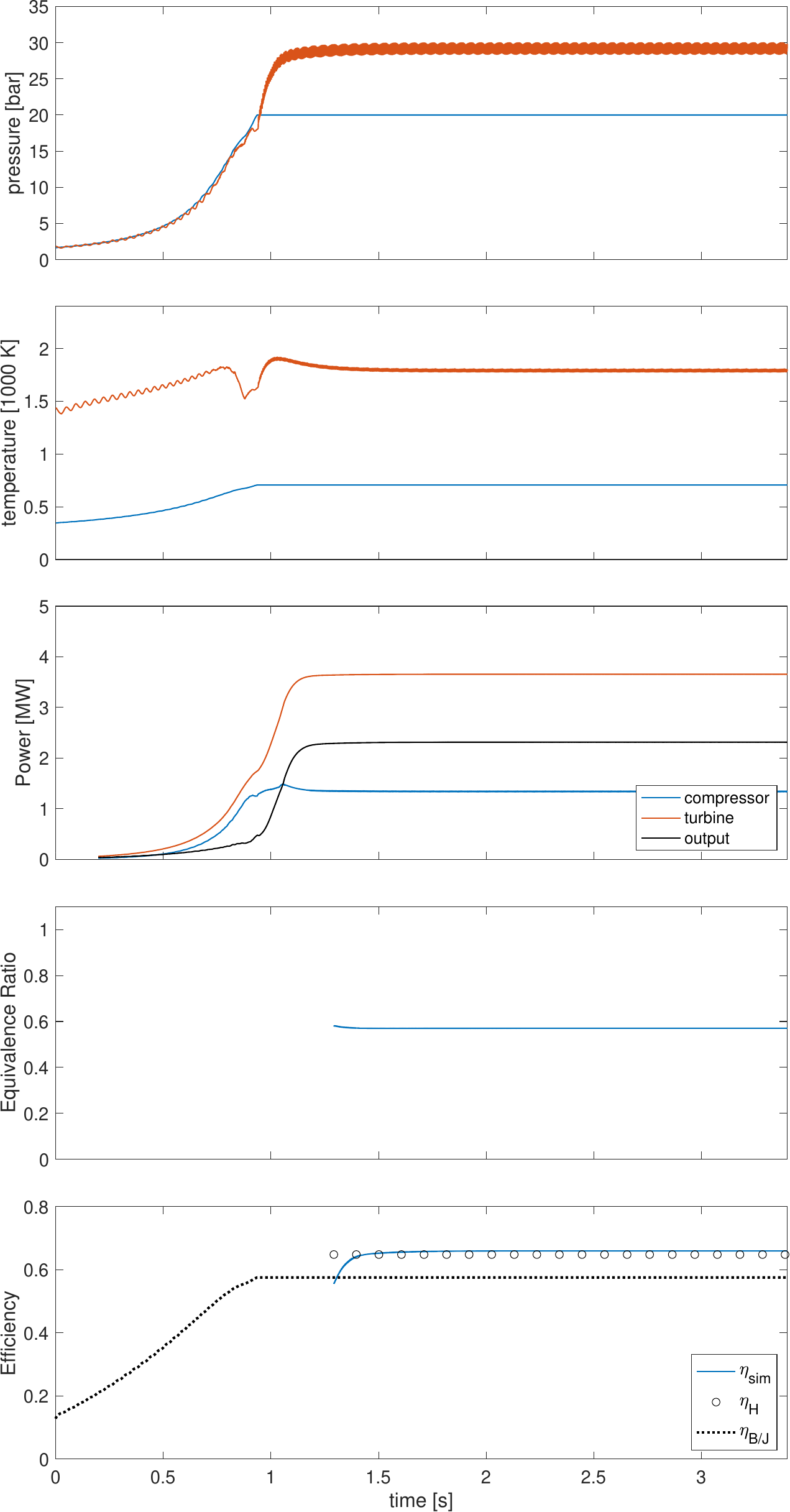}

\caption{Left: Selected time evolutions for deflagration combustion (left) and SEC (right) reference configurations from Table\,\ref{tab:CombustorModel2} (precompression: 20:1) at grid resolutions of 1024~cells. The SEC case is run with decreasing viscosity and diffusivity during the run according to \eqref{eq:DecreasingViscosityCase2}. }
\label{fig:EnginePerformanceAndPlenumStates_20-1}
\end{center}
\end{figure*}

Here we document simulations based on the second SEC engine reference configuration involving the higher compression ratio of 20:1 from Table\,\ref{tab:CombustorModel2}. Figure\,\ref{fig:EnginePerformanceAndPlenumStates_20-1} displays again, and in analogy with Fig.~\ref{fig:SEC_Evolutions_6-1}, from top to bottom the time histories of plenum pressures; plenum temperatures; compressor power uptake, turbine power delivery, and net power delivered; mean equivalence ratios; and estimated efficiencies.   
In these simulations, the targeted turbine plenum temperatures of $\unit{1775 \pm 5}{\kelvin}$ are achieved by chosing a fresh gas equivalence ratio of $\phi_{\text{DEFL}}^{\text{20:1}} = 0.58$ in deflagration mode, while in SEC mode we inject a mixture with $\phi_{\text{SEC}}^{\text{20:1}} = 0.8$  but use a larger air buffer flow time of $t_{\rm buff} = \unit{0.612}{\milli\second}$ to yield a cycle-mean equivalence ratio of $\overline{\phi}_{\text{SEC}}^{\text{20:1}} = 0.57$ in SEC mode (see also Table~\ref{tab:CombustorModel2}, entries ``chemical model/equivalence ratio'').

While the mean pressure gains across the combustors in the two cases are comparable with $p_{\tilde 3}/p_2 = 9.5:6$ for case~1 and with $p_{\tilde 3}/p_2 = 29:20$ for case~2, there are key differences: First we realize that the efficiency gain of the SEC engine over the deflagration-based one has reduced to 18\% from 30\%, which is consistent with estimates by Stathopoulos~\cite{Stathopoulos:2018,Stathopoulos:2020} who predicted reduced potential efficiency gains of pressure gain combustion over deflagrative combustion with increasing compressor pressure ratios. The remaining gain of 18\% is nevertheless still substantial. Secondly, and more remarkably, we find that now the SEC efficiency exceeds that of the associated Humphrey cycle based on the 20:1 compression ratio and mean equivalence ratio $\overline{\phi}_{\text{SEC}}^{20:1} = 0.57$ (compare blue solid line and open circles in the fifth row, right column of panels). As we will point out in section~\ref{sec:ThermodynamicCycleEfficiency} below, where we study efficiency estimates based on thermodynamic cycles along Lagrangian parcel trajectories in the combustor, this suggest that there is room for further improvements through optimization of the combustor and plenum designs.  

\begin{figure*}
\begin{center}

\includegraphics[width=\textwidth]{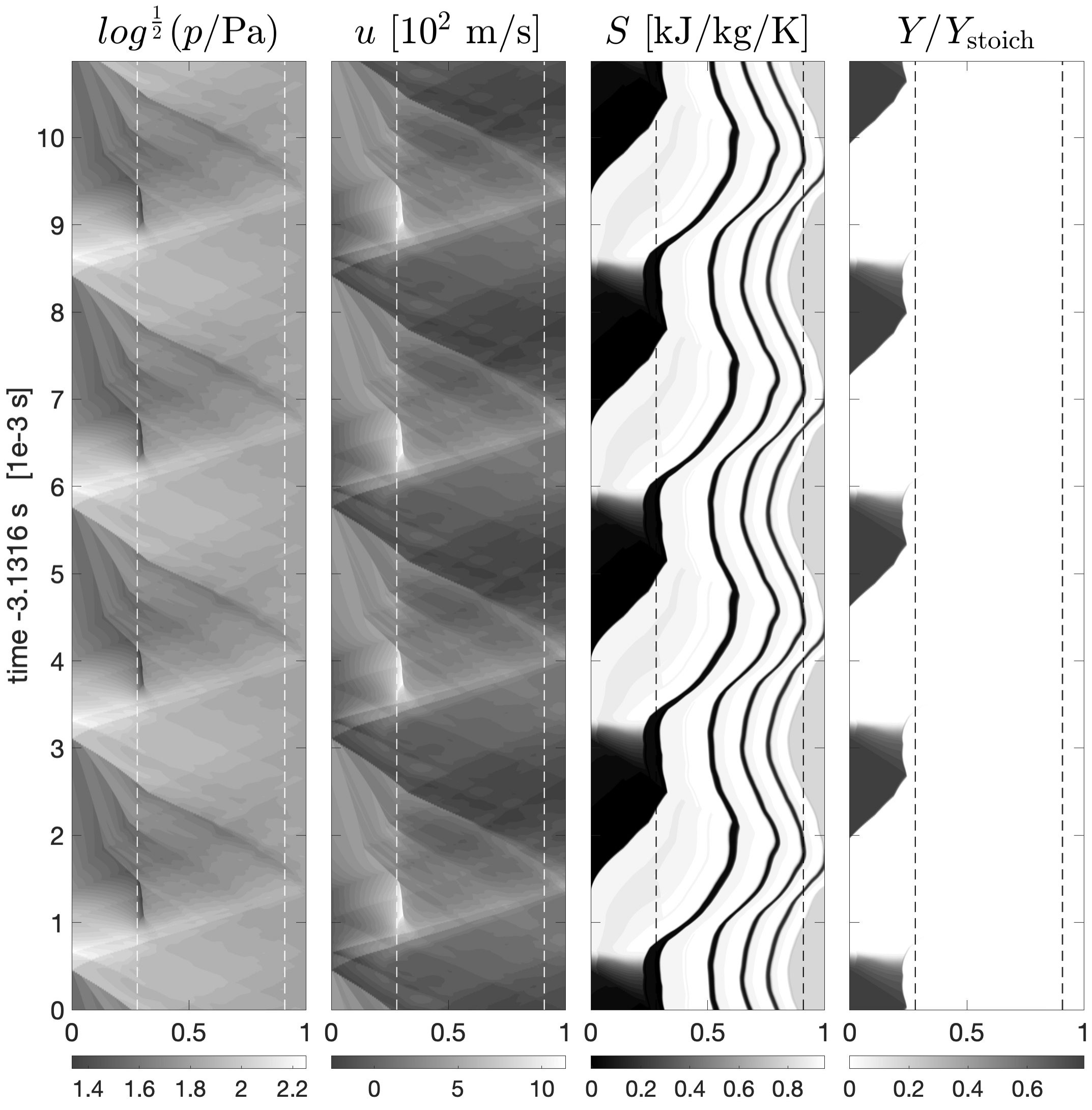}

\caption{Space-time diagrams for the SEC reference simulation for compression ratio 20:1 from Table\,\ref{tab:CombustorModel2} at 1024 gridpoint resolution. The vertical dashed lines indicate the beginning and end locations of the diffusor section.}
\label{fig:SEC_Evolutions_20-1}
\end{center}
\end{figure*}

The space-time evolutions in Fig.\,\ref{fig:SEC_Evolutions_20-1} reveal that, in this second configuration, the combustor settles into a different mode of operation than it did for configuration\,1 (Fig.\,\ref{fig:SEC_Evolutions_6-1}). Here, the firing time interval corresponds to one full forward-backward passage of the dominant pressure waves along the length of the combustor, while in configuration\,1 the combustor fires at twice this frequency. Correspondingly, the fresh charge induced into the combustor in each cycle covers a much larger fraction of the tube length (see third and fourth panel from the left), and the ensuing gasdynamic processes are more intense. An obvious indicator for the latter is the appearance of choked states at the beginning of the diffusor section as seen in the velocity panel (second from the left) in Fig.\,\ref{fig:SEC_Evolutions_20-1}. Although significant flow acceleration is also seen near the left edge of the diffusor in configuration\,1, Fig.\,\ref{fig:SEC_Evolutions_6-1}, choking does not arise there. 

The reader will also notice that features in the diagrams of Fig.\,\ref{fig:SEC_Evolutions_20-1} are somewhat less blurred than those seen in Fig.\,\ref{fig:SEC_Evolutions_6-1}, although the grid resolutions in the underlying simulations is the same (1024 gridpoints along the combustor). The reason is that the lower-frequency SEC mode seen here is much less sensitive to the level of turbulent transport than the double-frequency mode in reference case~1. In fact, by letting
\begin{equation}\label{eq:DecreasingViscosityCase2}
\begin{array}{r@{\ }c@{\ }l@{\quad}c@{\quad}l}
\nu_\text{eff} 
  & = 
    & \max(\nu_0\, a^k, \nu_{\text{min}})  
      & \text{for} 
        & t/\rfr{t} \in [t_k,t_{k+1}) 
          \\ 
t_k 
  & = 
    & (0.88 + 0.34 k) \second 
      & \text{for} 
        &  k = 0, ..., 5
          \\ 
t_6 
  & = 
    & \unit{3.4}{\second}\,,
\end{array}
\end{equation}
we have increased the model Reynolds number to $\Rey_t \approx 8.2\cdot 10^{6}$ here without affecting robust SEC operation. The ensuing decreased mixing intensity generally sharpens the flow features seen and, in particular, lets the cold air streaks in the entropy plot (Fig.\,\ref{fig:SEC_Evolutions_20-1}, third panel) dissipate only very slowly as they propagate down the combustor. Detailed investigations into the sensitivity of SEC operation with respect to turbulent transport and other key system parameters are deferred to future work.
%

% ===================================================================================
% ===================================================================================
% ===================================================================================

\section{Thermodynamic cycle analysis and efficiency}
\label{sec:ThermodynamicCycle}

% ===================================================================================
% ===================================================================================

\subsection{Evolution of flow states in the reaction region near the combustor entry}
\label{sec:ThermodynamicCycleNearTubeEntry}

\begin{figure*}[htbp]
\begin{center}

% \begin{minipage}[b]{0.49\columnwidth}
% \includegraphics[width=1.0\textwidth]{Fig07-left.pdf}
% \end{minipage}
% \hfil
% \begin{minipage}[b]{0.49\columnwidth}
% \includegraphics[width=1.0\textwidth]{Fig07-right.pdf}\hfil
% \end{minipage}

%\includegraphics[width=0.49\textwidth]{Graphics/ThermodynamicCycles}\hfil
\includegraphics[width=0.49\textwidth]{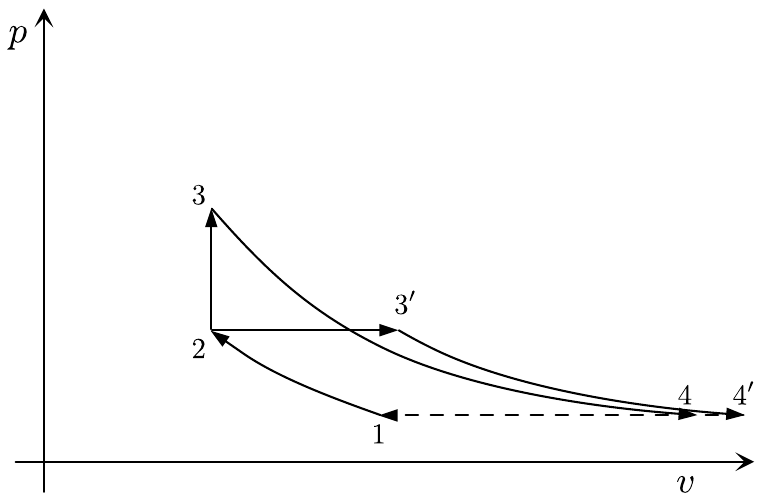}\hfil
\includegraphics[width=0.49\textwidth]{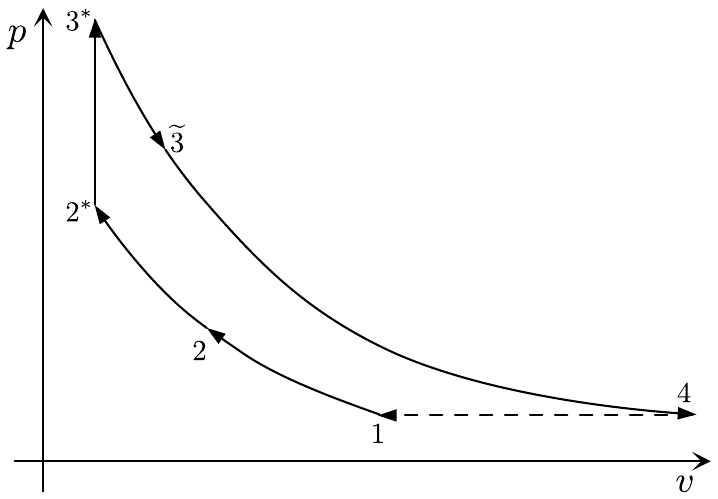}\hfil

\caption{Left: The classical Humphrey ($1$--$2$--$3$--$4$--$1$) and Brayton/Joule ($1$--$2$--$3'$--$4'$--$1$) cycles involving isochoric ($2$--$3$) and isobaric ($2$--$3'$) combustion.  Right: The Humphrey cycle ``$1$--$2$--$2^*$--$3^*$--$\tilde 3$--$4$--$1$'' modeling the thermodynamic process in a turbine with pressure gain combustion including internal nonstationary gasdynamic effects within the combustor device ($2$--2$^*$--3$^*$--$\tilde 3$). Figure*s adapted from \citep{Nalim2002}. }
\label{fig:ThermodynamicCycles}
\end{center}
\end{figure*}

The established reference thermodynamic cycle model for unsteady pressure gain combustion is the Humphrey cycle ``$1$--$2$--$3$--$4$--$1$'' (H) as sketched in the left panel of Fig.\,\ref{fig:ThermodynamicCycles}.  Ideally, in such a device isentropic compression (1-2) is followed by isochoric combustion ($2$--$3$), which is followed by isentropic expansion  ($3$--$4$) all the way to the base pressure $p_4 = p_1$. The same panel also shows the Brayton/Joule (B/J) cycle, which is the reference thermodynamic cycle for classical gasturbines with (quasi-)stationary low Mach number flow combustors. Here, after compression (1--2), the heat of reaction is added in an isobaric expansion process ($2$--$3'$), after which isentropic expansion to the base pressure is realized in the turbine ($3'$--$4'$). The thermodynamic efficiencies of these cycles depend on the initial compression ratio along ($1$--$2$) and, in the case of the Humphrey cycle, also on the total chemical energy supplied, see, e.g., \citep{Nalim2002,HeiserPratt2002} and appendix~\ref{app:Humphrey} below, so that
\begin{subequations}\label{eq:IdealCycleEfficiencies}
\begin{align}
\eta_{\text{B/J}}(T_1, T_2) 
  & = 1 - \frac{T_1}{T_2}
    \\
\eta_H(T_1, T_2, Q) 
  & = 1 - \frac{c_p T_1}{Q} \left(\left(1 + \frac{Q}{c_v T_2}\right)^{\frac{1}{\kappa}}  - 1\right)\,.
\end{align} 
\end{subequations} 
Of course, the temperature ratio can always be expressed in terms of the compressor pressure ratio by \eqref{eq:Verdichtertemperatur} with $\eta_C = 1$ in case of an ideally efficient compressor.

In a successful implementation of pressure gain combustion, the isentropic compression and expansion processes are therefore divided into (at least) two steps each as shown in the right panel of Fig.\,\ref{fig:ThermodynamicCycles}. After compression in the turbo-compressor ($1$--$2$) to, say, the same pressure ratio as that of the classical Brayton/Joule cycle in the left panel, internal non-stationary gasdynamic processes further compress the working fluid ($2$--$2^*$). Thus, the subsequent isochoric combustion ($2^*$--$3^*$) starts from a higher initial pressure and temperature, thereby providing enhanced thermodynamic efficiency of the combustion process. Isochoric combustion raises the pressure to very high levels and in a highly unsteady manner so that further appropriately designed gasdynamic processes must be foreseen to lower the pressure and minimize gasdynamic fluctuations ($3^*$--$\tilde 3$) before the burnt gas reaches the turbine \cite{HeiserPratt2002}. Isentropic expansion ($\tilde 3$--$4$) will then yield the main usable work of the entire device. This extended process may again be modelled by the Humphrey cycle, albeit with enhanced pre-compression before the onset of combustion. Using \eqref{eq:IdealCycleEfficiencies}, the ideal thermodynamic efficiency becomes $\eta_H(T_1, T_{2^*}, Q)$.

In the present context, the intermediate state $\tilde 3$ indicated in the right panel of Fig.\, \ref{fig:ThermodynamicCycles} should be thought of as the state in an intermediate ``turbine plenum'' or reservoir that receives the highly pulsating working fluid and, by its large volume relative to that of the combustor, attenuates the pressure fluctuations ultimately felt by the turbine to an acceptable level. Notably, this state is not constrained to having the same specific volume as the post-compressor state~2. 

The extended Humphrey cycle from Fig.\,\ref{fig:ThermodynamicCycles} (right panel) will be used repeatedly in our interpretation of the simulation results for the new SEC-engine design below. Before we move on to this analysis, let us report on a straightforward test of the physical consistency of our computational model: When the gasturbine model is operated with the quasi-stationary deflagration combustor only and with lossless compressor and turbine, it yields an efficiency -- computed straightforwardly as the ratio of energy delivered at the engine shaft divided by the chemical energy added (in our 1024 grid point simulation) -- of 
\begin{equation}
\eta_{\text{Defl}}\Bigm|_{\frac{p_2}{p_1} = 6} = 0.394 = 0.985\, \eta_{\text{B/J}}\Bigm|_{\frac{p_2}{p_1} = 6}
\qquad\text{for}\qquad 
\eta_{\verd} = \eta_{\turb} = 1\,.
\end{equation}
The difference to the Brayton/Joule efficiency may result from several sources, among them the facts that (i) the fluid dynamics model includes artificial/turbulent diffusion and friction and that (ii) a realistic deflagration is isobaric only in the zero Mach number limit whereas, with the thermodynamic parameters set for a typical hydrocarbon fuel, the pressure drops across the flame by about $0.25\%$. 

In any case, the present model approximates the efficiency of the Brayton/Joule cycle rather well and from below, as it should. Moreover, we can compare our results with realistic engine setups by taking into account the non-ideal efficiencies of compressor and turbine: Accounting for realistic efficiencies of both turbo-components of about 90\% yields overall machine efficiencies of 31\%. This value is close to that reported for real-life small gasturbines that operate with moderate compressor pressure ratios of $6:1$ \citep{MAN_GT}. 

% u = 0.69, p = 5.32, T = 7.86  ->  reacted gas Mach number: 0.21

%\begin{sidewaysfigure}%[htbp]
\begin{figure*}
\begin{center}
%\vspace{0.6\textheight}
%\includegraphics[width=1.0\textwidth]{Graphics/SEC_20-1_Evolution_u_periodic}
\includegraphics[width=1.0\textwidth]{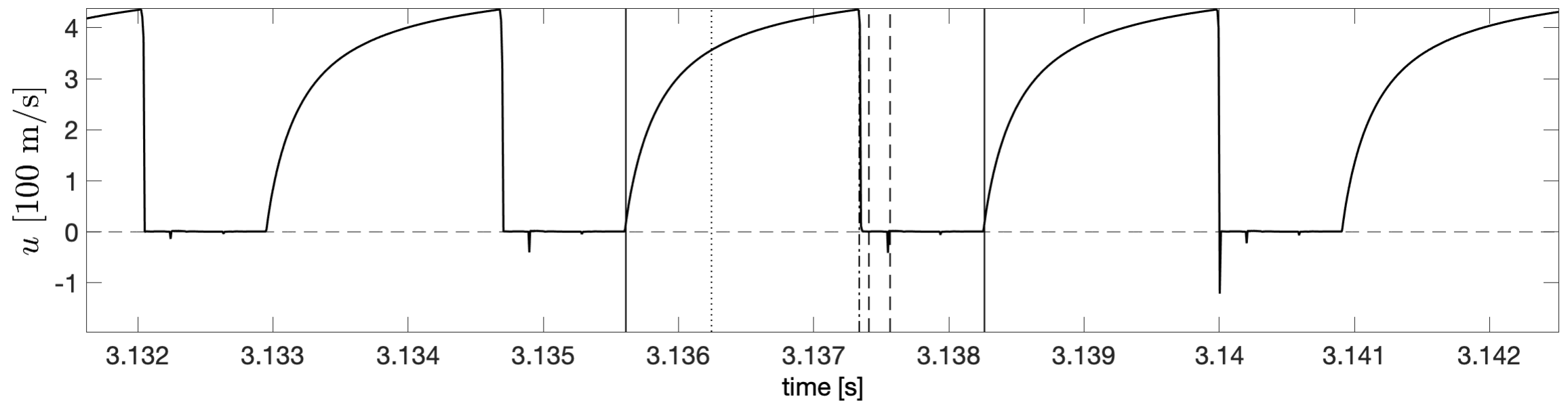}

\caption{Time evolution of the axial flow velocity in the first grid cell of the combustor under SEC operation of the model engine in the reference setup for pressure ratio 20:1, see Table\,\ref{tab:CombustorModel2} and Fig.\,\ref{fig:EnginePerformanceAndPlenumStates}.}
\label{fig:PeriodicCycleU}
\end{center}
\end{figure*}
\begin{figure*}
\begin{center}

\includegraphics[width=0.9\textwidth]{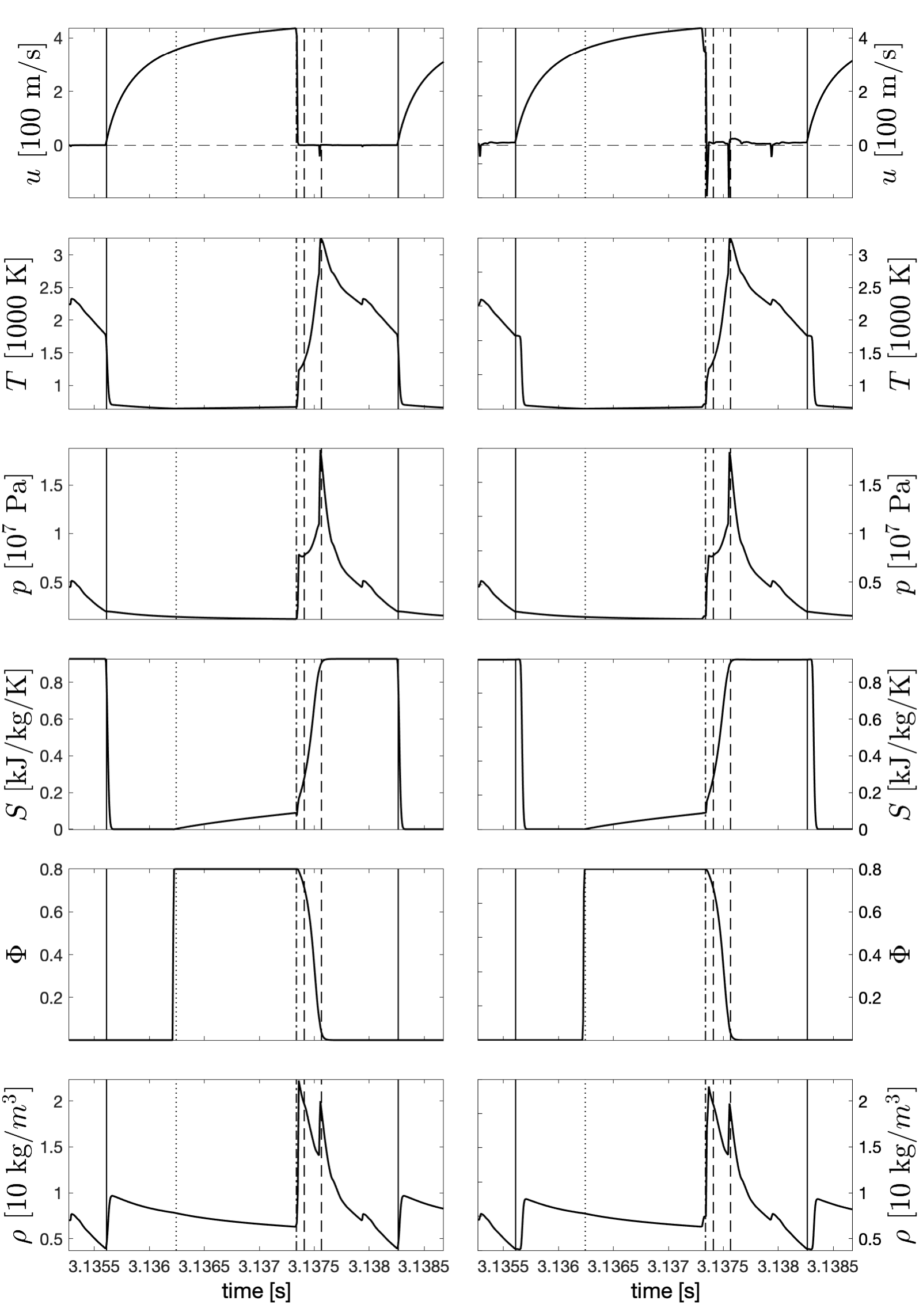}

\caption{Detailed time evolution of the state during one oscillation period in the $1^{\text{st}}$ (left) and $5^{\text{th}}$ (right) grid cells within the combustor under fully resonant operation of the model engine in the the reference setup for pressure ratio 20:1, see Table\,\ref{tab:CombustorModel2} and Fig.\,\ref{fig:EnginePerformanceAndPlenumStates}.}
\label{fig:PeriodicCycleAll}
\end{center}
\end{figure*}

Once the large amplitude resonant mode has established in the system, a rather regular oscillation pattern results as is documented in Figure\,\ref{fig:PeriodicCycleU} which shows the time history of the axial flow velocity in the first grid cell of the combustor for case~2 (20:1) between times $t \approx \unit{3.13}{\second}$ and $t \approx \unit{3.145}{\second}$ as a representative variable. The data are taken from a simulation of configuration\,2 (Table\,\ref{tab:CombustorModel2}) with spatial resolution of 1024 grid cells, see also Fig.\,\ref{fig:EnginePerformanceAndPlenumStates_20-1}. Also shown in the center of Fig.~\ref{fig:PeriodicCycleU} is a time window embracing one complete oscillation cycle that is now further discussed with reference to Fig.\,\ref{fig:PeriodicCycleAll}. This figure displays the time history of several thermodynamic variables in the 1$^{\text st}$ (left column of panels) and 5$^{\text th}$ (right column of panels) grid cells of the computational grid. Consider first the left column of graphs in the figure: 
\begin{enumerate}

\item The first vertical solid line in all the graphs indicates the time of valve opening as is seen in the top panel displaying the axial flow velocity. At this point, the axial velocity, $u$, begins to steeply rise from zero. At the same instance, the temperature, $T$, and entropy variable $S$ from \eq{eq:EntropyDefinition}, drop rapidly as fresh air from the compressor plenum replaces the burnt gas in the inlet section of the combustor. 

\item At the vertical dotted line fuel injection commences and the scaled fuel mass fraction, $Y$ (henceforth just called ``mass fraction''), increases swiftly towards its inflow value, \ie, equivalence ratio, of $Y_{\text{st}} = 0.8$. The inflow velocity, $u$, keeps rising during this process and the flow reaches a maximum inflow Mach number of $\Ma_{\text{in}} \approx 0.84$ just before the time of the dash-dotted line.

\item At the time marked by the dash-dotted line, the pressure, $p$, rises above the threshold value of $p_{\text{c}} = \unit{20.0}{\bbar}$, i.e., above the pressure in the compressor plenum. The one-way valve closes in response and the velocity quickly drops to zero accordingly. Immediately thereafter, a continuous marked rise of pressure, temperature, $T$, and density, $\rho$, are monitored. Early on, this compression is nearly isentropic as seen in the evolution of the entropy, $S$, which changes only marginally during this period. From this instance up until slightly beyond the second dashed line the fuel mass fraction, $Y$, exhibits the typical behavior of an autoignition process: reaction of the fuel species begins slowly but accelerates progressively and consumes the fuel ever more rapidly until the mass fraction, $Y$, eventually drops to zero. The first dashed line in the graphs indicates the time when the first significant reaction progress sets in, whereas at the second dashed line the fuel has essentially been consumed. In the present simulation, a backward facing shock arrives at roughly this point at the valve, indicated by the simultaneous sudden increase of pressure, temperature and density and by a negative spike of the flow velocity. This backward travelling wave is also clearly seen in the leftmost panel of Fig.~\ref{fig:SEC_Evolutions_20-1}, which shows the close-by arrival of two leftward travelling pressure waves between times $\unit{3.137}{\second}$ and $\unit{3.138}{\second}$.

Compression by gasdynamic waves and the autoignition with further pressure increase are superimposed here, but we can safely state that the combustible mixture undergoes substantial post-compression after it has entered the tube and before rapid chemical heat release arises. Thus the processes in this first grid cell realize approximately the loop ($2$-$2^*$-$\tilde 3$) of the idealized Humphrey cycle for pressure gain combustion of Fig.\,\ref{fig:ThermodynamicCycles}, left panel.
  
\item Scrutiny of the course of events between the two vertical dashed lines reveals that, adjacent to the inlet valve, the combustion process is not really isochoric but also far from isobaric: Within the time interval indicated by the dashed lines, 95\% of the fuel is consumed while the density drops by roughly $25\%$ whereas the pressure increases by about the same amount during the same period. 
% In fact, comparing the Humphrey cycle efficiency based on the density ratio \klein{$\rho/\rfr{\rho} = 20.4$ observed at the first dashed line we obtain $\eta_H = 70\%$} while in this simulation run the actual estimated engine efficiency is $67\%$. 

\item The rest of the cycle towards the next vertical solid line consists of isentropic gas expansion to very high accuracy as seen in the time trace of the entropy variable $S$.  

\end{enumerate}

The column of panels on the right of Fig.\,\ref{fig:PeriodicCycleAll} shows the evolution of flow states a short distance away from the tube entry in the fifth grid cell. Here, the combustion process is not as close to isochoric as it is in the first cell, but there is still considerable acoustic confinement (see behavior of the density $\rho$ between the dashed lines). The strong negative peaks of the flow velocity near the first and second vertical dashed lines indicate the arrival of compression waves travelling backwards towards the tube entry. The first of these waves is responsible for the additional precompression of the gas discussed in the context of the first grid cell above. Comparison of the traces of the fuel mass fraction, $Y$, at the two positions and within the time window given by the dashed lines reveals that the control of the inlet temperature actually succeeds in timing the autoignition of air parcels in the first and fifth cells such that the reaction processes are completed at essentially the same time. This is also corroborated by the nearly horizontal, \ie, instantaneous, termination of the dark spots in the right-most column of Fig.\,\ref{fig:SEC_Evolutions_20-1} which delineate the presence of fuel in the system.
\begin{figure*}
\begin{center}

\begin{minipage}{0.49\textwidth}
\centering
{\sffamily Case 1: $p_2/p_1 = 6$}

\includegraphics[width=0.95\textwidth]{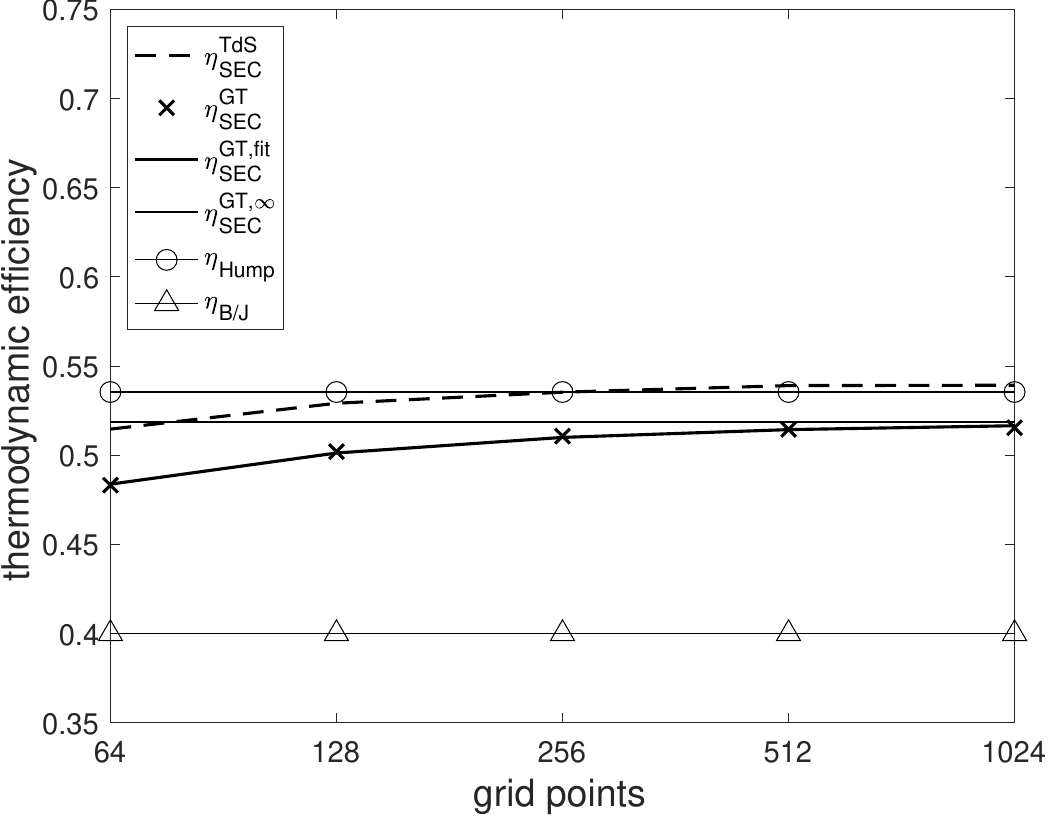}

\end{minipage}
\begin{minipage}{0.49\textwidth}
\centering
{\sffamily  Case 2: $p_2/p_1 = 20$}

\includegraphics[width=0.95\textwidth]{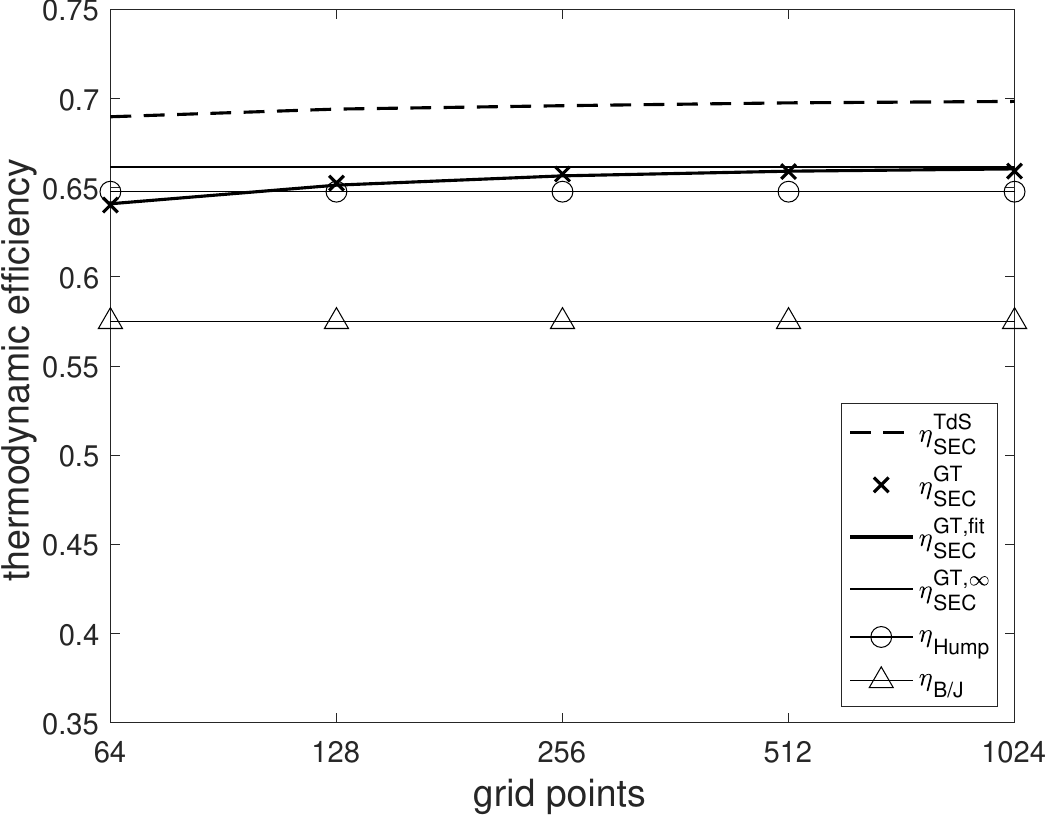}

\end{minipage}

\caption{Convergence of combustor efficiencies estimated from full engine simulations for varying grid resolution. Efficiencies estimated from: thermodynamic cycle integrals (dashed), shaft output vs.\ chemical energy input (crosses), fit of the latter (solid), grid-converged limit (thin solid) of the latter based on \eqref{eq:EfficiencyConvergenceFunction}. Efficiencies for given precompression ratio and mean equivalence ratio of the Humphrey (circles) and Brayton/Joule (triangles) cycles. Left: case~1 (compression ratio 6:1) from Table\,\ref{tab:CombustorModel1} with estimated limiting efficiency $\eta^{\text{\sf GT},\infty}_{\text{\sf SEC},1} = 0.52$, Humphrey cycle efficiency $\eta_{\text{\sf Hump},1} = 0.54$; Right: case~2 (compression ratio 20:1) from Table\,\ref{tab:CombustorModel2}, estimated limiting efficiency $\eta^{\text{\sf GT},\infty}_{\text{\sf SEC},2} = 0.66$, Humphrey cycle efficiency $\eta_{\text{\sf Hump},2} = 0.65$.}
\label{fig:EfficiencyConvergence}
\end{center}
\end{figure*}
%

% ===================================================================================
% ===================================================================================

\subsection{Estimates of thermodynamic efficiency}
\label{sec:ThermodynamicCycleEfficiency}

Since the exact time traces of the state variables vary from grid cell to grid cell during a period, it is very difficult to extract a meaningful estimate of the machine efficiency from time evolutions of the state variables in the Eulerian frame. Instead we provide here some efficiency estimates that are based directly upon the full engine simulations and upon the Humphrey cycle theory based on the mean pressures in the compressor plenum, and  we provide a more detailed estimate based on the evolution of the gasdynamic state variables along Lagrangian trajectories.  

% ===================================================================================

\subsubsection{Efficiency estimates from full-engine simulations}
\label{sec:FullEngineEfficiency}

Here we assess the overall efficiency, $\eta^{\text{\sf GT}}_{\text{\sf SEC}}$, of the SEC-based Gas Turbine model directly from simulations by the ratio of net power output to net chemical energy input over a gliding averaging time window of ca.~$\unit{0.34}{\second}$ (blue curves in Figs.\,\ref{fig:EnginePerformanceAndPlenumStates} and \ref{fig:EnginePerformanceAndPlenumStates_20-1}, fifth rows of panels), \ie,
\begin{equation}\label{eq:EfficiencyFromSimulation}
\eta^{\text{\sf GT}}_{\text{\sf SEC}}(t) = 
\frac{W_{\text{shaft}}\bigm|_{t-\delta t}^{t}}{Q \cdot M_{\text{fuel}}\bigm|_{t-\delta t}^{t}} 
\qquad
(\delta t \approx \unit{0.34}{\second})\,.
\end{equation}
As a convergence test for the numerical model, we display in Fig.\,\ref{fig:EfficiencyConvergence} the results of these efficiency estimates towards the end of the simulation time for both reference engine configurations and for a sequence of grid resolutions with grid cell numbers $N \in \{64, 128, 256, 512, 1024\}$ (crosses). The heavy solid lines in Fig.\,\ref{fig:EfficiencyConvergence} represent curve fits of the form
\begin{equation}\label{eq:EfficiencyConvergenceFunction}
\eta^{\text{\sf GT,fit}}_{\text{\sf SEC}}(N) = \eta^{\text{\sf GT},\infty}_{\text{\sf SEC}} + \frac{\delta\eta}{N^q}\,,
\end{equation}
where $\eta^{\text{\sf GT},\infty}_{\text{\sf SEC}}$ is the expected limiting efficiency at infinite grid resolution, the error coefficient $\delta\eta$ measures the deviation from the limit after rescaling by the convergence order, and $q$ is the estimated order of consistency of the estimate. In practice we calculate for fixed values of $q \in \{\frac{1}{2},1,\frac{3}{2}, 2\}$ the parameters $\eta^{\text{\sf GT},\infty}_{\text{\sf SEC}}$ and $\delta\eta$ by least-squares fits, and then select the value of $q$ that yields the smallest residual error across the resolutions. In both cases, a value of $q = 1$ corresponding to first order convergence clearly gives the best results (others not shown), and the corresponding curves are presented in the panels. With estimated limiting efficiencies of $\eta^{\text{\sf GT},\infty}_{\text{\sf SEC},1} = 0.52$ and $\eta^{\text{\sf GT},\infty}_{\text{\sf SEC},2} = 0.66$, we find efficiency gains over the classical deflagration-based gas turbine model of $30\%$ and $18\%$, respectively. See the end of section~\ref{sec:LagrangianPressureGainEfficiency} for further discussion of Fig.~\ref{fig:EfficiencyConvergence}.

% ===================================================================================

\subsubsection{Efficiency estimates from Lagrangian trajectories}
\label{sec:LagrangianPressureGainEfficiency}

The efficiency estimates based on the overall external energy in- and outputs from \eqref{eq:EfficiencyFromSimulation} are encouraging, but we would like to complement them by detailed internal thermodynamic process analyses along Lagrangian parcel paths as they pass through the combustor. This analysis is motivated by the fact that parts of the full engine model involve modelling assumptions whose overall effect on thermodynamic consistency is difficult to check. This concerns, in particular, the zero-dimensional representation of the plena and of their coupling to the one-dimensional combustor model. For thermodynamic consistency we require that the mass-averaged mean efficiencies based on individual Lagrangian paths should not come out less than the overall efficiency estimates, because the compressor, plena and turbine models themselves must not artificially increase or re-establish exergy that is lost in the combustor. For both reference cases the Lagrangian efficiency estimates indeed exceed the overall ones by several percentage points, which is reassuring in this sense, but also suggests the possibility of further efficiency improvements through optimized designs of inlet valve, combustor, and plena. 

% -----------------------------------------------------------------------------------

\medskip\noindent
\textbf{Parcel tracking}

During a simulation we monitor the total mass, $M_0^n = M_0(t^n)$, of working fluid that has entered the combustor by completion of the $n$th time step by adding up all mass fluxes across the inlet valve during the previous time steps. Within some time window close to the final simulation time, we use these total mass values as  labels of parcel trajectories that are to be tracked in the sequel. Let $M_0^{n_k}$ be this mass value associated with the $k$th trajectory. We find the position $x_k^l = x_k(t^l)$ of that parcel of working fluid at some later time $t^l > t^{n_k}$ by solving the following interpolation problem for given density $\rho(x,t)$ and combustor crossectional area $A(x)$: 
Let
´\begin{equation}\label{eq:MassBasedTracking}
M(t,x) = M_0(t) - \int\limits_0^{x} \rho(\xi,t) A(\xi) \, \textnormal{d}\xi \,, 
\qquad \text{then find $x_k^l$ such that} \qquad
M(t^l, x_k^l) = M_0^{n_k}\,.
\end{equation}
Since the density is positive everywhere in our simulations, $M(t,x)$ is monotonously decreasing as a function of $x$. Since the ideal inflow valve is diodic, $M(t) = M(t,0)$ is monotonously increasing with time, and therefore $M^{n_k} \leq M(t^l)$ when $t^l > t^{n_k}$. To account for slightly non-ideal behavior of the discrete valve model, we start a new Lagrangian trajectory, say $k+1$, only if $M^{n_{k+1}} > M^{n_k}$. As a consequence, the problem in \eqref{eq:MassBasedTracking} either has a unique solution $x_k^l \in [0,L]$ or the $k$th Lagrangian parcel has already left the combustor and cannot be tracked any further. 

For a numerical representation of the problem in \eq{eq:MassBasedTracking}, $\rho(x,t^l)$ is constructed as a piecewise constant approximation of the area-weighted density based on the cell averages $\rho_i^l A_i$ computed in the $i$th grid cell at time level $t^l$, \ie, 
\begin{equation}
\rho(x,t^l) = \frac{1}{2}\left( \rho_{i+1}^l A_{i+1} + \rho_i^l A_i \right) 
\qquad\text{for}\qquad 
x_i \leq x \leq x_{i+1} \quad \text{and} \quad  0 \leq i \leq I\,.
\end{equation}
Here the smallest and largest indices that appear, \ie, $i=0$ and $i = I+1$, denote auxiliary ghost cells placed just before the combustor entry and just behind the combustor exit, respectively, see also Fig.~\ref{fig:SystemDesign}. The flow states in these cells are set in the course of formulating the inflow and exit boundary conditions (see app.~\ref{app:BoundaryConditions}). This piecewise constant approximation of the density renders the integral in the problem \eqref{eq:MassBasedTracking} piecewise linear in $x$, yields a first-order accurate approximation of the interpolation problem, and solving it becomes straightforward. 

Once the positions of the tracked parcels have been localized this way, we compute the parcels' temperatures, $T$, pressures, $p$, velocities, $u$, and fuel mass fraction, $\phi$, by linear interpolation in terms of the mass coordinate between the neighboring cell centers. That is, when at time $t^l$ the $k$th trajectory with inflow mass $M_0^{n_k}$ is found between, say, cells $i-1$ and $i$ with mass values $M(t^l,x_{i-1})$ and $M(t^l,x_{i})$ according to \eqref{eq:MassBasedTracking}, then for one of said specific state variables $\psi \in \{T, p, u, \phi\}$ we let 
\begin{equation}
\psi_k^l = w_k^l \, \psi_{i-1}^l + (1-w_k^l) \, \psi_{i}^l 
\qquad\text{where}\qquad
w_k^l = \frac{M_0^{n_k} - M(t^l,x_{i})}{M(t^l,x_{i-1}) - M(t^l,x_{i})} \,.
\end{equation}
The parcels' entropies are then obtained via the equation of state from pressures and temperatures. In this fashion, we obtain time series $(t^l, x_k^l, T_k^l, S_k^l, ...)$, based on which we can subsequently analyse the thermodynamic cycles which the parcels undergo upon passage through the combustor and into the turbine plenum. Once a parcel has passed beyond the end of the combustor at time $t^{L_k}$, one last entry is added to its time series consisting of $t^{L_k}$ and the state variables taken from turbine plenum state at that time. This completes the parcel time series $(t^l, x_k^l, T_k^l, S_k^l, ...)_{l=n_k}^{L_k}$ for all parcels $k = 1, ..., K$. 

% -----------------------------------------------------------------------------------

\medskip\noindent
\textbf{Thermodynamic cycles along Lagrangian paths}

\begin{figure*}
  \centering
  \includegraphics[width=0.9\textwidth]{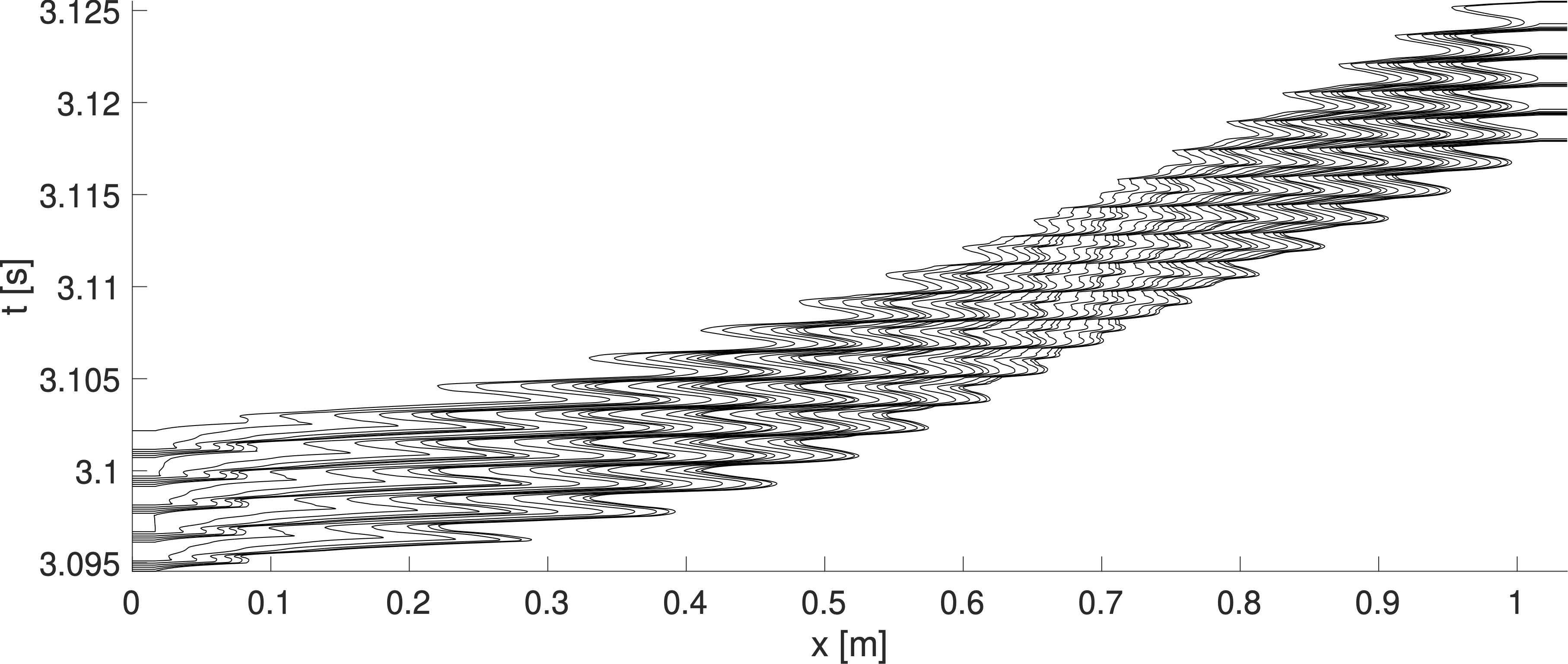}

  \medskip

  \includegraphics[width=0.9\textwidth]{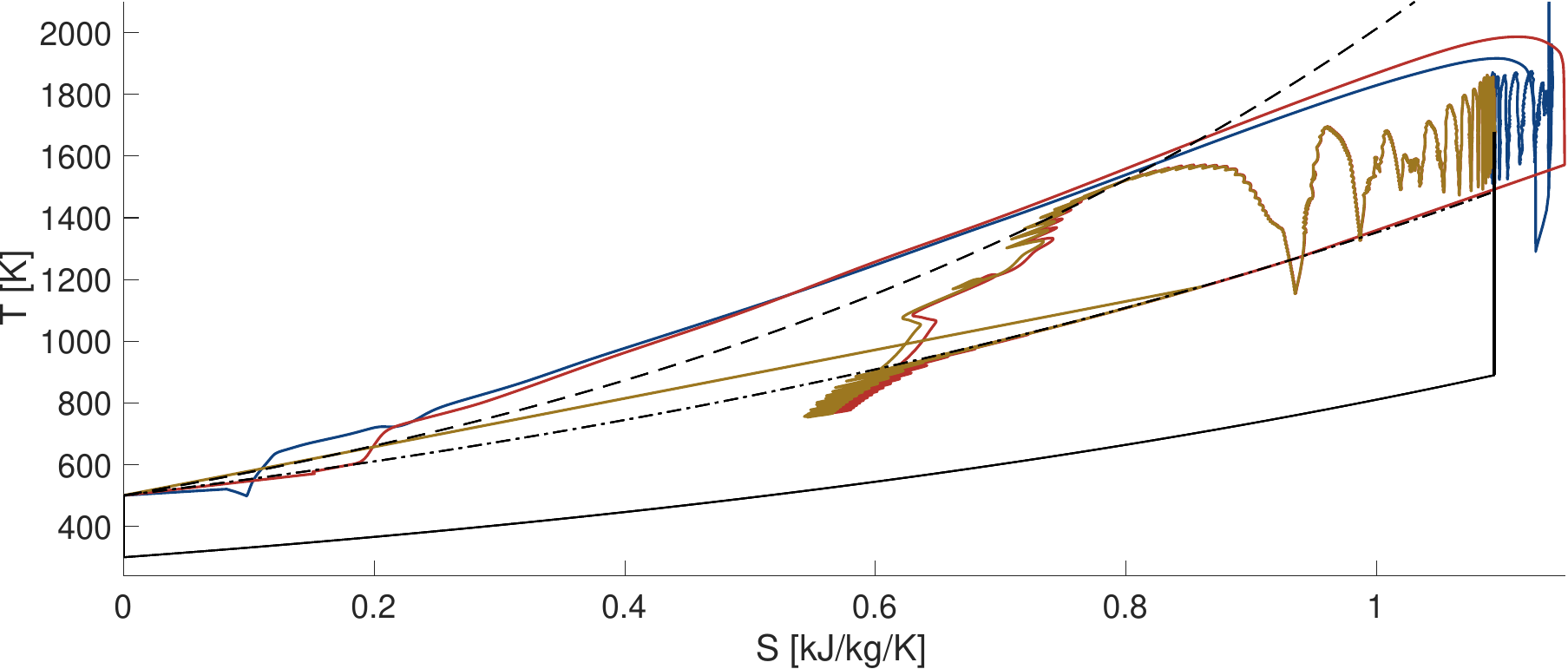}

  \medskip

  \includegraphics[width=0.9\textwidth]{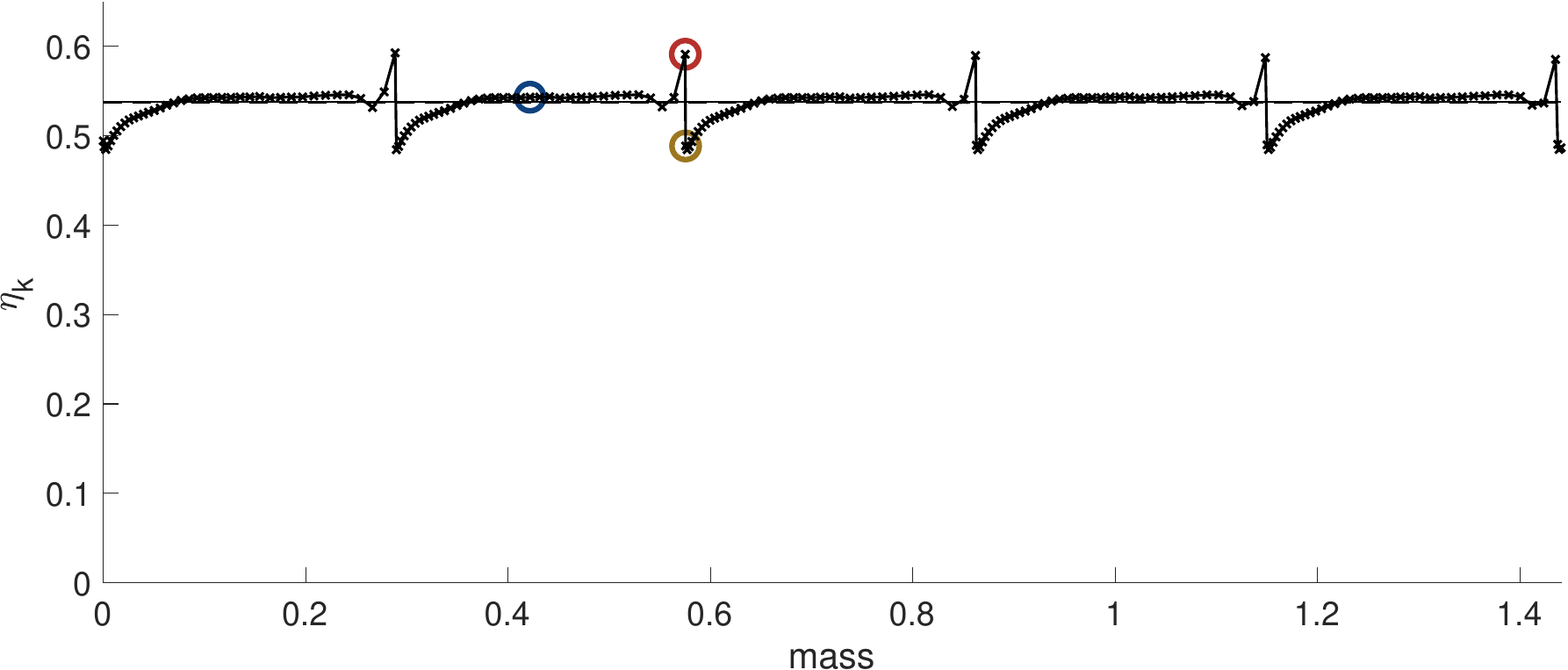}
  
  \caption{Top: Every seventh Lagrangian trajectory out of 200 trajectories collected over a time span covering six valve openings towards the end of the SEC-simulation for case~1 ($p_2/p_1 = 6:1$). Middle: $T$-$S$ (temperature-entropy) diagrams for selected parcel trajectories. Blue indicates a trajectory from somewhere in the middle and red the last trajectory from one valve opening time window, whereas dark-yellow designates the first (pure air) trajectory tracked from the next one. The black solid lines outline the idealized lossless compression and expansion in the compressor and turbine as well as the isobaric heat removal that closes the thermodynamic cycles. We also show the curves for isochoric (dashed) and isobaric (dash-dotted) combustion corresponding to the Humphrey and Brayton/Joule cycles starting from the compressor plenum state, respectively. Bottom: Efficiencies computed on the basis of thermodynamic cycle analyses along the trajectories from five full valve openings. $\rule[2.55pt]{7.5pt}{1pt} \times \rule[2.55pt]{7.5pt}{1pt}$: Efficiencies for individual trajectories; $\rule[2.55pt]{15pt}{1pt}$: mass-weighted mean value.}
\label{fig:LagrangeTrajectories}
\end{figure*}

To illustrate this, we discuss results from a simulation of reference case~1 (6:1 compression ratio) at 256-cell grid resolution in Fig.~\ref{fig:LagrangeTrajectories}. The top panel shows every seventh parcel trajectory from a total of 200 tracked parcels in an $x$-$t$-diagram. It is seen how mass parcels are pushed along the tube by the sequence of strong pressure waves that emanate from the auto-igniting charges while they oscillate more or less in place in between those pulses. The graphics also illustrates how tracked parcel paths are terminated once parcels have moved beyond the combustor exit.

The second panel shows the thermodynamic cycles along three parcel trajectories in terms of a $T$-$S$ diagram as relevant for the subsequent efficiency estimates. The cycles correspond to some parcel from the middle of a charge (blue), the last tracked parcel of the same charge (red), and the first tracked parcel (dark-yellow) of the next charge. Starting from the compressor plenum state $(S,T) = (\unit{0}{\kilo\joule\per\kilo\gram\per\kelvin}, \unit{500}{\kelvin})$ the typical blue path shows first the relatively small temperature and entropy increases induced by the the temperature control at the combustor inlet (cf.~App.~\ref{sapp:InjectionControlSEC}), which is followed by a sudden increase of temperature and entropy that is induced by the arrival of a backward running shock. This temperature rise ignites the mixture and the blue curve then follows that of isochoric combustion rather well up until about $S = \unit{0.7}{\kilo\joule\per\kilo\gram\per\kelvin}$ -- compare with the isochore (dashed line) emanating from the compressor plenum state. Beyond this point, the slope of the curve decreases until, towards the right edge of the plot, it resembles more an isobaric than an isochoric reaction process -- compare with the isobar (dashed-dotted line) emanating from the compressor plenum state. Finally, the temperature drops strongly and almost adiabatically ($S = \text{const.}$) along the blue curve which signals the arrival of a strong backward running suction wave that reduces the pressure near the combustor entry sufficiently to open the valve again for the next recharging process. The remaining oscillations of the blue curve at nearly constant entropy trace its state as it travels along the combustor and towards the turbine plenum and is subject to the effect of the resonant pressure waves and turbulence in the tube. It finally ends up in the turbine plenum state marked by the top of the black vertical line at about $S = \unit{1.1}{\kilo\joule\per\kilo\gram\per\kelvin}$. Its thermodynamic cycle is closed by adiabatic expansion in the turbine (represented by said black vertical line) and isobaric cooling back to zero entropy and to the environmental temperature of $T_1 = \unit{300}{\kelvin}$ (lower isobar closing the cycle (black solid)). 

Consider next the red and dark-yellow curves in the same graph. The red path closely resembles the blue one up to the end of its strong temperature drop induced by the suction wave and visible at the right edge of the graph. From here, the parcel's thermodynamic state essentially jumps along an isobar to meet with the first state along the dark-yellow curve that follows immediately after its emergence from the compressor plenum state. These coupled rapid changes indicate strong diffusive mixing of two adjacent mass parcels at constant pressure: In fact, both starting points and the new joint mixed state lie on the same dash-dotted isobar. In the sequel, both parcels get to be cooled further by mixing to some extent with the rest of the fresh air buffer before being heat up again along a steeper path that brings the two trajectories close to the dashed isochore. The increase of entropy along this part of the path is due to mixing with the hot gas from the last charge (to which the red-marked parcel belongs). That the temperature increases rather rapidly during the process, with a slope of the curve substantially exceeding that of the isochoric path (dashed), is due to strong compression of the gas by backward propagating pressure waves when the diodic valve has closed for the next combustion cycle. The remaining oscillatory path from a tangent to the dashed isochore towards the turbine plenum state is due to further mixing, the mean pressure rise along the diffusor section of the combustor, and the oscillatory pressure changes induced by the resonant large amplitude acoustics in the tube.   

The third panel of Fig.~\ref{fig:LagrangeTrajectories} displays the individual efficiencies (crosses) for all of the parcels that have entered the combustor during five successive full SEC cycles during a short time window around $t = \unit{3.14}{\second}$. These efficiencies all lie around the mean value of $\overline{\eta}_{\text{\sffamily SEC}}^{\text{\sffamily TdS}} = 0.537$ indicated by the solid line. The sharp down- and upturns in the sequence of individual efficiencies mark the first (bottom of downturn) and last (top of upturn) parcels of a fresh charge which correspond to the dark-yellow and red paths in Fig.~\ref{fig:LagrangeTrajectories}, respectively. It may seem remarkable that mass parcels which entered the combustor as part of the fresh air buffer, i.e., the next couple of tracked parcels following one of the steep downturns, are associated with an individual efficiency just about 10\% lower than those seen for parcels from the bulk of the charge. At the same time, why would the parcels that have last entered the combustor come with an efficiency that stands out and is about 10\% larger than the mean? These effects are understood following the thermodynamic cycles of the red and dark-yellow marked parcels in the middle panel of the figure, whose efficiencies are encircled by the same colors in the bottom panel: Similar to the blue-marked mass element from the core of the charge in question, the red-marked parcel first accumulates a positive contribution to its $T$-$S$-integral while undergoing near constant volume combustion along the upper branch of its $T$-$S$-trace. Its backward jump in entropy is due to mixing with the fresh air yellow-marked parcel from the next charge. The subsequent further cooling of both parcels corresponds to further mixing with the fresh charge and represents a loss (negative contribution to the $T$-$S$-integral). This loss is more than compensated, however, by the subsequent strong recompression and re-heating by mixing with the hot reacted gas that lies ahead of the red-marked parcel along the tube. As a consequence, the red curve again adds a strong positive contribution to its $T$-$S$-integral while moving along the oscillatory path in the diagram towards the turbine plenum state. The fate of parcels from the fresh air buffer is exemplified here by the dark-yellow curve and circle in the middle and bottom panels of the figure. These parcels are not at all heated by chemical reaction, but they do receive heat energy by mixing with the burnt gas. As a consequence, they also accumulate substantial positive contributions to their $T$-$S$-integrals and deliver positive work throughout their cycle accordingly. 

Going back to Fig.~\ref{fig:EfficiencyConvergence}, we observe that the efficiency estimates based on the full engine performance (thick solid line) lie below the mean values obtained from the present thermodynamic cycle analysis by several percentage points. This is reassuring: While the computational full-engine model is mass and energy conserving by design of the numerical method, we do not have an \emph{a priori} guarantee that all of its components -- especially the combustor-plenum coupling schemes -- are also consistent with the second law of thermodynamics.  The present results indicate that the combustor alone is actually more efficient than the full engine, which means that the other engine model components do not artificially enhance the engine performance, \ie, that the rather favorable full engine efficiencies are not an artefact. They also indicate that there even is potential for further improvements by optimized designs of all engine components. Such optimization is left for future work.

% ===================================================================================
% ===================================================================================
% ===================================================================================

\section{Multidimensional wave propagation in the turbine plenum}
\label{sec:MultiDimensionalWavePropagation}
%\zenker{The compressor plenum is still 0D, so we cannot say anything about the multidimensional wave propagation.}
%\zenker{TODO replace tube with combustor...}

Strong pressure waves leaving the combustor tube should be mitigated to an acceptable level at turbine entry by means of a large volume turbine plenum. In this section we evaluate the potential impact of multidimensional wave propagation within the turbine plenum on engine efficiency and confirm the validity of the zero-dimensional representation of the plenum in the whole engine model above, which considers thermodynamic changes of state only. We conduct this with a simplified two-dimensional representation of the turbine plenum coupled to the one-dimensional combustor model. More precisely, we extend the representation of the turbine plenum by only one cell, for which we balanced mass and energy (cf. figure~\ref{fig:SystemDesign}), to a representation by a structured two-dimensional spatial grid. For solving the two-dimensional Euler equations we use the same numerical schemes as before (see appendix~\ref{app:Quasi1DEulerDiscretization}) together with operator splitting in the spatial directions.

To couple the one-dimensional combustion tube and the two-dimensional turbine plenum, we define an overlap region in which the data of the conserved quantities are exchanged, and we fill the ghost cells from one domain with the corresponding data from the other.
This means that we obtain the ghost cell states of the 1D tube from the averaged data of the 2D plenum. 
Conversely, we distribute the states from the tube to the 2D plenum ghost cells in such a way that the sum of the conservation variables remains the same. The numerical scheme and the domain coupling have been presented in a number of previous works, see,  e.g., \cite{Nadolski2019,Haghdoost2020,Haghdoost2021,Thethy2022}.
%\oevi{In case the cell boundaries in axial direction of the one-dimensional combustor tube and the two-dimensional plenum do not coincide in the overlap region, data is linearly interpolated to the corresponding cell center position.}

\begin{figure*}
\centering
\includegraphics[width=\textwidth]{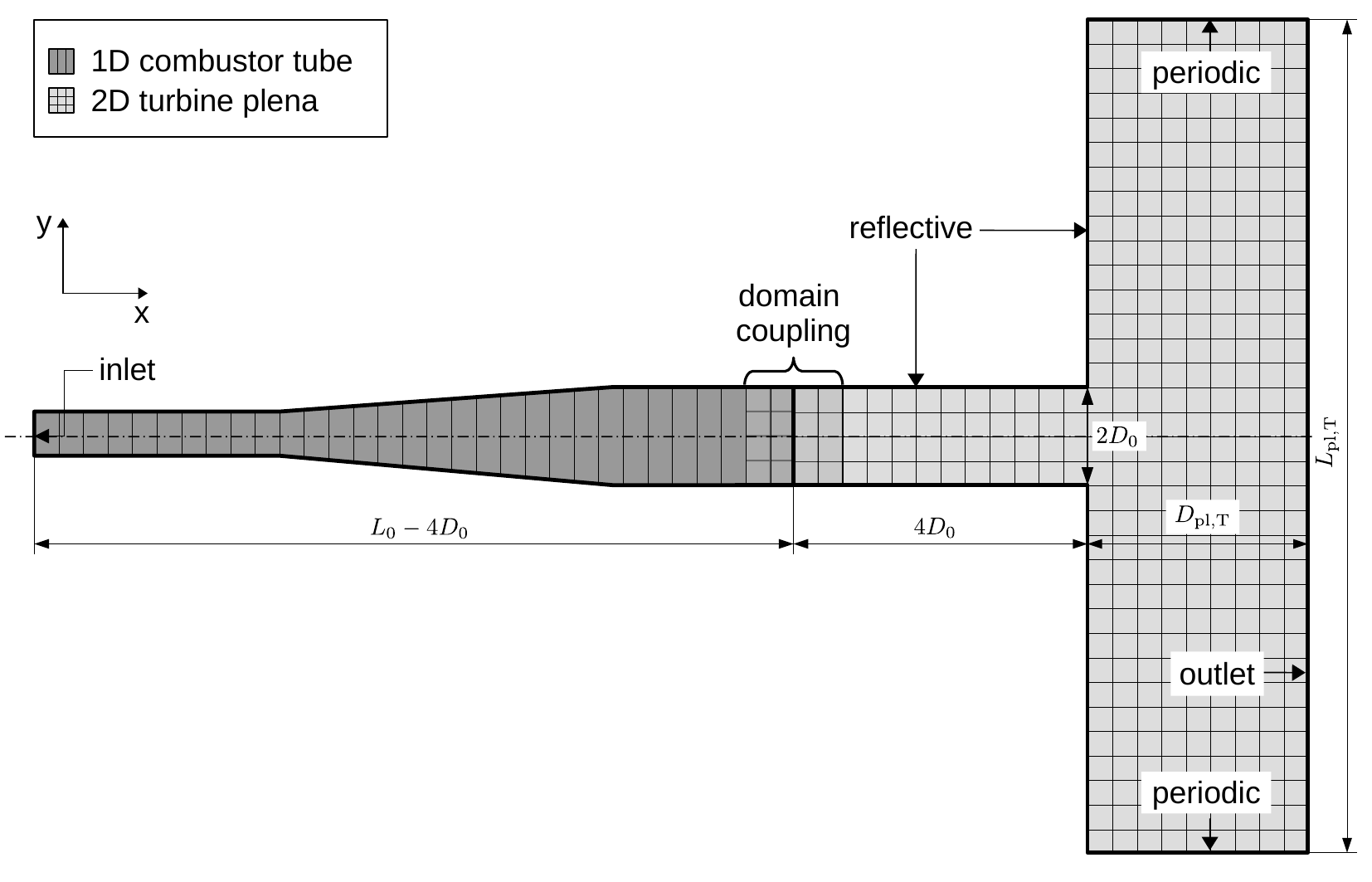}
\caption{\label{fig:2DPlenaCFDMesh}Overview of the simulation domain. The used boundary conditions and geometrical details are also shown in the sketch.}
\end{figure*}
The geometry of the simplified planar representation of the turbine plenum is shown in figure \ref{fig:2DPlenaCFDMesh}. 
As mentioned above, the plenum is idealized as a torus with area and length ratio relative to the combustor tube according to tables \ref{tab:CombustorModel1} and \ref{tab:CombustorModel2} with 
$D_{\text{pl,T}} = \sqrt{A_{\text{pl,T}} / A_0} \, D_0$  and 
$D_0 = 0.05 \, \rfr{l}$ from equation (\ref{eq:TurbulenceEstimates}~b). For compression ratio 1:6, which we consider here only, this leads to $D_0 = \unit{5}{cm}$, $D_{\text{pl,T}} = \sqrt{8} \, D_0 = \unit{14.14}{cm}$ and $L_{\textnormal{pl,T}} = 5\, \rfr{l} = \unit{5}{m}$. However, in order to realize a uniform Cartesian mesh without cut-cells we resolve the combustor exit diameter of $\unit{10}{cm}$ with grid resolutions of
$\Delta x = \Delta y = \unit{1}{cm}$ and  $\unit{0.5}{cm}$ leading to $7200$ and $14400$ grid cells for the 2D domain, respectively. To obtain a stable SEC mode in conjunction with the 2D plenum setting we needed to increase the turbine mass flux parameter $C_\text{T}$ by a factor of $1.75$.

The applied boundary conditions are shown in figure \ref{fig:2DPlenaCFDMesh} as well. Consistent with solving the inviscid Euler equations in the two-dimensional domain, fixed walls are modeled with a reflective, i.e. inviscid, boundary condition whereas across the upper and lower plenum boundaries periodicity is assumed. The right boundary of the plenum, which corresponds to the turbine inlet, is modelled as an outflow boundary condition. The fluxes across this boundary are determined from exact solutions of local Riemann problems with prescribed mass fluxes. The local mass fluxes are determined in a two step process: First, we calculate the total mass flux across the turbine plenum outlet by integrating the local mass fluxes in each cell via equation (\ref{eq:TurbineEquation}) and second, an area weighted distribution of the total mass flux is applied to the outlet cell faces.

\begin{figure*}
\centering
\includegraphics[width=1.\textwidth]{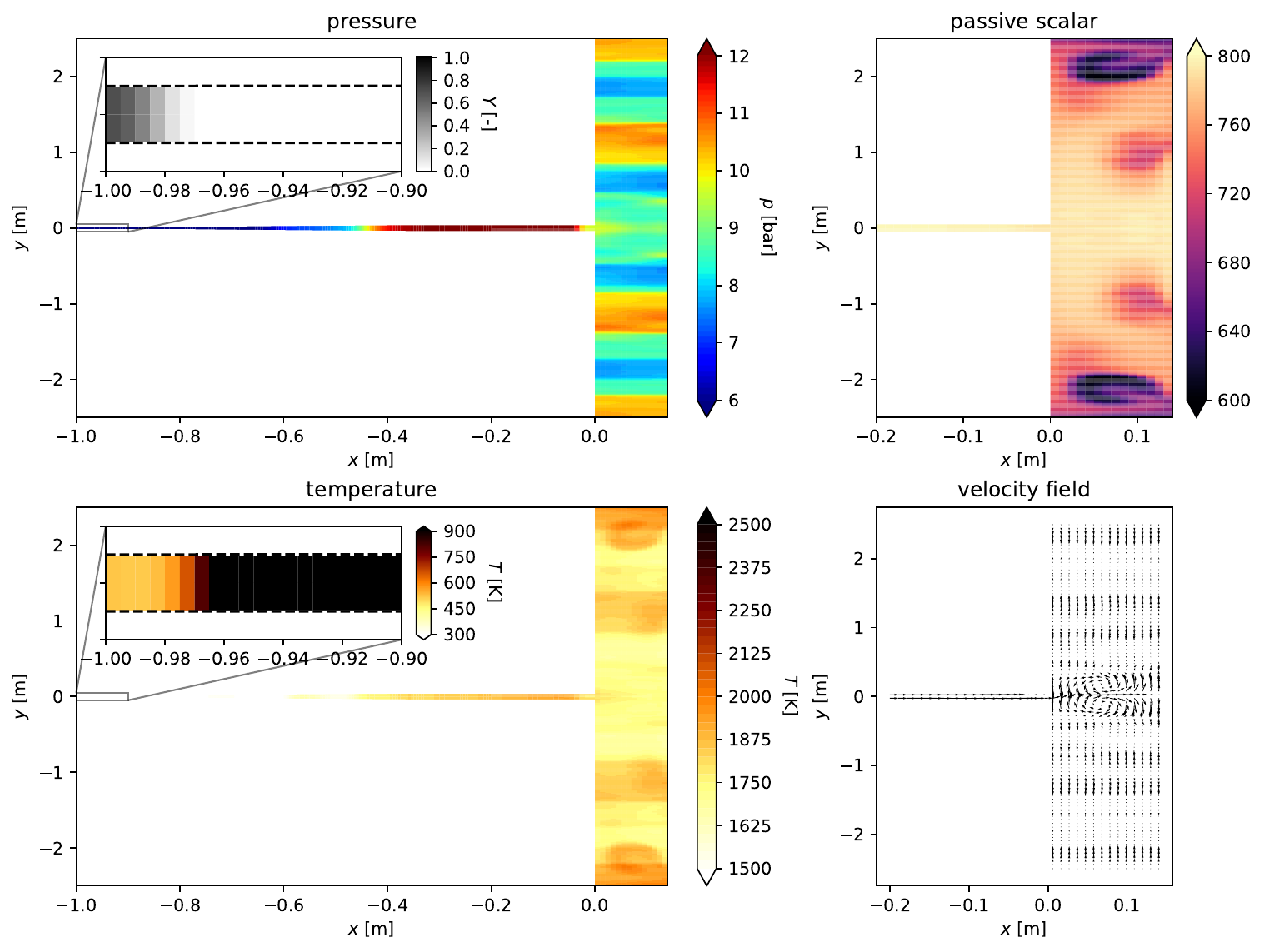}
\caption{\label{fig:2DPlena}Snapshots from the second order simulation on the coarse mesh (case C3, table~\ref{tab:2DPlenaFluctuations}) at $t=\unit{2.725}{s}$ for pressure, temperature, passive scalar and velocity vectors. The passive scalar shown in upper right panel corresponds to the scalar used for the Lagrangian parcel tracking inside the combustor, cf.\ section \ref{sec:LagrangianPressureGainEfficiency}.}
\end{figure*}

Snapshots of pressure, velocity and temperature at $t = \unit{2.723}{s}$ in figure \ref{fig:2DPlena} provide a qualitative picture of the flow structure in SEC mode. Strong axial pressure waves inside the combustor are mitigated to pressure waves of lower amplitude in the circumferential direction in the plenum, \ie, in the vertical direction in Fig.~\ref{fig:2DPlena}. The pressure wave amplitudes directly correlate with the observed temperature variations seen in the lower left panel. The plot in the upper right of the figure shows the passive scalar used for the Lagrangian parcel tracking described in section \ref{sec:LagrangianPressureGainEfficiency}, which is evolved in two space dimensions as a passively advected scalar. Together with the velocity vectors in the lower right panel, large scale vortex structures inside the plenum can be identified which need to be investigated qualitatively and quantitatively with a full 3D simulation left for the future.

Table \ref{tab:2DPlenaFluctuations} summarizes key parameters and results for simulations with 0D and 2D models for the plena for different grid resolutions. The results for the 0D plenum confirm that a resolution of the 1D combustor domain with 256 grid cells is sufficient here and that (time) averaged values for pressure, temperature, mass flow and efficiency do not change significantly any more under further grid refinement. Simulations with the coupled 2D plenum have been performed for each grid resolution with two different numerical schemes for the advective fluxes: a first order upwind scheme and a second order scheme using the Van Leer flux limiter. Whereas results obtained with the second order upwind scheme on both grids and the first order scheme using the fine grid resolution show similar averaged mean values for the shown quantities, the first order simulation on the coarse grid features larger deviations from the other simulations. Interestingly, the efficiencies obtained from the 2D plenum are actually higher than those from simulations with the 0D plenum. This is also reflected in higher pressures and temperatures. A possible explanation for these higher efficiencies is that part of the kinetic energy of the flow leaving the combustor and work done by pressure waves inside the plenum can be recovered in two space dimensions, whereas neither of them can be accounted for  
% have not been considered not with the simplifying assumption of an adiabatic change of state of the combustor exit state to the thermodynamic state of the plenum 
in the 0D case, cf.\ section \ref{sec:TurbinePlenumAndTurbineModel}. In this respect, the efficiencies presented in the previous sections of the paper should be regarded as conservative estimates.

\begin{table*}[]
  \centering
  \caption{\label{tab:2DPlenaFluctuations}Summary of key outcomes for various simulations with 0D and 2D turbine plenum. Pressure and temperature values are (time) averaged mean values over the whole 2D turbine plenum. $\dot{m}$ is the mean mass-flow over the entire outlet boundary of the 2D plena. $\Delta x$ denotes the grid spacing for the 2D plenum simulations. Note that the length of the 1D combustor domain is $L_0$ for the 0D plenum and $L_0 - 4D_0$ for the 2D plenum simulation, cf.\ figure \ref{fig:2DPlenaCFDMesh}.}
  \begin{tabular}{c|c|c|c|c|c|c|c}
   & case & $n_{1D}$& $\Delta x$~[cm] & $p$ [bar] & $T$ [K] & $\dot{m}$ [kg/s] &  $\eta$ \\ \hline
  & 0D turbine plenum & 256 & & 8.5153 & 1753.02 & 10.358 & 0.5143 \\ 
  & 0D turbine plenum & 512 & & 8.5402 & 1754.95 & 10.383 & 0.5153 \\ 
  & 0D turbine plenum & 1024 & & 8.5530 & 1756.62 & 10.393 & 0.5160 \\ \hline
  C1 & 2D turbine plenum 1st or. & 160 & 1 & 8.2798 & 1757.80 & 10.052 & 0.5084 \\
  C2 & 2D turbine plenum 1st or. & 320 & 0.5 & 8.985 & 1783.16 & 10.826 & 0.532 \\ \hline
  C3 & 2D turbine plenum 2nd or. & 160 & 1 & 9.1471 & 1795.46 & 10.979 & 0.5379 \\
  C4 & 2D turbine plenum 2nd or.  & 320 & 0.5 & 8.893 & 1797.07 & 10.607 & 0.5334 \\
  \end{tabular}
\end{table*}
\begin{figure*}
\centering
\includegraphics[width=1.\textwidth]{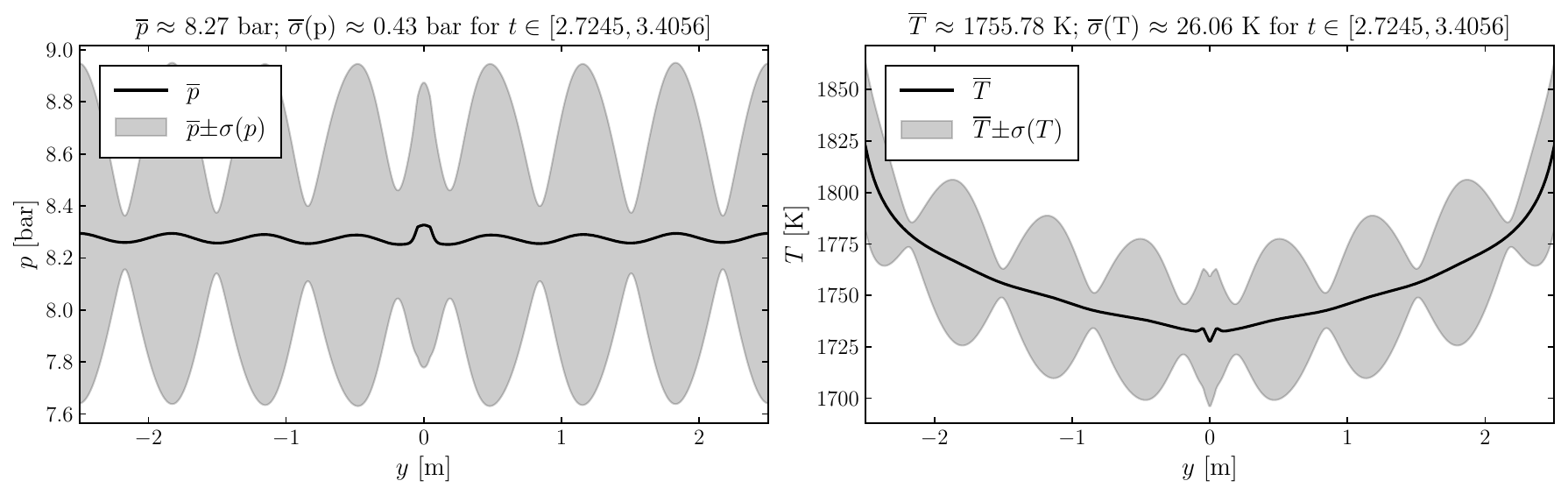}
\includegraphics[width=1.\textwidth]{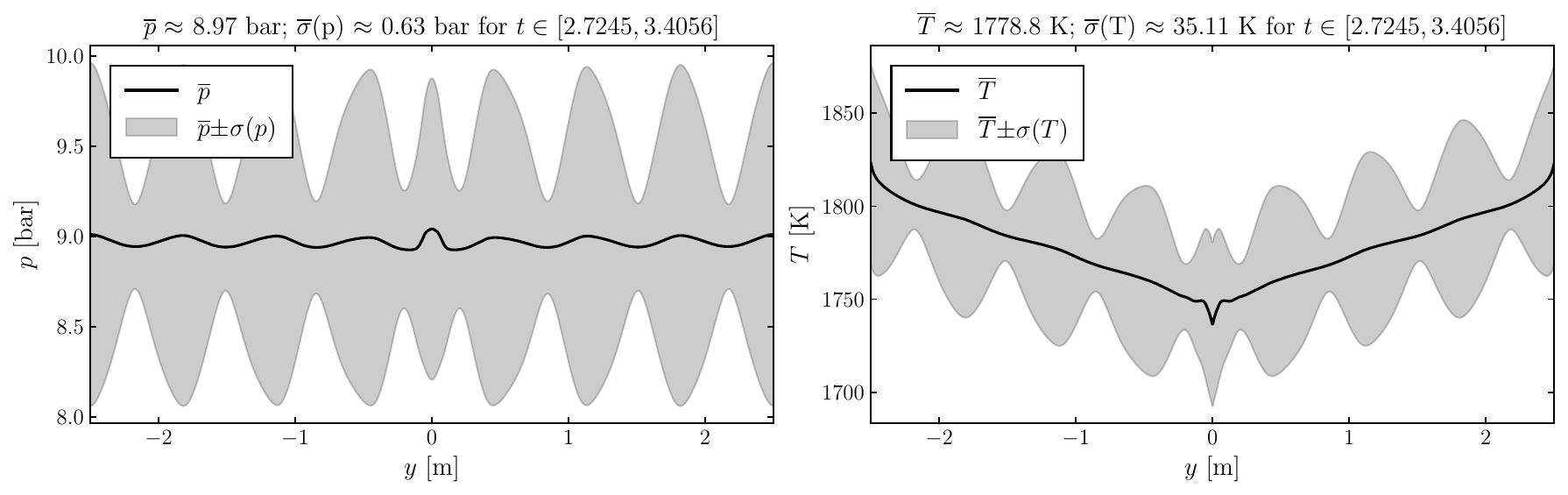}
\includegraphics[width=1.\textwidth]{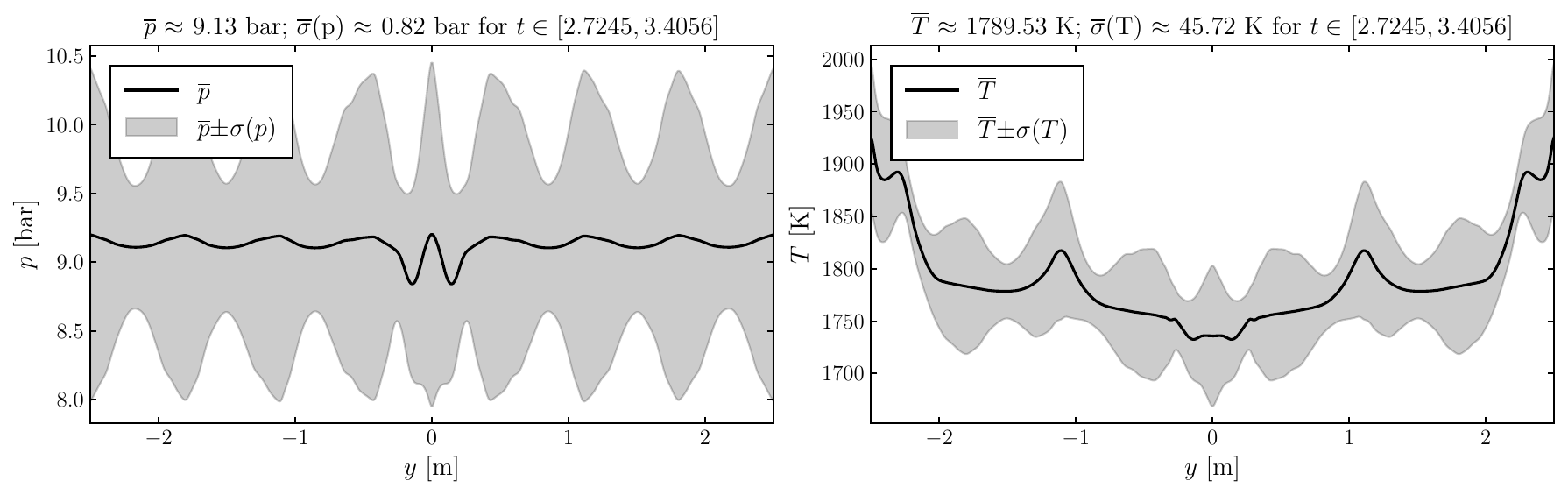}
\includegraphics[width=1.\textwidth]{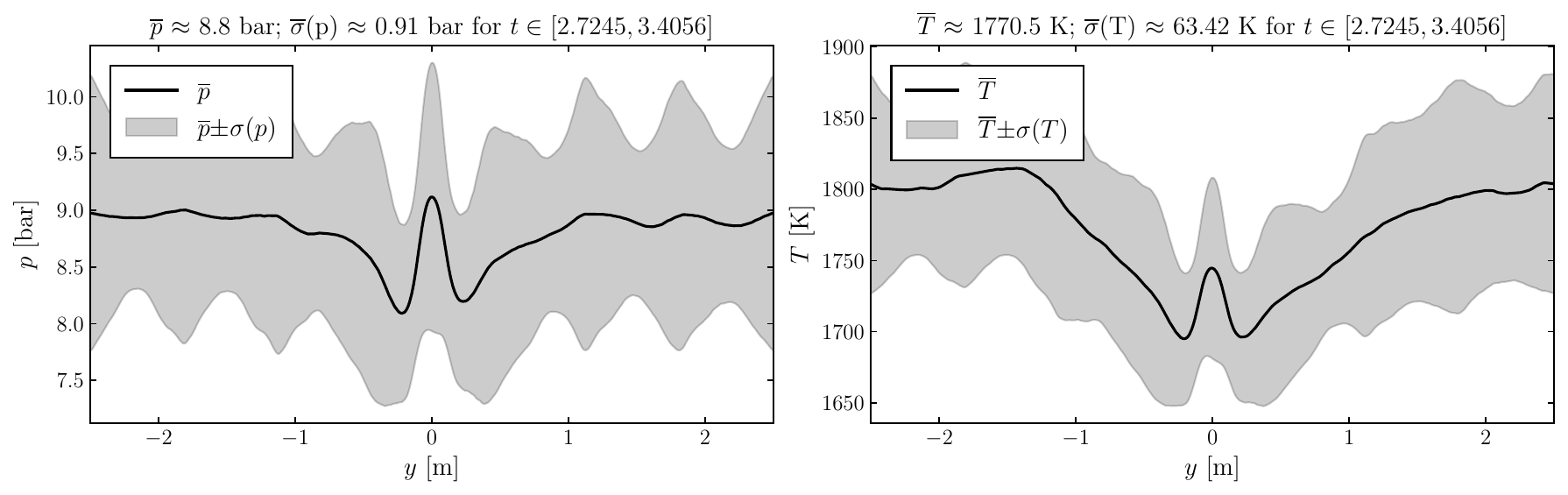}
\caption{\label{fig:2DPlenaFluctuations}Pressure and temperature profiles at turbine plenum outlet for cases C1 to C4 (top to bottom), cf.\ table~\ref{tab:2DPlenaFluctuations}.
Solid lines represent time averaged mean values and the shaded areas the corresponding standard deviations.}
\end{figure*}

Figure~\ref{fig:2DPlenaFluctuations} displays time-averaged spatial profiles with standard deviations of pressure (left) and temperature (right) at the turbine plenum outlet for cases C1 to C4, see table \ref{tab:2DPlenaFluctuations}. As expected, the simulations with the second order scheme reveal substantially more spatial structure than the corresponding first order simulations on the same grids. Moreover, the first order simulations are perfectly symmetric relative to the centerline $y = \unit{0}{cm}$. The symmetry is also still present for the second order scheme on the coarse grid, however on the fine grid slightly asymmetric profiles are visible. This effect is often seen in analytically symmetric flow simulations when slope limiters are used in the context of higher-order upwind methods. The origin is the nonlinear clipping of extrema that arises consistently along lines of symmetry and induces random fluctuations in response to round-off errors. Except for the region around $y=\unit{0}{cm}$ where the jet-like flow exiting the combustor directly interacts with the plenum outlet, the (time) averaged pressure features little spatial variations with comparatively low standard deviation. 
The observed temperature stratification with peak values at the upper and lower end of the plenum outlet is a result of the large vortex structures in the turbine plenum, compare figure \ref{fig:2DPlena}. It is expected that this strong temperature stratification will disappear in case the combustor tubes are not oriented perpendicularly to the turbine plenum anymore and thus induce a circumferential mean flow inside the plenum. In-depth studies of these multi-dimensional effects are deferred to future work.

% ===================================================================================
% ===================================================================================
% ===================================================================================

\section{Conclusions and outlook}
\label{sec:Conclusions}

This paper proposes an idealized computational model representing a single-shaft gasturbine driven by a combustor that optionally operates in either the classical turbulent premixed deflagration or in the SEC (shockless explosion combustion) mode \citep{BobuschEtAl2014}. Combustion is based on one-step Arrhenius kinetics with a lower cut-off temperature \citep{BerkenboschEtAl1998}. The model assumes quasistationary characteristics for the (lossless) compressor and turbine, a quasi-one-dimensional (1D) representation of the combustor, and zero-dimensional (0D) budgets of mass and energy for two plena connecting these components. An optional two-dimensional representation of the combustor exit and turbine plenum has been used both to validate the idealized coupling of the 1D combustor and the 0D turbine plenum in the simpler model version and to yield estimates of pressure fluctuation amplitudes near the turbine entry port.

This model allowed us to simulate the engine's acceleration from right after its start-up in deflagration mode to the targeted compressor pressure ratio, a subsequent switch to SEC mode, and the further approach to quasi-stationary resonant operation. Deflagration-only runs as well as simulations that included the transition to SEC mode have been performed for two reference compressor pressure ratios of 6:1 and 20:1, and with global equivalence ratios tuned to guarantee turbine plenum temperatures of $\unit{1775 \pm 5}{\kelvin}$. Chemical parameters and the combustor's diffusor dimensions were set such that for the 20:1-case the combustor oscillations locked into the lowest-possible resonant frequency (single-frequency mode), and that it attained twice that frequency in the 6:1-case (double-frequency mode). Our comparisons yielded efficiency gains of the SEC over the deflagration-based full engine of $30\%$ and $18\%$, while efficiency gains relative to the classical Brayton/Joule cycles were $29\%$ and $15\%$ for the reference compressor pressure ratios of 6:1 and 20:1, respectively. These efficiencies have been calculated directly as the ratio of mean rate of chemical energy input and engine shaft power output. To generate corroborating evidence that these high efficiencies are not artefacts of our computational model, we have tracked individual fluid parcels on their way through the combustor and calculated parcel-wise efficiencies based on their traces in a $T$-$S$-diagram. The parcel-based efficiencies systematically exceed the full engine efficiencies by another few percentage points. This is not only reassuring in that the full engine model does not unphysically yield higher efficiency than the combustor can provide in the first place, but it also hints at further room for improvement by optimization of all engine components.

``Pressure gain'' is achieved in two ways within the SEC-based engine. On the one hand, borrowing from known pulsed combustor designs, see, e.g., \citep{PutnamEtAl1986}, large amplitude resonant acoustic oscillations in the combustor enable its recharging against a mean pressure increase between the compressor and turbine plena. An across-combustor pressure ratio of about 1.5 is achieved for both reference precompression ratios. This mean pressure increase is essential for the observed combustor efficiencies as it guarantees that backward propagating pressure waves, upon returning from the turbine plenum to the combustor inlet, substantially post-compress the fresh charge before it auto-ignites and releases its chemical energy. The observed SEC combustor efficiency can be explained qualitatively by an idealized Humphrey cycle with a total pre-reaction gas compression composed of the compressor pressure ratio and this gasdynamic post-compression. 

The simulations utilizing the two-dimensional representation of the turbine plenum further corroborate the results from the simpler model in that the observed overall efficiencies are slightly higher the 2D case. In addition, the 2D simulations indicate that turbine inlet pressure amplitudes can effectively be mitigated by appropriate geometrical system designs. Even for the present rectangular geometry with the combustor firing directly towards the turbine inlet section, relative pressure fluctuations show a relative variance of 10-15\%. 

Extensive parameter sensitivity studies for the SEC-operated engine are deferred to future work, but our tests have already revealed some interesting dependencies. While turbulent transport can be beneficial in stabilizing the double-frequency SEC mode, it does not seem to be crucial and was not needed in the single-frequency mode. Both the single- and double-frequency SEC-modes could be realized for both the 6:1 and 20:1 precompression ratios with suitably adjusted chemical parameters and combustor geometries but, for brevity, only one mode was documented for each reference case above. 

In the sense of a proof of concept, the present study demonstrates that a concrete SEC-based gas turbine design may be achieveable that delivers much of the known theoretical efficiency gains of pulsating constant volume combustion. Yet, the suggested computational model involves several assumptions that call for further research in upcoming studies:
\begin{itemize}

\item Realistic chemistry: Here we have adopted a simple one-step Arrhenius reaction model determined essentially by the ignition delay time and its (exponential) temperature sensitivity. Current and future gasturbine fuels come with complex chemistry, however, that involves several additional independent time scales, such as different delay times during the overall autoignition process or the time scale of the subsequent chemical heat release (``excitation time'') \citep{CaiPitsch2014,DjodjevicEtAl2019,VinkeloeEtAl2020}. Whether this will turn out to be an obstacle for efficient tuning of SEC resonance or whether, in contrast, it can be utilized constructively will have to be investigated. 

% \item The full-engine setup, including the deflagration run-up, was implemented to demonstrate the possibility of a hybrid deflagration/SEC engine that can run in deflagration mode at low and partial loads while utilizing the SEC mode as a highly efficient ``overdrive''. 

\item One-way inlet valve: Here we have assumed the availability of an ideal diodic valve without spelling out its concrete design. Clearly, in a gasturbine a mechanical valve (with oscillating parts) should be avoided, so that some intelligent fluidics-based flow diode should be developed. A promising design was suggested by \citet{BobuschEtAl2014,MairEtAl2019}. A very preliminary test based on a low-order numerical representation of this design by \citet{Berndt2016} showed that losses in the inlet valve can significantly affect resonant SEC operation. A related study based on computational simulation should be in reach utilizing modern computational fluid dynamics methods. 

\item Three-dimensional effects: A fully three-dimensional computational model, including the combustor and plena, will be of high interest eventually as it would enable more realistic estimates of the influence of turbulence on the combustor dynamics or of the geometrical design of the system. The latter may be important, e.g., for further efficiency gains through improved recovery of the remaining kinetic energy of the burnt gas when entering the turbine plenum.

\item Theoretical model: Finally, a combination of the present idealized estimates based on an extended Humphrey cycle with theoretical descriptions of the large-amplitude resonant waves in the combustor may lead to a predictive theory describing the nearly stationary equilibrium pulsations seen towards the end of the present SEC-mode simulations.

% \item Lowering turbine entry temperatures by sidetracking compressed air will not straightforwardly be possible, owing to the significant pressure gain across the combustor. One might think of enlarging the inert air buffer in the SEC so as to make up for this. We should try how far down we can drive the turbine plenum temperature without loosing the edge over the classical turbine. 

% \item Noisy curves in engine performance plots. 

% \item Design of a fluidics-based unidirectional valve.

% \item Design of more sensitive and optimezed injection control for minimizing the danger of local detonation-runup.  

\end{itemize}

\section*{Acknowledgements}
The authors thank Deutsche Forschungsgemeinschaft for funding under grant CRC~1029 ``Substantial efficiency increase in gas turbines through direct use of coupled unsteady combustion and flow dynamics'', Project No.~200291049, subprojects A03 and C01.

\section*{Declaration of Interests}
The authors report no conflict of interest.

\appendix

\section{Charge stratification for SEC cycles}
\label{app:InjectionControl}

% ===================================================================================
% ===================================================================================

\subsection{Control of ignition delay times by inflow temperature adjustments}
\label{sapp:InjectionControlSEC}

Here we describe how to vary the inflow temperature into the combustor so as to achieve nearly homogeneous auto-ignition of the charge in an SEC cycle. The key motivation for this approach lies in the fact that typical gasturbine fuels exhibit rather large activation energies over most of the presently relevant temperature range. Therefore, relatively weak inflow temperature variations allow us to significantly influence the stratification of a charge packet's ignition delay times. The basis for the control algorithm is the following approximation of the ignition delay time for the Arrhenius kinetics from \eqref{eq:SwitchedArrheniusKinetics} as a function of the initial temperature, $T_0$, and fuel mass fraction, $Y_0$. For $T < \Tsw$ no reaction takes place and the ignition delay is formally infinite, whereas for $T > \Tsw$,
\begin{equation}\label{eq:AsymptoticIgnitionDelay}
\tign(T_0,Y_0) = \frac{1 + \frac{RT_0}{E} \left[2 + \frac{c_v T_0}{Q Y_0}\right]}{B \frac{E}{R T_0} \exp\left(-\frac{E}{RT_0}\right) \frac{Q Y}{c_v T_0}}\,.
\end{equation}
%
%%
%\begin{equation}\label{eq:AsymptoticIgnitionDelay}
%\tign(T_0,Y_0) = \frac{\dss 1 + \frac{RT_0}{E} \left[2 + \frac{c_v T_0}{Q Y_0}\right]}{\dss B \frac{E}{R T_0} \exp\left(-\frac{E}{RT_0}\right) \frac{Q Y}{c_v T_0}} \qquad (T > \Tsw)\,.
%\end{equation}
%%
This formula is (re-)derived to leading and first order in the small parameter $\eps = \frac{RT_0}{E} \ll 1$ in section\,\ref{sapp:LAEForAutoignition} via large activation energy asymptotics. For the general approach and the original derivation see \cite{Kassoy1975}. 

\begin{figure}
\centering
\includegraphics[width=0.5\textwidth]{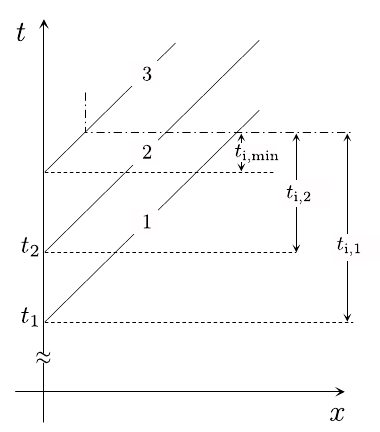}
\caption{\label{fig:ChargeControl}Illustration of the charge stratification control strategy for the SEC modus as described in section\,\ref{sapp:InjectionControlSEC}. The upper dash-dotted line indicates the location in space-time at which incoming parcels would autoignite under the idealized assumption of a constant flow velocity near the combustor entry at $x=0$. ``1'' denotes the first Lagrangian path along which the gas mixture is reactive (with stoichiometric equivalence ratio), and which entered the combustor at time $t_1$; ``2'' indicates some generic Lagrangian path of a parcel that entered the domain at time $t_2$ with $t_3 > t_2 > t_1$; ``3'' numbers the first path for which the minimum ignition delay time is prescribed to the fixed value of $t_{\text{i,3}} = t_{\text{i,min}}$.}
\end{figure}

Given the formula in \eqref{eq:AsymptoticIgnitionDelay}, we develop a control strategy for the combustor inlet temperature during the opening time interval of an SEC cycle. The reaction progress along Lagrangian paths of the  gas mixture is modelled as a near-constant volume combustion process, so that \eqref{eq:AsymptoticIgnitionDelay} would apply. Suppose the inlet valve has opened during an SEC cycle, and the inert air buffer has already passed into the combustor. At time $t_1$ the first Lagrangian path, ``$1$'', carrying a reactive gas mixture enters the tube (see Fig.\,\ref{fig:ChargeControl} for illustration). The ignition delay time for the associated first parcel of reactive mixture entering the tube during this cycle is $t_{\text{i},1} = \tign(T_1,1)$, where $T_1$ is the temperature in the inflow crossection of the combustor at time $t_1$. 

The injection control is to ensure that any parcel ``$2$'', which enters the combustor at some later time $t_2 > t_1$ but before the valve closes again, autoignites at approximately the same time as parcel ``$1$''. Assuming that the charge comes with a stoichiometric mixture, \ie, $Y_1 = Y_2 = 1$, we require that
\begin{equation}
\tign(T_2,1) = t_1 + t_{\text{i},1} - t_2\, .
\end{equation}
Aside from one exception, this determines the inlet temperature $T_2$ at time $t_2$. The exception results from the observation that, with the given formula, we would reach $\tign(T_2,1) \leq 0$ if the valve stayed open long enough, \ie, for sufficiently large $t_2$. This is avoided by introducing a lower cut-off, $t_{\text{i},\text{min}}$, for the ignition delay. In practice, the inlet temperature at time $t_2$ is therefore determined by solving 
\begin{equation}
F(T_2) :=  \tign(T_2,1) - t_{\text{i},2} = 0\,,
\qquad\text{where}\qquad
t_{\text{i},2} = \max\Bigl(t_{\text{i},\text{min}}, \,t_1 + t_{\text{i},1} - t_2\Bigr) = 0\,,
\end{equation}
using a damped Newton iteration. In all the simulations reported above we let $t_{\text{i},\text{min}} =  t_{\text{i},1}/10$. 

%
% ===================================================================================

\subsection{Large activation energy asymptotics for the auto-ignition delay time}
\label{sapp:LAEForAutoignition}

Here we provide a self-contained derivation of the ignition delay time formula in \eq{eq:AsymptoticIgnitionDelay}. When $T = T_0 > \Tsw$ initially, the homogeneous autoignition for the reactive gas model from \eq{eq:Q1DNS}--\eq{eq:Thermo} is governed by 
\begin{subequations}
\begin{align}
\frac{dY}{dt} 
  & = - B Y \exp\left(-\frac{E}{RT}\right)\,,
    &\hspace{-1cm}  Y(0) = Y_0 \,,
    \label{eq:Yeqn}\\
\frac{dT}{dt} 
  & = - \frac{Q}{c_v}\frac{dY}{dt}\,,
    & \hspace{-1cm} T(0) = T_0\,.
    \label{eq:Teqn}
\end{align}
\end{subequations}
Integrating the temperature equation \eqref{eq:Teqn}, we obtain
\begin{equation}
T - T_0 = - \frac{Q}{c_v} (Y-Y_0)\,,
\end{equation}
which we can use, together with \eqref{eq:Teqn}, to eliminate $Y$ from \eqref{eq:Yeqn}, so that
\begin{equation}
\frac{dT}{dt} 
= B \frac{Q}{c_v} \left(Y_0 - \frac{c_v}{Q} (T - T_0)\right) \exp\left(-\frac{E}{RT}\right)
\end{equation}
with $T(0) = T_0$ becomes a closed-form problem for the temperature alone. 

This problem is non-dimensionalised by introducing the following dimensionless parameters and the transformed dependent and independent variables $\theta$ and $\tau$, respectively, by
\begin{equation}
\begin{array}{r@{\ }c@{\ }l}
\dss \frac{T(t)}{T_0} 
  & = 
    & \dss 1 + \eps\theta(\tau) 
      \\
\dss B t
  & =
    & \dss \eps \exp\left(\frac{1}{\eps}\right) q \, \tau
\end{array}\,,
\ \ \text{where} \ \
\begin{array}{r@{\ }c@{\ }l}
\dss q 
  & = 
    & \dss \frac{c_v T_0}{Q Y_0}
      \\[10pt]
\dss \eps
  & =
    & \dss \frac{R T_0}{E} \ll 1 
\end{array}\,.
\end{equation}
These definitions emphasize that we are interested in identifying a time at which the temperature will deviate significantly from its initial value, \ie, at which $\theta \gg 1$, and they lead to the transformed problem
\begin{equation}
\frac{d\theta}{d\tau}
= \left(1 - \eps q \, \theta\right) \exp\left(\frac{\theta}{1+\eps\theta}\right)\,,
\qquad \theta(0) = 0\,. 
\end{equation}
As long as $(1 - \eps q \theta) > 0$, the dependence of $\theta$ on $\tau$ is monotonous, and by exchange their roles we obtain
\begin{equation}
\frac{d\tau}{d\theta}
= \frac{1}{1 - \eps q \, \theta} \exp\left(-\frac{\theta}{1+\eps\theta}\right)\,,
\qquad \tau(0) = 0\,. 
\end{equation}
Asymptotic expansion for $\eps \ll 1$ using
\begin{equation}
\tau(\theta) = \tau\order{0}(\theta) + \eps \tau\order{1}(\theta) + \littleoh{\eps}\, ,
\end{equation}
and 
\begin{equation}
\frac{1}{1 - \eps q \, \theta} e^{-\frac{\theta}{1+\eps\theta}}
= \Bigl(1 + \eps \left[q \theta +  \theta^2\right] \Bigr) e^{-\theta}  + \littleoh{\eps}\,,
\end{equation}
and collecting like powers of $\eps$ in the expanded equations as usual yields, 
\begin{subequations}
\begin{align}
\frac{d\tau\order{0}}{d\theta} 
  & = \exp(-\theta)
    && \hspace{-1cm}\tau\order{0}(0) = 0\,,
    \\
\frac{d\tau\order{1}}{d\theta} 
  & = \Bigl(q\theta + \theta^2\Bigr)\exp(-\theta)
    && \hspace{-1cm}\tau\order{1}(0) = 0\,.
\end{align}
\end{subequations}
These equations can be integrated analytically to find
\begin{subequations}
\begin{align}
\tau\order{0}(\theta) 
  & = 1 - e^{-\theta}\,,
    \\
\tau\order{1}(\theta) 
  & = (q + 2)\Bigl(1 - (1+\theta) e^{-\theta}\Bigr) - \theta^2 e^{-\theta} \,.
\end{align}
\end{subequations}
Clearly, for $\theta \gg 1$ these solutions approach finite limits, so that
\begin{equation}
\tau_{\text{ign}} = 1 + \eps (2+q) + \littleoh{\eps}\,,
\end{equation}
which after re-introducing dimensional expressions yields the approximation of the ignition delay time from \eqref{eq:AsymptoticIgnitionDelay}.

% ===================================================================================
% ===================================================================================

%\subsection{One-step-Arrhenius with switching temperature}
%\label{sapp:ArrheniusBerkenboschModel}

% ===================================================================================
% ===================================================================================

%\subsection{Other kinetics ... ?}
%\label{sapp:OtherKinetics}

% ===================================================================================
% ===================================================================================
% ===================================================================================

\section{Discretization of the quasi 1D Euler equations}
\label{app:Quasi1DEulerDiscretization}

% ===================================================================================
% ===================================================================================

\subsection{Finite volume scheme for compressible gasdynamics}

The following discretization of the momentum source term due to cross-sectional area variation of a flow tube in a quasi one-dimensional Euler model was - to the best of our knowledge - first spelled out by \citet{Tornow2021}. Here we provide a brief description for the non-reacting case. The extension to reactive flow is trivial within the present scheme as the reaction source terms and diffusive terms are separated from the Euler flow step by operator splitting according to \citet{Strang1968}, see Tab.~\ref{tab:CombustorNumerics} for a summary of the applied numerical discretization schemes.

Thus, we consider the quasi one-dimensional Euler equations, 
\begin{subequations}\label{eq:Q1DEuler}
\begin{align}
(\rho A)_t + (\rho u A)_x
  & = 0
    \\ 
\label{eq:Q1DEulerMomemtum}
(\rho u A)_t + ([\rho u^2 +p] A)_x
  & = p \, A_x \,,
    \\ 
(\rho E A)_t + ([\rho E + p] u A)_x
  & = 0
    \\
 E = U  + \frac{u^2}{2} = \frac{1}{\kappa-1} T + \frac{u^2}{2}\,,
 \ \  
 T 
  & = \frac{p}{\rho}\,,
\end{align}
\end{subequations}
where $(\rho, u, p, E, U)$ are density, velocity, pressure, total specific energy, and specific internal Energy, respectively, and $\kappa$ is the isentropic exponent of the gas. Our second order MUSCL solver utilizes standard piecewise linear slope-limited reconstruction in the characteristic (perturbation) variables as described, e.g., in \citet[section 16.3.3]{LeVequeBook1990}. This yields reconstructed states $\mathbf{U}_l, \mathbf{U}_r$ at each grid cell interface, where $\mathbf{U} = (\rho, \rho u, \rho e)$. The HLLE-M approximate Riemann solver by B.~Einfeldt \emph{including}, in particular, his elegant and efficient anti-diffusive correction in the linear degenerate characteristics \citep[section 3, eq.~(3.15)]{Einfeldt1988b}, is then used as a numerical flux function. This version of Einfeldt's solver, by the way, predates similar updates of the original HLL solver by \citet{HartenEtAl1983}, such as the HLLC \citep{Toro1992}, by at least four years and seems to have been largely ignored in the literature.

In the present context of discretizing the non-conservative source term on the right hand side of the momentum equation \eq{eq:Q1DEulerMomemtum} there is one interesting aspect of the HLL family of schemes: The definition of the numerical flux ${\widehat f}$, including Einfeldt's correction for the advective characteristic as the last term, reads
\begin{equation}\label{eq:EinfeldtFlux}
{\widehat f}(\mathbf{U}_l,\mathbf{U}_r) = \frac{b^+ f(\mathbf{U}_l) - b^- f(\mathbf{U}_r)}{b^+ - b^-}
 + \frac{b^+ b^-}{b^+ - b^-} \left(1 - \eta\,  R^2 \circ l^2\right) (\mathbf{U}_r - \mathbf{U}_l)\,,
\end{equation}
with $b^{\pm}$ Einfeldt's estimates of the fastest/slowest signal speeds emerging from the Riemann problem with $\mathbf{U}_r, \mathbf{U}_l$ as the right and left initial states of a Riemann problem, $f(\mathbf{U}) = (\rho u, \rho u^2 + p, u [\rho e + p])^T$ the exact flux function of the system, and $R^2$ and $l^2$ the right and left eigenvectors of a local approximation of the exact flux Jacobian matrix, such as the Roe-matrix \citep{Roe1981a} for the states $(\mathbf{U}_r,\mathbf{U}_l)$. Close inspection of this formula shows that the momentum component of ${\widehat f}(\mathbf{U}_l,\mathbf{U}_r)$ additively decomposes into two terms that represent the advective momentum flux $\rho u^2$ and the pressure $p$ of the exact momentum flux. Thus, in particular, an explicit expression for the pressure at each grid cell interface is available, and reads
\begin{equation}
{\widehat p}(\mathbf{U}_l,\mathbf{U}_r) = \frac{b^+ p(\mathbf{U}_l) - b^- p(\mathbf{U}_r)}{b^+ - b^-}\, .
\end{equation}
The second component of the second additive term in \eq{eq:EinfeldtFlux} only contributes to the advective momentum transport and thus does not feature here. With an explicit value of the cell interface pressure at hand, there is a natural way of discretizing the momentum source term for the $i$-th grid cell, \emph{viz.}
\begin{equation}
(p A_x)_i = {\widehat p}_{i}
\frac{A_{i+\half} + A_{i-\half}}{(\Delta x)_i}\, ,
\qquad\text{where}\qquad
{\widehat p}_{i} = \frac{{\widehat p}_{i+\half} + {\widehat p}_{i-\half}}{2}\, ,
\end{equation}
and this is straightforwardly integrated in the full momentum update of the solver by letting
\begin{equation}
\frac{(\rho u A)^{n+1}_{i} - (\rho u A)^n_{i}}{(\Delta t)^{n+\half}}
=
- \frac{F^{\rho u}_{i+\half} - F^{\rho u}_{i-\half}}{(\Delta x)_i}\, ,
\qquad\text{where}\qquad
F^{\rho u}_{i+\half} = \left[(\widehat{\rho} \widehat{u}^2 + {\widehat p})_{i+\half} - {\widehat p}_{i}\right] A_{i+\half}
\end{equation}
and an analogous formula holds for $F^{\rho u}_{i-\half}$. 
This implementation has the advantage of automatically maintaining a state of rest in a tube with crossectional area variations \citep{Tornow2021}.

%to represent the contributing physical processes are listed in Tab.~\ref{tab:CombustorNumerics}. Standard Strang splitting \citep{Strang1968} is used to couple the physical process representations. 
%
\begin{table*}
  \begin{center}
  \caption{\label{tab:CombustorNumerics} Numerical methods used to represent participating physical processes}
  \begin{tabular}{l|l}
  physical process\rule[-8pt]{0pt}{25pt}
    & numerical method 
      \\
  \hline
  inviscid gasdnamics \rule[-8pt]{0pt}{25pt}
    & 2nd order Godunov-Type MUSCL-scheme
      \\
  (incl.\ tube area variation)
    & (see Appendix~\ref{app:Quasi1DEulerDiscretization}) \rule[-8pt]{0pt}{15pt}
      \\
  molecular transport \rule[-8pt]{0pt}{15pt}
    & 2nd order central in space, explicit mid-point in time 
      \\
  reaction \rule[-8pt]{0pt}{15pt}
    & semi-analytic/frozen temperature \citep{BourliouxMajda1992}
      % (Arrhenius-K.) oder Radau-Verf.
  \end{tabular}
  \end{center}
  \end{table*}

% ===================================================================================
% ===================================================================================

\subsection{Boundary conditions}
\label{app:BoundaryConditions}

Boundary conditions for the quasi one-dimensional model are implemented by means of ghost cells with the aim to make as much use as possible of the capabilities of the second order MUSCL solver of representing local unsteady flow evolutions through its numerical flux function based on an approximate Riemann-solver.
\subsubsection{Combustor inlet state}
The velocity in the ghost cell adjacent to the first cell of the combustor is determined under the assumption of an isotropic expansion from the compressor plenum pressure to the pressure of the first grid cell of the combustor:
\begin{equation}
  \label{eq:ghostU}
   u_{gh} = \sqrt{\frac{2 \kappa}{\kappa - 1} R T 
   \left[ 
      1 - \left(\frac{p_{\gh}}{p_{2}}\right)^{\frac{\kappa -1}{\kappa}}
   \right]},  
\end{equation}
where $p_2$ is the compressor plenum pressure and $p_{\gh} = p_{1^{\textnormal{st}}}$ with 
$1^{\textnormal{st}}$ indicating states in the first grid cell. The remaining ghost cell values (density, temperature, mass-fractions) are set according to the operating mode of the engine:

\medskip\noindent
\textbf{Deflagration mode}\medskip

In deflagration mode, density and temperature are determined asssuming an adiabatic expansion from the compressor plenum state to the pressure of the first grid cell:
\begin{equation}
   \frac{T_{\gh}}{T_2} = \left(\frac{p_{\gh}}{p_2}\right)^{\frac{\kappa -1}{\kappa}},
   \quad
   \frac{\rho_{\gh}}{\rho_2} = \left(\frac{p_{2}}{p_{\gh}}\right)^\kappa,
\end{equation}
mass fractions are set according to the prescribed equivalence ratio, which is assumed to be one (stoichiometric conditions) for this study: $Y_{F,gh} = Y_{F,st}$.

\medskip\noindent
\textbf{SEC mode}\medskip

As long as the compressor plenum pressure is higher than the pressure in the first cell of the combustor domain we have an inflow boundary condition. Velocity and pressure in the ghost cell are determined as in deflagration mode. Otherwise, we implement a fixed wall boundary condition representing a perfectly closed valve at the inlet. When the pressure in the first cell of the combustor domain drops below the compressor plenum pressure, the valve opens and we are back in inflow condition. Directly after valve opening an air buffer is introduced by setting $Y_{F,gh} = 0$ while the temperature $T_{\gh}$ of the inflowing fuel-air stream in the ghost cell is adjusted such that the mixture ignites almost homogeneously as described in App.\,\ref{app:InjectionControl}.

\subsubsection{Combustor exit/turbine plenum}
The turbine plenum adjacent to the exit of the combustor is represented as a single additional grid cell with corresponding volume. Fluxes between combustor and turbine plenum are calculated by the Riemann solver setting cell states in the last cell of the combustor and the plenum cell according to Table \ref{tab:TurbinePlenumGhostCellData}. The physical reasoning for this procedure is outlined in section \ref{sec:TurbinePlenumAndTurbineModel}.

\begin{table*}
\caption{Case-dependent models for the flow state in the turbine plenum adjacent to the combustor port. $(p,T,u)$: pressure, temperature, and velocity in the last grid cell representing the combustor, $p_2, T_2$: current turbine plenum pressure and temperature. Perfect gas relations are used to describe adiabatic processes. Chemical species mass fractions are extrapolated from the last combustor cell values $Y_i$ when the ghost cell velocity $u_{\text{gh}} > 0$, while turbine plenum mean values  $Y_{i,2}$ are imposed otherwise. \label{tab:TurbinePlenumGhostCellData}}
\hfil
\begin{tabular}{l|l}
 \rule[-8pt]{0pt}{25pt}
  & Ghost cell flow states $(\rho, p, u)_{\text{gh}}$ representing the turbine plenum 
    \\
\hline
$p > p_2, \ u > 0$ \rule[-8pt]{0pt}{25pt}
  & $u_{\text{gh}} = u, \ p_{\text{gh}} = p, \ \rho_{\text{gh}} = \rho (p_2/p)^{\kappa}$
    \hfill (adiabatic expansion; no acceleration)
    \\
$p > p_2, \ u \leq 0$ \rule[-8pt]{0pt}{15pt}
  & $u_{\text{gh}} = 0, \ p_{\text{gh}} = p_2, \ \rho_{\text{gh}} = \rho_2$
    \hfill (ghost cell state = plenum state)
    \\
$p \leq p_2, \ u > 0$ \rule[-8pt]{0pt}{15pt}
  & $u_{\text{gh}} = \max(0,u_{\text{ad}}(u,p,T,p_2)), \ p_{\text{gh}} = p_2, \ \rho_{\text{gh}} = \rho_2$
    \hfill (adiabatic deceleration)
    \\
$p \leq p_2, \ u \leq 0$ \rule[-8pt]{0pt}{15pt}
  & $u_{\text{gh}} = u_{\text{ad}}(0,p_2,T_2,p), \ p_{\text{gh}} = p_2, \ \rho_{\text{gh}} = \rho_2$
    \hfill (adiabatic expansion)
    \\
  & $u_{\text{ad}}(u_0,p_0,T_0,p_1) = \pm \sqrt{u_0^2 - \frac{2 \kappa}{\kappa-1} T_0 \left(1-\left(\frac{p_1}{p_0}\right)^{\frac{\kappa-1}{\kappa}}\right)}$\rule[-15pt]{0pt}{15pt}
    \\
\hline
 \rule[-8pt]{0pt}{25pt}
  & Ghost cell values for species mass fractions $Y_{i,\text{gh}}\ \ (i = 1, ..., i_{\text{spec}})$ 
    \\
\hline
\hfill $u_{\text{gh}} > 0$ \rule[-8pt]{0pt}{25pt}
  & $Y_{i,\text{gh}} = Y_i$
    \hfill (extrapolation from combustor exit)
    \\
\hfill $u_{\text{gh}} \leq 0$ \rule[-8pt]{0pt}{15pt}
  & $Y_{i,\text{gh}} = Y_{i,2}$
    \hfill (extrapolation from combustor exit)
\end{tabular}
\end{table*}

% ===================================================================================
% ===================================================================================
% ===================================================================================

\section{Efficiency of the Humphrey cycle}
\label{app:Humphrey}

\begin{figure}[htbp]
\begin{center}
\includegraphics[width=0.5\columnwidth]{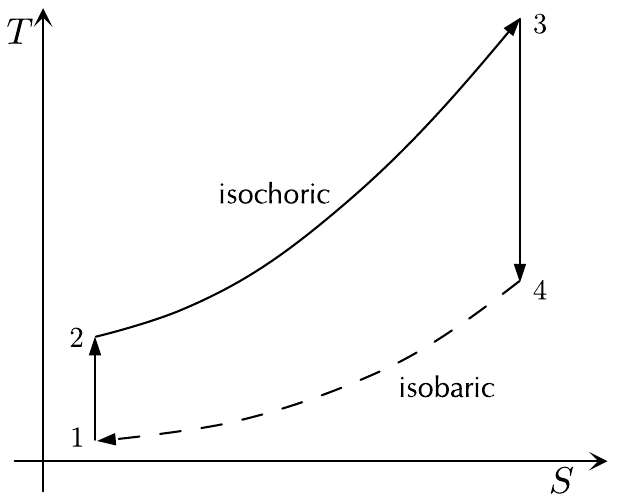}\hfil
\caption{The Humphrey cycle ($1$--$2$--$3$--$4$--$1$) in the temperature-entropy ($T$-$S$) diagram. }
\label{fig:HumphreyCycleTS}
\end{center}
\end{figure}

For the readers' convenience, we summarize here the well-known derivation of an explicit formula for the efficiency of the Humphrey cycle (see fig.~\ref{fig:ThermodynamicCycles}). The process combines isochoric, isobaric, and isentropic sub-processes. To prepare for the evaluation of the efficiency using the trace of the cycle in the $T$-$S$-diagram (see fig.~\ref{fig:HumphreyCycleTS}) as 
\begin{equation}
\eta_H = \frac{\oint T\, \textnormal{d}S}{\int_{S_1}^{S_4} T\, \textnormal{d}S} = \frac{\text{work done}}{\text{heat added}}\,,
\end{equation}
we recall that, for the perfect gas assumed throughout this paper, 
\begin{equation}
\frac{S-S_0}{c_v} = \kappa \ln\left(\frac{T/T_0}{(p/p_0)^{\Gamma}}\right) = \ln\left(\frac{T/T_0}{(\rho/\rho_0)^{\kappa\Gamma}}\right),
\end{equation}
where $c_p$ and $c_v$ are the specific heat capacities at constant pressure and at constant volume, respectively, $\kappa = c_p / c_v$, $\Gamma = \frac{\kappa-1}{\kappa}$, and the subscript $(\cdot)_0$ indicates some reference state. Thus, along the isochoric ($2\to 3$) and isobaric ($4\to 1$) branches of the $T$-$S$-diagram in fig.~\ref{fig:HumphreyCycleTS} and noting that $S_1 = S_2$ and $S_3 = S_4$, we have
\begin{equation}\label{eq:HeatAdded}
Q 
= \int_{S_1}^{S_4} T\, \textnormal{d}S 
= T_0 \left(\frac{\rho_2}{\rho_0}\right)^{\kappa-1} \int_{S_2}^{S_3} e^{\frac{S-S_0}{c_v}} \textnormal{d}S
= c_v T_0 \left(\frac{\rho_2}{\rho_0}\right)^{\kappa-1} \left(e^{\frac{S_3-S_0}{c_v}} - e^{\frac{S_2-S_0}{c_v}}\right)
= c_v (T_3 - T_2)\,,
\end{equation}
%
%%
%\begin{equation}\label{eq:HeatAdded}
%\begin{array}{r@{\ }c@{\ }l}
%Q 
%  & = 
%    & \dss \int_{S_1}^{S_4} T\, \textnormal{d}S 
%      = T_0 \left(\frac{\rho_2}{\rho_0}\right)^{\kappa-1} \int_{S_2}^{S_3} e^{\frac{S-S_0}{c_v}} \textnormal{d}S
%      \\
%  & = 
%    & \dss c_v T_0 \left(\frac{\rho_2}{\rho_0}\right)^{\kappa-1} \left(e^{\frac{S_3-S_0}{c_v}} - e^{\frac{S_2-S_0}{c_v}}\right)
%      \\
%  & =
%    & \dss c_v (T_3 - T_2)\,,
%\end{array}
%\end{equation}
%%
where $Q$ is the added chemical energy. Next, taking into account that $4 \to 1$ is isobaric instead of isochoric, we have
\begin{equation}
L = \int_{S_4}^{S_1} T\, \textnormal{d}S = c_p (T_1 - T_4)\,,
\end{equation}
with $L$ the heat lost to the environment. As an intermediate result, this yields
\begin{equation}\label{eq:EtaH-Tbased}
\eta_H 
= \frac{Q - L}{Q} = \frac{c_v (T_3 - T_2) - c_p (T_4 - T_1)}{c_v (T_3 - T_2)}
= 1 - \kappa \frac{T_4 - T_1}{T_3 - T_2}\,.
\end{equation}

For the purposes of this paper, we express $\eta_H$ as a function of the pre-compression ratio $p_2/p_1$ and the dimensionless chemical energy 
\begin{equation}\label{eq:Qstar}
Q^* = \frac{Q}{c_v T_1} = \frac{T_3}{T_1} - \frac{T_2}{T_1}\,. 
\end{equation}
To this end, we observe that (using again $\Gamma = \frac{\kappa-1}{\kappa}$)
\begin{subequations}
\begin{align}
\frac{T_2}{T_1} 
  & = \left(\frac{p_2}{p_1}\right)^{\Gamma}\,,
    \\
\frac{T_4}{T_3} 
  & = \left(\frac{p_4}{p_3}\right)^{\Gamma} 
    = \left(\frac{p_1}{p_3}\right)^{\Gamma}
    = \left(\frac{p_1}{p_2}\right)^{\Gamma} \left(\frac{p_2}{p_3}\right)^{\Gamma}
    = \frac{T_1}{T_2} \left(\frac{T_2}{T_3}\right)^{\Gamma}.
\end{align}
\end{subequations}
The first two equalities on this line just recall the pressure-temperature relation for isentropes, the third uses that $4\to 1$ is isobaric, the fourth comes from multiplication by unity, the fifth utilizes that $1\to 2$ is an isentrope and that $2\to 3$ is isochoric. Multiplying the relation for $T_4/T_3$ by $T_3/T_1$ and using that \eqref{eq:HeatAdded} amounts to $T_3 = T_2 + Q/c_v$, we obtain
\begin{equation}
\frac{T_4}{T_1}
= \left(\frac{T_3}{T_2}\right)^{1- \frac{\kappa-1}{\kappa}}
= \left(\frac{T_3}{T_2}\right)^{\frac{1}{\kappa}}
= \left(1 + \frac{Q}{c_v T_2}\right)^{\frac{1}{\kappa}}
= \left(1 + Q^*\frac{T_1}{T_2}\right)^{\frac{1}{\kappa}}.
\end{equation}
Finally, inserting into \eqref{eq:EtaH-Tbased}, we have the following expression for the efficiency of the Humphrey cycle,
\begin{equation}
\eta_H = 1 - \frac{\kappa}{Q^*} \left(\left(1 + Q^* \left(\frac{p_1}{p_2}\right)^{\frac{\kappa-1}{\kappa}}\right)^{\frac{1}{\kappa}} - 1\right).
\end{equation}
%

% ===================================================================================
% ===================================================================================
% ===================================================================================

\section{Engine efficiency estimate from states along parcel trajectories}
The equation defining the specific entropy, $S$, and temperature, $T$, the latter as an integrating factor, reads
\begin{align}
\textnormal{d}U + p\, \textnormal{d}V = T\, \textnormal{d}S\,,
\end{align}
where $U$ is the specific internal energy, $p$ the pressure and $V = 1/\rho$ the specific volume. Integration of any state variable, say $U$, along a cyclic process we have $\oint_C \textnormal{d}U \equiv 0$, and therefore
\begin{equation}
\oint_C T\,\textnormal{d}S = \oint_C \textnormal{d}U + p\,\textnormal{d}V = \oint_C p\,\textnormal{d}V \,.
\end{equation}
Thus, $\oint_C T\,dS$ is the work \emph{per unit mass} which a parcel of the fluid delivers to its environment in the course of one cycle of the process. Interpreting the fresh charge loaded into the combustor during a single opening of the unidirectional valve as a sequence of parcels with total mass $M$, we calculate the total work delivered by the charge taken up during the $k$th valve opening upon passage through the engine as
\begin{equation}
W_k^{\text{out}} = \int_0^{M_k} \left(\oint_C T\,\textnormal{d}S\right)(m) \, \textnormal{d}m\, .
\end{equation}
The uptake of chemical energy by that charge is 
\begin{equation}
E_k^{\text{in}} = \int_0^{M_k} Q\Phi(m) \, \textnormal{d}m\, ,
\end{equation}
where $Q$ is the chemical energy per unit mass of stoichiometric mixture and $\Phi(m)$ is the equivalence ratio of the fluid parcel labelled by the total mass $m$ that has entered the combustor through the inlet valve since the $k$th valve opening.

The parcel-based efficiency estimate of the engine, neglecting losses in compressor and turbine as before, then becomes
\begin{equation}
\eta_k^{\text{pcl}} 
= \frac{W_k^{\text{out}}}{E_k^{\text{in}}}
= \frac{\int_0^{M_k} \left(\oint_C T\,\textnormal{d}S\right)(m) \, \textnormal{d}m}{\int_0^{M_k} Q\Phi(m) \, \textnormal{d}m}\,.
\end{equation}
This is to be distinguished from the average of the individual parcel efficiencies, which would be
\begin{equation}
\overline{\eta_k^{\text{pcl}}} 
= \frac{1}{M_k} \int\limits_0^{M_k} \frac{\left(\oint_C T\,\textnormal{d}S\right)(m)}{Q\Phi(m)}  \, \textnormal{d}m\,.
\end{equation}
%

\bibliographystyle{abbrvnat}
\bibliography{literature}   % name your BibTeX data base

\end{document}